\documentclass[pre,groupedaddress,twocolumn,floatfix]{revtex4}
\usepackage[reqno,intlimits]{amsmath}
\usepackage{graphicx}
\usepackage{amsmath}
\usepackage{amssymb}
\usepackage{texdraw}
\usepackage{epic}
\usepackage{eepic}
\usepackage{color}
\usepackage{fancybox}

\newcommand{\erf}{\ensuremath{\text{erf}}}    
\newcommand{\half}{\ensuremath{\frac{1}{2}}}  

\newcommand{\ave}[1]{\ensuremath{\langle #1 \rangle}}

\newcommand{\floatfig}[4][]{
  \begin{figure}[#1]
	#4
	\caption{#3}
	\label{#2}
  \end{figure}
}

\newcommand{\ea}{et al.}
\newcommand{\at}{\ensuremath{n^{+1}_t}}        
\newcommand{\genat}[1]{\ensuremath{n^{+1}_{#1}}} 
\newcommand{\jea}{Johnson \ea }
\newcommand{\ind}[1]{\ensuremath{#1_{i;t}}}    
\newcommand{\f}{\text{if }}  		       
\newcommand{\pbar}{\ensuremath{\overline p_t}} 
\newcommand{\pval}{gene value}                 
\newcommand{\fol}{\ensuremath{n^{+h_t}_t}}     
\newcommand{\mc}{\mathcal}
\newcommand{\mv}{\mathbf}                      
\newcommand{\opt}{\ensuremath{\text{opt}}}     

\renewcommand{\floatfig}[5][]{
  \begin{figure}[#1]
        #5
        \caption[#3]{#4}
        \label{#2}
  \end{figure}
}
\newcommand{\bigfloatfig}[5][]{
  \begin{figure*}[#1]
        #5
        \caption[#3]{#4}
        \label{#2}
  \end{figure*}
}

\newcommand{\floatplace}{htb}

\begin{document}

\title{Memory and self-induced shocks in an evolutionary population competing
for limited resources}
\author{Roland Kay}\email{roland.kay@physics.ox.ac.uk}
\author{Neil F. Johnson}
\affiliation{Physics Department, Oxford University, Parks Road, Oxford, OX1 3PU,
U.K.}

\begin{abstract}
We present a detailed discussion of the role played by memory, and the nature
of self-induced shocks, in an evolutionary population competing for limited
resources. Our study builds on a previously introduced multi-agent system [Phys.
Rev. Lett {\bf 82}, 3360 (1999)] which has attracted significant attention in
the literature. This system exhibits self-segregation of the population based on
the `gene' value $p$ (where $0\leq p\leq 1$), transitions to `frozen'
populations as a function of the global resource level, and self-induced large
changes which spontaneously arise as the dynamical system evolves. We find that
the large, macroscopic self-induced shocks which arise, are controlled by
microscopic changes within extreme subgroups of the population (i.e. subgroups
with `gene' values $p\sim 0$ and $p\sim 1$).
\end{abstract}

\date{\today}
\maketitle

%
\newcommand{\gauss}{\ensuremath{\Phi_\mu^\sigma(x)}}    
\newcommand{\set}[1]{\ensuremath{\{#1_{i;t}\}_i}}  
\newcommand{\gu}{\ensuremath{\mc U_t}}       
\newcommand{\std}[2]{\ensuremath{\sigma(#1,#2)}} 
\newcommand{\autoc}[2]{\ensuremath{\mathcal C_{#2}\!\!\!\:\left(#1\right)}}

\section{Introduction}

The dynamical behavior of a population of objects or `agents' (e.g. software or
hardware modules, cellular organisms such as bacteria or viruses, human beings,
animals) is of interest across a range of disciplines. Physics is arguably
luckier than most disciplines in that the `agents' of interest (i.e. particles)
do not adapt their behavior according to past failure, hence evolving new sets
of rules as time progresses. Nor do the agents in question have any individual
memory. Biological and social disciplines are not so lucky. Through a desire to
develop a minimal model which could incorporate such features into a manageable
yet non-trivial system, Arthur introduced the so-called `El Farol' Bar Problem,
which concerns the repeated competition between bar-goers to attend a popular
bar with limited seating \cite{bar-orig}. Challet and Zhang
\cite{challet-orig-mg} subsequently introduced a binary version of this bar
problem for the case where the amount of resource (e.g. number of seats) is just
less than half the number of agents (e.g. possible attendees). This system is
referred to as the Minority Game. 

The Minority Game does not allow an agent to continuously evolve new strategies
and hence explore the entire strategy space. The Minority Game is also
essentially deterministic, apart from occasional coin-tosses which are used to
break ties in strategy scores. Furthermore the resource level is set at just
less than half the number of agents, so that there are always more losers than
winners. To help overcome these limitations, \jea\ introduced a stochastic
version of the Minority Game\cite{self-seg} which is subsequently referred to as
the Genetic Model, in which an agent's strategy (characterized by a `gene' value
$p$) can evolve indefinitely in time, and is in principle allowed to access the
entire space of strategies (i.e. all $p$ values). The resulting Genetic Model
has provoked much interest in the literature (for example see Refs.
\cite{evol-freeze,ceva-mem,ceva-thermal,ceva-quench,emg-cluster,emg-phase}).  As
commented in the original paper of Johnson \ea\cite{self-seg} and confirmed by
Burgos \ea\cite{ceva-mem,ceva-quench}, the self-segregation observed in the
Genetic Model is insensitive to changes in an agent's memory length $m$.  

In this paper we  present a detailed discussion of the role of memory in the
Genetic Model. We also explain the origin of the remarkable step-like structure
in the global output time series as a function of the resource level, which was
first observed in by \jea\ in Ref. \cite{evol-freeze}.  We then introduce (Sec.
\ref{ch:large-changes}) a new variant of the Genetic Model in which the number
of agents competing at a given time step, is allowed to fluctuate. Because of
the analogy with the Grand Canonical Ensemble in physics, we shall refer to this
model as the `Grand Canonical Genetic Model' (GCGM) By considering versions of
the GCGM both with and without memory, we shall investigate the
endogenous (i.e. self-induced) large changes which arise in the system. These
large changes represent abrupt macroscopic `shocks', and occur with a greater
probability than would be expected based on random walk statistics.  We provide
a detailed analysis of the mechanism that generates these large changes.

\section{Effects of memory in the Genetic Model}
\label{ch:gm-memory}

Various papers\cite{evol-freeze,self-seg,ceva-mem,ceva-quench} have made claims
with regard to the role of memory in the Genetic Model. To date though, no one
has performed a detailed analysis of this problem. In this section we present
such an analysis which involves comparing the behavior of the original model
with that of a memoryless variant.
The results presented here for the Genetic Model are reminiscent of earlier
results for the Minority Game. In particular, Hart \ea\cite{crowd-effect} showed
that a crowd-anticrowd theory which assumes {\em random} history, provides a
quantitative description of the time-averaged fluctuations in the Minority Game.
Subsequently Cavagna \cite{cavagna-mgmem} demonstrated numerically that the
time-averaged fluctuations were indeed largely unaffected if the global history
was replaced with randomly generated data.

Ceva and Burgos \cite{ceva-mem} investigated explicitly the role of memory in
the Genetic Model, however the results are restricted to a comparison of the
\pval\ distributions in the minority case, in which the amount of resource is
just less than half the number of agents. In contrast, in this section we shall
treat the general case, in which the amount of resource is unrestricted, and
provide some theoretical analysis to explain the differences that we shall
observe between the two models. We shall show in Secs.  \ref{sec:hbar-3a} and
\ref{sec:dev-3a} that the observation that the \pval\ distributions are
identical only holds in the special case considered in Ref. \cite{ceva-mem}.
Furthermore, we shall show that the feedback
introduced
by the existence
of memory does influence the behavior of the model in that it controls the
time-average of the prediction (Sec. \ref{sec:hbar-3a}) and introduces
autocorrelations into the mean attendance time series (Sec.
\ref{sec:autoc-3a}).
By comparing and contrasting the memoryless Genetic Model with
the original model we shall be able to make some important observations about
the memory's true significance. We shall also answer some important questions as
to the extent to which memory is of benefit to the agents and the system as a
whole.

\begin{table}
\begin{center}
\begin{tabular}{|l l c l l|}
	\hline
	Number of agents: &$N=501$ & \hspace{3mm} &
	Memory length:    &$m=4$ \\
	Death score:      &$D=4$ & &
	Mutation range:   &$r=0.2$ \\
	\hline
\end{tabular}
\end{center}
\caption{Parameters used to generate numerical data.
}
\label{tab:params-3a}
\end{table}

Before going on to describe the Genetic Model, we note that all of the numerical
results presented in this paper were obtained using the model parameters listed
in Table \ref{tab:params-3a} unless otherwise stated. Similarly all time
averages are taken over the period $10000<t<60000$. The first 10 000 time steps
being neglected to allow any transients due to the initial conditions to die
away.

\subsection{Original Genetic Model}
\label{sec:tradintro-3a}

In this section we present expressions for some of the most basic quantities
in the original Genetic Model, which includes the memory. In Sec. 
\ref{sec:memless-3a} we shall consider the equivalent expressions in the 
memoryless variant.
We start with a brief summary of the Genetic Model. Fuller details are given in
Ref. \cite{self-seg}.

The Genetic Model consists of a population of agents who must decide at
every time step between two possible choices. We shall refer to the decision of
an agent $i$ as its \emph{action}, \ind{a}. Each agent is defined by a
\emph{gene value}, \ind{p}, which can take any value $0 \leq \ind{p} \leq 1$. At
each time step the model makes a \emph{prediction}, $h_t$, of the outcome of the
time step available to all of the agents.  Each agent chooses its action to be
equal or opposite to this prediction with probabilities $\ind{p}$ and
$1-\ind{p}$ respectively. $h_t$ is calculated based on a global memory that
the model maintains of the outcome of the previous $m$ time steps and the
assumption that patterns that have occurred in the time series of these outcomes
in the past will recur in the future. The parameter $m$ is known as the
\emph{memory length}.
The outcome of each time step is determined based on the actions of all of the
agents. From now on we shall refer to the outcome as the \emph{global action}
at time $t$, $A_t$.

Agent gene values are not constant with time. Each agent maintains a record of
its \emph{score}, \ind{s}, which determines when it changes its gene value. At
every time step, \ind{s} increases by one unit if $\ind{a}=+A_t$ and decreases
otherwise. If $\ind{s}=-D$ then the agent \emph{mutates}. The parameter $D$ is
known as the \emph{death score}. When an agent mutates it chooses a new gene
value at random from a range of values of width $2r$ centered on the old gene
value. The parameter $r$ is known as the \emph{mutation range}.

Let the possible values taken by $h_t$ be $-1$ and $+1$.  Agent $i$ makes a
choice to follow the prediction ($\ind{a}=+h_t$) with probability $p_{i;t}$ or
to refute it ($\ind{a}=-h_t$) with probability $1-p_{i;t}$. 
There are two opposite definitions of the global action $A_t$ used in the
literature. For example, Ref. \cite{evol-freeze} defines $A_t$, by analogy with
Zhang and Arthur's Bar model, to be the state of the bar at time $t$. Thus,
$A_t=+1$ would denote an \emph{overcrowded} bar and the optimal action of each
agent would be to stay at home (i.e. $a_{i;t}=-1$). However, in this paper we
shall adopt the convention of Ref. \cite{self-seg} whereby $A_t$ represents the
optimal decision of each agent at time $t$.
The global action $A_t$ is given, in terms of a model parameter $l$, which can
take values $0\leq l\leq 1$,  by:
\begin{equation}
\label{eqn:Acond-3a}
	A_t = \begin{cases}
	        +1  &  \at \le Nl \\
		-1  &  \at >   Nl
	      \end{cases}
\end{equation}
where:
\begin{equation}
\label{eqn:at-3a}
	\at = \half\left( \sum\limits_i a_{i;t} + N \right) \;.
\end{equation}
In other words, \at\ is the number of agents for which $a_{i;t}=+1$. We shall
refer to $l$ as the \emph{resource level}.

Let \set p denote the set of values of $\ind p$ for all agents at time $t$.
It is easy to show that the ensemble average number of agents following the
prediction $\ave{\fol}$ is given by:
\begin{align}
\label{eqn:avefol_en-3a}
  \ave{\fol}=N\overline p_t && \text{where: } &&
  \pbar=\frac{1}{N} \sum\limits_{i=1}^N \ind p \; .
\end{align}
In equilibrium, where the population evolves such that $\overline p_t$ is
approximately constant, the ensemble average $\ave{\fol}$ and the time average
$\ave{\fol}_t$ will coincide. Thus:
\begin{equation}
\label{eqn:avefol-3a}
  \ave{\fol}_t \approx N \ave{\overline p_t}_t  \; .
\end{equation}

\subsection{Memoryless Genetic Model}
\label{sec:memless-3a}

In this section we introduce a memoryless variant of the 
original
Genetic Model.
In contrast to the 
original
model described above, where the \pval\ of
the $i$th agent $p_{i;t}$ gives the probability of it choosing to follow
the prediction ($a_{i;t}=+h_t$), in the memoryless model $p_{i;t}$ gives the 
probability that $a_{i;t}=+1$ directly. With this modification the prediction
$h_t$ and hence the global memory that produce it become redundant and can be
removed from consideration. The agents in this variant are memoryless, by which
we mean that their actions $a_{i;t}$ at time $t$ are independent of the state of
the model at earlier times. \ind a is dependent only on \ind p.
An equivalent way of considering this is to take $h_t=+1\;\forall\;t$.
The global action $A_t$ and the number of agents attending the bar \at\ are
unchanged and so are given by Eqs. \eqref{eqn:Acond-3a} and \eqref{eqn:at-3a} as
before.
However, Eq. \eqref{eqn:avefol-3a} becomes:
\begin{align}
\label{eqn:aveat-ml-3a}
	\ave{\at}_t = N \ave{\overline p_t}_t  \; .
\end{align}

\subsection{Comparison of the performance of the original and memoryless
models} 
\label{sec:comptradml-3a}


We shall now compare the performance of the two models. In order to quantify
performance we define \gu\ to be the total number of points scored by all of the
agents at time $t$. Therefore we are considering the performance of the system
as a whole, rather than that of individual agents.  If we consider the models to
be analogous to an economic system then the question that we are investigating
becomes, to what extent can the agents in this system exploit the potential
wealth available to them as a population.  Note that in Ref. \cite{ceva-thermal}
Burgos \ea\ treat the memoryless Genetic Model in terms of a cost function given
by the second moment of \at\ with respect to $Nl$. However this cost function is
symmetric in that it assigns an equal cost to deviations of \at\ from $Nl$ of
opposite signs. As we shall see, \gu\ is not symmetric about $Nl$ and hence can
distinguish between positive and negative deviations.

\subsubsection{Original Genetic Model}
\label{sec:perftrad-3a}

First we shall derive expressions for \gu\ in the
model with memory. Later we will see how these expressions are modified in the
absence of the memory.
From Eq. \eqref{eqn:Acond-3a} the condition that $A_t=+1$ is:
\begin{align}
\label{eqn:folcond-3a}
	    \fol\ &\le Nl      & \text{if } h_t &= +1 \notag\\
	    \fol\ &\ge N(1-l)  & \text{if } h_t &= -1 \; .
\end{align}
Now consider the total number of points scored by the agents, \gu.
Agents for which $\ind a=+A_t$ will gain one point whereas agents for 
which $\ind a=-A_t$ will lose. If $h_t=+1$ and $A_t=+1$ it will be the \fol\
agents who choose to follow $h_t$ who will gain. If $A_t=-1$, then the
$N-\fol$ agents who choose to refute $h_t$ will gain. 
Thus, using Eq. \eqref{eqn:folcond-3a}, we have:

\smallskip
\noindent
For $h_t=+1$:
\begin{equation}
\label{eqn:guplus-3a}
	\gu(\fol) = 
	  \begin{cases}
	    2\fol - N		& \text{if } \fol \le Nl \\
	    -2\fol + N	    	& \text{if } \fol > Nl
	  \end{cases} \\  \;. 
\end{equation}
When $h_t=-1$ and $A_t=+1$ it will be the $N-\fol$ agents who choose to refute
the prediction who will gain and vice-versa for $A_t=-1$. Thus, for $h_t=-1$
the above expression becomes:

\smallskip
\noindent
For $h_t=-1$:
\begin{equation}
\label{eqn:guminus-3a}
	\gu(\fol) = 
	  \begin{cases}
	    -2\fol + N		& \text{if } \fol \ge N(1-l) \\
	     2\fol - N	    	& \text{if } \fol < N(1-l)
	  \end{cases} \\ \;.
\end{equation}

\bigfloatfig[\floatplace]{fig:tradutil-3a}{
Global utility as a function of \fol\ in the original Genetic Model.
}{
Global utility as a function of \fol\ in the original Genetic Model with
memory. The dashed lines indicate \gu\ when $h_t=-1$, the narrow lines \gu\ 
when $h_t=+1$. The emboldened lines represent those part of the line
defined by \gu\ which are invariant under $h_t\rightarrow -h_t$. The black
and white circles represent the values of \fol\ at which \gu\ is a maximum when
$h_t = +1$ and $-1$ respectively.
}{
\begin{center}
\resizebox{0.96\textwidth}{!}{
\begin{picture}(0,0)%
\includegraphics{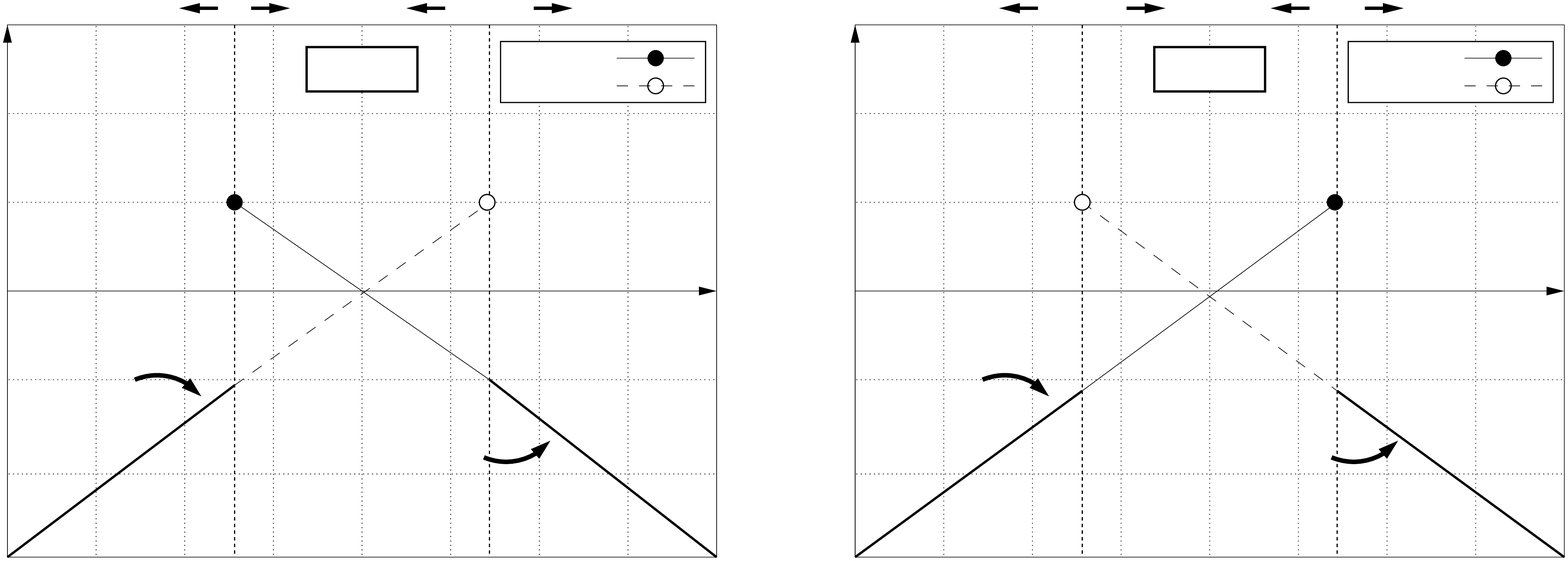}%
\end{picture}%
\setlength{\unitlength}{3947sp}%
\begingroup\makeatletter\ifx\SetFigFont\undefined%
\gdef\SetFigFont#1#2#3#4#5{%
  \fontsize{#1}{#2pt}%
  \fontfamily{#3}\fontseries{#4}\fontshape{#5}%
  \selectfont}%
\fi\endgroup%
\begin{picture}(21750,7604)(601,-7036)
\put(1126,164){\makebox(0,0)[rb]{\smash{\SetFigFont{20}{24.0}{\sfdefault}{\mddefault}{\updefault}{\color[rgb]{0,0,0}$+N$}%
}}}
\put(1126,-7036){\makebox(0,0)[rb]{\smash{\SetFigFont{20}{24.0}{\sfdefault}{\mddefault}{\updefault}{\color[rgb]{0,0,0}$-N$}%
}}}
\put(976,-3436){\makebox(0,0)[lb]{\smash{\SetFigFont{20}{24.0}{\sfdefault}{\mddefault}{\updefault}{\color[rgb]{0,0,0}$0$}%
}}}
\put(4276,389){\makebox(0,0)[b]{\smash{\SetFigFont{20}{24.0}{\sfdefault}{\mddefault}{\updefault}{\color[rgb]{0,0,0}$Nl$}%
}}}
\put(7801,-5536){\makebox(0,0)[rb]{\smash{\SetFigFont{20}{24.0}{\sfdefault}{\mddefault}{\updefault}{\color[rgb]{0,0,0}$\mc U_t=N-2\fol$}%
}}}
\put(10876,-3436){\makebox(0,0)[lb]{\smash{\SetFigFont{20}{24.0}{\sfdefault}{\mddefault}{\updefault}{\color[rgb]{0,0,0}$\fol$}%
}}}
\put(10276,-3736){\makebox(0,0)[lb]{\smash{\SetFigFont{20}{24.0}{\sfdefault}{\mddefault}{\updefault}{\color[rgb]{0,0,0}$N$}%
}}}
\put(7726,389){\makebox(0,0)[b]{\smash{\SetFigFont{20}{24.0}{\sfdefault}{\mddefault}{\updefault}{\color[rgb]{0,0,0}$N(1-l)$}%
}}}
\put(3526,-4936){\makebox(0,0)[rb]{\smash{\SetFigFont{20}{24.0}{\sfdefault}{\mddefault}{\updefault}{\color[rgb]{0,0,0}$\mc U_t=2\fol-N$}%
}}}
\put(751,-3361){\rotatebox{90.0}{\makebox(0,0)[b]{\smash{\SetFigFont{20}{24.0}{\sfdefault}{\mddefault}{\updefault}{\color[rgb]{0,0,0}$\mc U_t$}%
}}}}
\put(12601,164){\makebox(0,0)[rb]{\smash{\SetFigFont{20}{24.0}{\sfdefault}{\mddefault}{\updefault}{\color[rgb]{0,0,0}$+N$}%
}}}
\put(12601,-7036){\makebox(0,0)[rb]{\smash{\SetFigFont{20}{24.0}{\sfdefault}{\mddefault}{\updefault}{\color[rgb]{0,0,0}$-N$}%
}}}
\put(12451,-3436){\makebox(0,0)[lb]{\smash{\SetFigFont{20}{24.0}{\sfdefault}{\mddefault}{\updefault}{\color[rgb]{0,0,0}$0$}%
}}}
\put(15751,389){\makebox(0,0)[b]{\smash{\SetFigFont{20}{24.0}{\sfdefault}{\mddefault}{\updefault}{\color[rgb]{0,0,0}$N(1-l)$}%
}}}
\put(19276,-5536){\makebox(0,0)[rb]{\smash{\SetFigFont{20}{24.0}{\sfdefault}{\mddefault}{\updefault}{\color[rgb]{0,0,0}$\mc U_t=N-2\fol$}%
}}}
\put(22351,-3436){\makebox(0,0)[lb]{\smash{\SetFigFont{20}{24.0}{\sfdefault}{\mddefault}{\updefault}{\color[rgb]{0,0,0}$\fol$}%
}}}
\put(21751,-3736){\makebox(0,0)[lb]{\smash{\SetFigFont{20}{24.0}{\sfdefault}{\mddefault}{\updefault}{\color[rgb]{0,0,0}$N$}%
}}}
\put(19201,389){\makebox(0,0)[b]{\smash{\SetFigFont{20}{24.0}{\sfdefault}{\mddefault}{\updefault}{\color[rgb]{0,0,0}$Nl$}%
}}}
\put(12226,-3361){\rotatebox{90.0}{\makebox(0,0)[b]{\smash{\SetFigFont{20}{24.0}{\sfdefault}{\mddefault}{\updefault}{\color[rgb]{0,0,0}$\mc U_t$}%
}}}}
\put(15001,-4936){\makebox(0,0)[rb]{\smash{\SetFigFont{20}{24.0}{\sfdefault}{\mddefault}{\updefault}{\color[rgb]{0,0,0}$\mc U_t=2\fol-N$}%
}}}
\put(9226,-286){\makebox(0,0)[rb]{\smash{\SetFigFont{20}{24.0}{\sfdefault}{\mddefault}{\updefault}{\color[rgb]{0,0,0}$h_t=+1$:}%
}}}
\put(9226,-661){\makebox(0,0)[rb]{\smash{\SetFigFont{20}{24.0}{\sfdefault}{\mddefault}{\updefault}{\color[rgb]{0,0,0}$h_t=-1$:}%
}}}
\put(20701,-286){\makebox(0,0)[rb]{\smash{\SetFigFont{20}{24.0}{\sfdefault}{\mddefault}{\updefault}{\color[rgb]{0,0,0}$h_t=+1$:}%
}}}
\put(20701,-661){\makebox(0,0)[rb]{\smash{\SetFigFont{20}{24.0}{\sfdefault}{\mddefault}{\updefault}{\color[rgb]{0,0,0}$h_t=-1$:}%
}}}
\put(6001,-511){\makebox(0,0)[b]{\smash{\SetFigFont{29}{34.8}{\familydefault}{\mddefault}{\updefault}{\color[rgb]{0,0,0}$l<0.5$}%
}}}
\put(17476,-511){\makebox(0,0)[b]{\smash{\SetFigFont{29}{34.8}{\familydefault}{\mddefault}{\updefault}{\color[rgb]{0,0,0}$l>0.5$}%
}}}
\end{picture}
}
\end{center}
}

The expressions in Eqs. \ref{eqn:guplus-3a} and \ref{eqn:guminus-3a} are plotted
in Fig. \ref{fig:tradutil-3a}. The black and white circles represent the value
of \fol\ at which \gu\ is a maximum for $h_t=+1$ and $h_t=-1$ respectively.
We shall call the value of \fol\ at which \gu\ is a maximum the \emph{optimal}
value and denote it by $\opt[\fol]$. From Fig. \ref{fig:tradutil-3a} we can see
that $\opt[\fol]$ is given by:
%

\smallskip
\noindent
For $l<0.5$:
\begin{align}
\label{eqn:optfolraw-3a}
	\opt[\fol] &= \begin{cases}
	               Nl+1     & \text{if } h_t = +1 \\
		       \makebox[23mm][l]{$N(1-l)-1$} & \text{if } h_t = -1
		     \end{cases}
\intertext{For $l>0.5$:}
	\opt[\fol] &= \begin{cases}
	               Nl       & \text{if } h_t = +1 \\
		       \makebox[23mm][l]{$N(1-l)$}   & \text{if } h_t = -1
		     \end{cases} \; .
\end{align}
The most important feature of this equation to recognize is that, in general,
there is no unique value of $\opt[\fol]$ independent of $t$. Only if
$h_t=+1\text{ or }-1\;\forall\;t$ would such a unique solution exist.

We note in passing that since $N=\at+n_t^{-1}$ the optimal value of \at\
in the original model is:
\begin{equation}
\label{eqn:optattrad-3a}
	\opt[\at] = Nl \; .
\end{equation}
In contrast to $\opt[\fol]$, the optimal value of $\at$ is independent of $t$.
Nevertheless the most important quantity for the analysis that we present here
is $\opt[\fol]$ in the case of the original model since, as we shall see
later, it is the value of \fol\ that the agents can directly control and not
that of \at.

Ref. \cite{evol-freeze} demonstrated the existence of so-called
\emph{frozen} regimes which exist when $l$ lies outside of the region bounded by
two critical values, which we shall label here $l_{c1}$ and $l_{c2}$. These
regimes were described as \emph{quenched} by Burgos \ea\ in Ref.
\cite{ceva-quench}. The frozen regime obtains when $l<l_{c1}$ or $l>l_{c2}$. The
behavior of the Genetic Model in these regimes is well
understood\cite{evol-freeze,ceva-quench}, therefore we shall restrict ourselves
to a consideration of the dynamic regime. In what follows, unless otherwise
stated, $l$ takes values on the interval $l_{c1}<l<l_{c2}$.  Therefore,
inequalities such as $l<0.5$ should be taken as shorthand for $l_{c1}<l<0.5$.

Compared with the variation of $h_t$, agent mutation is a slow process. The
value of $h_t$ changes on a timescale of $\Delta t \sim 1$ while agents mutate
on a time scale $\Delta t \gg D$. The result of this is that we should not
expect the agents to be sensitive to the instantaneous value of \gu\ given in
Eqs. \eqref{eqn:guplus-3a} and \eqref{eqn:guminus-3a}. They will only be
sensitive to the time average $\ave{\gu}_t$. Translated to the conventions used
here, Ref. \cite{evol-freeze} found that:
\begin{equation}
\label{eqn:hvals-3a}
	\ave{h_t}_t = \begin{cases}
	                +0.5 & l > 0.5 \\
			-0.5 & l < 0.5
		      \end{cases}  \;.
\end{equation}
Therefore, for $l < 0.5$, $h_t=+1$ for a fraction
$0.25$ of the time steps whereas the fraction is $0.75$ for $l>0.5$.
Thus we can calculate the following expression for $\ave{\gu(\fol)}_t$:
%
%
%
%

\smallskip\noindent
For $l<0.5$:
\begin{align}
\label{eqn:guave_a-3a}
	\ave{\mc U_t}_t =
	  \begin{cases}
 	    2\fol-N            & \fol \le Nl \\
	    \fol - \frac{N}{2} & Nl < \fol < N(1-l) \\
	    -2\fol+N           & \fol \ge N(1-l)
	  \end{cases}
\intertext{For $l>0.5$:}
\label{eqn:guave_b-3a}
	\ave{\mc U_t}_t =
	  \begin{cases}
 	    2\fol-N            & \fol < N(1-l) \\
	    \fol - \frac{N}{2} & N(1-l) \le \fol \le Nl \\
	    -2\fol+N           & \fol > Nl
	  \end{cases} \; .
\end{align}
This expression is plotted in Fig. \ref{fig:utilmeanplusmemless-3a}a. From the
figure we can see that the optimal value of \fol\ that maximizes 
$\ave{\gu(\fol)}_t$ is given by:
\begin{equation}
	\opt[\fol] = \begin{cases}
			N(1-l)-1  &  l<0.5 \\
			Nl     	  &  l>0.5 \\
		     \end{cases} \;.
\label{eqn:optfol-3a}
\end{equation}
Note that although we have assumed the values given in Eq. \eqref{eqn:hvals-3a}
for $\ave{h_t}_t$, the values of $\opt[\fol]$ given in the above equation 
in fact only depend upon the signs of the values of $\ave{h_t}_t$. The
significance of this will become apparent in Sec. \ref{sec:hbar-3a}.

Let $\ind{\Pi}$ be the probability that $\fol=i$. Lo \ea\ \cite{emg-theory}
demonstrated that $\ind{\Pi}$ will be approximately Gaussian with mean
$\mu=\ave{\fol}_t=N\overline p_t$ and standard deviation $\sigma=\sqrt{\sum_i
\ind p(1-\ind p)}$. From Eq. \eqref{eqn:optfol-3a} it follows that the optimal
form of $\ind{\Pi}$ will obtain if $\overline p_t$ is as given by the following
equation and $\sigma=0$:
\begin{equation}
\label{eqn:optp-3a}
	\opt[\overline p_t] = \begin{cases}
			          1-l-\frac{1}{N} & l <0.5 \\
	                          l   & l >0.5
				\end{cases} \; .
\end{equation}
The term $\frac{1}{N}$ results from the asymmetry of the condition in Eq.
\eqref{eqn:Acond-3a} which determines $A_t$ in terms of \at.  For $l<l_{c2}$ in
the case of $N\gg1$ considered here $\frac{1}{N}\ll1-l$ and so we can neglect
this term .  $\sigma=0$ if \set p contains only the values $0$ and $1$. Let
$P(x)$ be the distribution of \set p such that $NP(x)dx$ is the probability that
if an agent $i$ is chosen at random from the set \set p then $x\le\ind p\le
x+dx$. The optimal form for $P(x)$ is then:
\begin{equation}
	P(x) = (1-\opt[\overline p_t]) \delta(x) 
	     + \opt[\overline p_t] \delta(1-x) \;.
\label{eqn:idealP-3a}
\end{equation}
This represents a distribution which is zero everywhere except for peaks at
$x=0$ and $x=1$. The relative heights of the peaks being such that 
$\mu=\opt[\overline p_t]$.

It is well known \cite{seg,self-seg,lo-thesis} that in the long time limit
where $t\rightarrow \infty$ the population of agents
evolves such that $P(x)$ is strongly peaked about $x=0$ and $x=1$ and $\overline
p_t$ takes the value given by Eq. \eqref{eqn:optp-3a}. Although the agents never
manage to achieve a form such that the standard deviation $\std{\fol}t$ is
exactly zero, they do approach the optimal distribution represented by Eq.
\eqref{eqn:idealP-3a}. Thus, we can see that the population of agents is capable
of evolving such that $\ind{\Pi}$ is close to its optimal form and the time
series of values of \fol\ contains values clustered around the optimal value
$\opt[\fol]$ given by Eq. \eqref{eqn:optfol-3a}.

\bigfloatfig[\floatplace]{fig:utilmeanplusmemless-3a}{
Average global utility as a function of \fol\ in the original Genetic Model
and the global utility in the memoryless model.
}{
a, The average global utility $\ave{\gu(\fol)}_t$ in the original Genetic
Model with memory. b, The global utility $\gu(\at)$ in the memoryless model.
}{
\begin{center}
\resizebox{0.96\textwidth}{!}{
\begin{picture}(0,0)%
\includegraphics{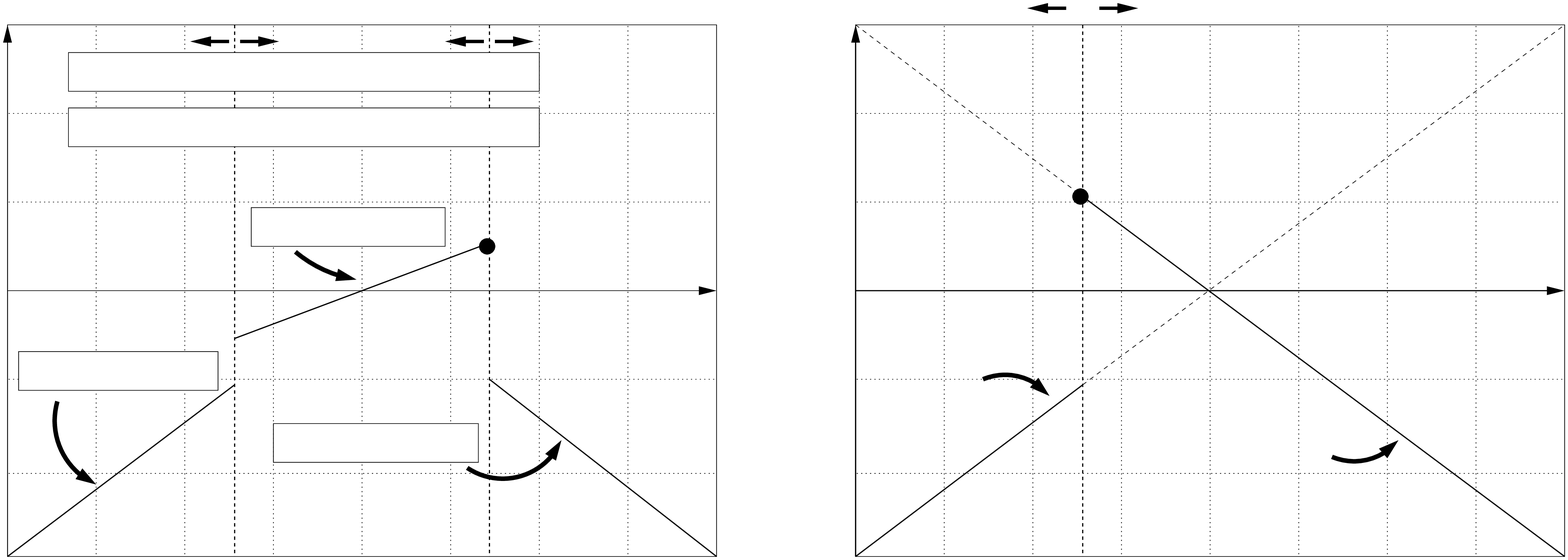}%
\end{picture}%
\setlength{\unitlength}{3947sp}%
\begingroup\makeatletter\ifx\SetFigFont\undefined%
\gdef\SetFigFont#1#2#3#4#5{%
  \fontsize{#1}{#2pt}%
  \fontfamily{#3}\fontseries{#4}\fontshape{#5}%
  \selectfont}%
\fi\endgroup%
\begin{picture}(21920,7604)(439,-7036)
\put(1126,164){\makebox(0,0)[rb]{\smash{\SetFigFont{20}{24.0}{\sfdefault}{\mddefault}{\updefault}{\color[rgb]{0,0,0}$+N$}%
}}}
\put(1126,-7036){\makebox(0,0)[rb]{\smash{\SetFigFont{20}{24.0}{\sfdefault}{\mddefault}{\updefault}{\color[rgb]{0,0,0}$-N$}%
}}}
\put(976,-3436){\makebox(0,0)[lb]{\smash{\SetFigFont{20}{24.0}{\sfdefault}{\mddefault}{\updefault}{\color[rgb]{0,0,0}$0$}%
}}}
\put(10876,-3436){\makebox(0,0)[lb]{\smash{\SetFigFont{20}{24.0}{\sfdefault}{\mddefault}{\updefault}{\color[rgb]{0,0,0}$\fol$}%
}}}
\put(10276,-3736){\makebox(0,0)[lb]{\smash{\SetFigFont{20}{24.0}{\sfdefault}{\mddefault}{\updefault}{\color[rgb]{0,0,0}$N$}%
}}}
\put(751,-3361){\rotatebox{90.0}{\makebox(0,0)[b]{\smash{\SetFigFont{20}{24.0}{\sfdefault}{\mddefault}{\updefault}{\color[rgb]{0,0,0}$\ave{\mc U_t}_t$}%
}}}}
\put(12609,164){\makebox(0,0)[rb]{\smash{\SetFigFont{20}{24.0}{\sfdefault}{\mddefault}{\updefault}{\color[rgb]{0,0,0}$+N$}%
}}}
\put(12609,-7036){\makebox(0,0)[rb]{\smash{\SetFigFont{20}{24.0}{\sfdefault}{\mddefault}{\updefault}{\color[rgb]{0,0,0}$-N$}%
}}}
\put(12459,-3436){\makebox(0,0)[lb]{\smash{\SetFigFont{20}{24.0}{\sfdefault}{\mddefault}{\updefault}{\color[rgb]{0,0,0}$0$}%
}}}
\put(19284,-5536){\makebox(0,0)[rb]{\smash{\SetFigFont{20}{24.0}{\sfdefault}{\mddefault}{\updefault}{\color[rgb]{0,0,0}$\mc U_t=N-2\at$}%
}}}
\put(22359,-3436){\makebox(0,0)[lb]{\smash{\SetFigFont{20}{24.0}{\sfdefault}{\mddefault}{\updefault}{\color[rgb]{0,0,0}$\at$}%
}}}
\put(21759,-3736){\makebox(0,0)[lb]{\smash{\SetFigFont{20}{24.0}{\sfdefault}{\mddefault}{\updefault}{\color[rgb]{0,0,0}$N$}%
}}}
\put(15009,-4936){\makebox(0,0)[rb]{\smash{\SetFigFont{20}{24.0}{\sfdefault}{\mddefault}{\updefault}{\color[rgb]{0,0,0}$\mc U_t=2\at-N$}%
}}}
\put(15759,389){\makebox(0,0)[b]{\smash{\SetFigFont{20}{24.0}{\sfdefault}{\mddefault}{\updefault}{\color[rgb]{0,0,0}$Nl$}%
}}}
\put(12234,-3361){\rotatebox{90.0}{\makebox(0,0)[b]{\smash{\SetFigFont{20}{24.0}{\sfdefault}{\mddefault}{\updefault}{\color[rgb]{0,0,0}$\mc U_t$}%
}}}}
\put(10246,-511){\makebox(0,0)[b]{\smash{\SetFigFont{34}{40.8}{\familydefault}{\mddefault}{\updefault}{\color[rgb]{0,0,0}a)}%
}}}
\put(21696,-511){\makebox(0,0)[b]{\smash{\SetFigFont{34}{40.8}{\familydefault}{\mddefault}{\updefault}{\color[rgb]{0,0,0}b)}%
}}}
\put(4276,-1261){\makebox(0,0)[b]{\smash{\SetFigFont{20}{24.0}{\sfdefault}{\mddefault}{\updefault}{\color[rgb]{0,0,0}$N(1-l)$}%
}}}
\put(7801,-1261){\makebox(0,0)[b]{\smash{\SetFigFont{20}{24.0}{\sfdefault}{\mddefault}{\updefault}{\color[rgb]{0,0,0}$Nl$}%
}}}
\put(2176,-1261){\makebox(0,0)[lb]{\smash{\SetFigFont{20}{24.0}{\sfdefault}{\mddefault}{\updefault}{\color[rgb]{0,0,0}If $l>0.5$:}%
}}}
\put(4276,-511){\makebox(0,0)[b]{\smash{\SetFigFont{20}{24.0}{\sfdefault}{\mddefault}{\updefault}{\color[rgb]{0,0,0}$Nl$}%
}}}
\put(7801,-511){\makebox(0,0)[b]{\smash{\SetFigFont{20}{24.0}{\sfdefault}{\mddefault}{\updefault}{\color[rgb]{0,0,0}$N(1-l)$}%
}}}
\put(2176,-511){\makebox(0,0)[lb]{\smash{\SetFigFont{20}{24.0}{\sfdefault}{\mddefault}{\updefault}{\color[rgb]{0,0,0}If $l<0.5$: }%
}}}
\put(7051,-2611){\makebox(0,0)[rb]{\smash{\SetFigFont{20}{24.0}{\sfdefault}{\mddefault}{\updefault}{\color[rgb]{0,0,0}$\ave{\mc U_t}_t=\fol-\frac{N}{2}$}%
}}}
\put(4051,-4486){\makebox(0,0)[rb]{\smash{\SetFigFont{20}{24.0}{\sfdefault}{\mddefault}{\updefault}{\color[rgb]{0,0,0}$\ave{\mc U_t}_t=2\fol-N$}%
}}}
\put(7501,-5536){\makebox(0,0)[rb]{\smash{\SetFigFont{20}{24.0}{\sfdefault}{\mddefault}{\updefault}{\color[rgb]{0,0,0}$\ave{\mc U_t}_t=N-2\fol$}%
}}}
\end{picture}
}
\end{center}
}

\subsubsection{Memoryless Genetic Model}

In this section we shall see how the above analysis applies to
the memoryless variant of the model.
\gu\ in the memoryless model is given by Eq.
\eqref{eqn:guplus-3a} above. Thus:
\begin{equation}
\label{eqn:guml-3a}
	\gu(\at) = 
	  \begin{cases}
	    2\at - N		& \text{if } \at \le Nl \\
	    -2\at + N	    	& \text{if } \at > Nl
	  \end{cases} \\ \; .
\end{equation}
Equation \eqref{eqn:guml-3a} is plotted in Fig.
\ref{fig:utilmeanplusmemless-3a}b. Once again the black circle represents the
value of \at\ at which \gu\ is a maximum. From the figure we can see that the
optimal value of \at\ is now:
\begin{equation}
\label{eqn:optatml-3a}
	\opt[\at] = \begin{cases}
	              Nl+1 & \f l<0.5 \\
		      Nl   & \f l>0.5
		    \end{cases} \;.
\end{equation}
Unlike the original Genetic Model, this optimal value of \at\ is independent of $t$.
As before, the optimal form of $\ind{\Pi}$ will be that for which 
$\sigma=\std{\at}t=0$ and $\mu=\ave{\at}t=N\overline p_t$ with $\overline p_t$
given by:
\begin{equation}
\label{eqn:optpml-3a}
	\opt[\overline p_t] = \begin{cases}
	                        l + \frac{1}{N} & \f l<0.5 \\
				l               & \f l>0.5
			      \end{cases} \;.
\end{equation}
Note that, from Eqs.  \eqref{eqn:optp-3a} and \eqref{eqn:optpml-3a}, this will
mean that the \pval\ distribution of the agents for $l<0.5$ in the
original and memoryless models will not be identical, but will be related by
the transformation $\{\ind p\}_t \rightarrow \{1-\ind p\}_t$.

\subsubsection{Direct comparison of the models}
\label{sec:directcomp-3a}

In this section we shall compare the value of $\gu(\opt[\fol])$ in the original
Genetic Model with that of $\gu(\opt[\at])$ in the memoryless Genetic Model in
order to establish what effect the memory has on the performance of the model.
In both the original Genetic Model with memory and the memoryless variant, the
agents are rewarded based on the value of \at. This is because the value of the
global action is determined from the condition on \at\ in Eq.
\eqref{eqn:Acond-3a} and an agent $i$ gains or loses one point depending on
whether $\ind a = \pm A_t$.  There is, however, one difference between the two
models that will be extremely important in what follows. The population of
agents can control $\ind{\Pi}$ through their effect on $\overline p_t$ and
$\sigma=\sqrt{\sum_i \ind p(1-\ind p)}$. In the memoryless model $\ind{\Pi}$
represents the probability distribution for \at\ whereas in the original model
$\ind{\Pi}$ represents the distribution function for \fol. The result of this is
that in the memoryless model the population of agents can directly control the
values that occur in the time series of \at\ whereas in the original model
they can only control the values of \fol. In the latter case \at\ will also
depend on the value of $h_t$ over which the agents have no direct control.

From Eqs. \eqref{eqn:guave_a-3a}, \eqref{eqn:guave_b-3a} and \eqref{eqn:guml-3a}
it follows that the maximum values of \gu\ (which obtain at $\opt[\fol]$ and
$\opt[\at]$ given by Eqs. \eqref{eqn:optfol-3a} and \eqref{eqn:optatml-3a}) are:

\smallskip\noindent
memory:
\begin{align}
\label{eqn:gumag-3a}
	\gu(\opt[\fol])&= 
		\begin{cases}
	          \frac{N}{2} (1-2l) -1 & \f l<0.5 \\
	          \frac{N}{2} (2l-1)    & \f l\ge0.5
		\end{cases}\\
\intertext{no memory:} 
        \gu(\opt[\at]) &=
	        \begin{cases}
		  N(1-2l) -2 & \f l<0.5 \\
		  N(2l-1)    & \f l\ge0.5 \\
		\end{cases} \; .
\end{align}
Thus the optimal value of \gu\ in the original model is exactly half that
achieved by the memoryless model. As we suggested above, the reason for this 
is because in the original model \at\ is a function of both $h_t$ and
$\ind{\Pi}$. The instantaneous optimal value of \fol\ will therefore depend on
$h_t$ (see Fig. \ref{fig:tradutil-3a} and Eq. \eqref{eqn:optfolraw-3a}). Note
that the value of \fol\ that maximizes the time average 
$\ave{\gu(\fol)}_t$, $\opt[\fol]$ given by Eq. \eqref{eqn:optfol-3a} will always
be one of the instantaneous optimal values given in Eq. 
\eqref{eqn:optfolraw-3a}. Thus the agents cannot improve the global utility
by varying $\overline p_t$. They adopt the value of $\overline p_t$ that is
optimal for the most common value of $h_t$, but they must pay the penalty when 
$h_t$ takes the opposite value.
In contrast, in the memoryless model the instantaneous optimal value of $\at$
in Eq. \eqref{eqn:optatml-3a} is independent of $t$. Thus, by evolving such that
$\overline p_t = \opt[\overline p_t]$ the agents can insure that \at\ is close
to the optimal value at each time step.

\subsection{Analytical expressions for $\ave{\at}_t$ and $\std{\at}t$}
\label{sec:consistency-3a}

We can use the same method that we used to derive the expression for
$\ave{\gu(\fol)}_t$ in Eqs. \eqref{eqn:guave_a-3a} and \eqref{eqn:guave_b-3a} to
obtain expressions for $\ave{\at}_t$ and the standard deviation $\std{\at}t$
of the \at time series.
From Eq. \eqref{eqn:avefol_en-3a}:
\begin{equation}
	\ave{\fol} = N \overline p_t \; .
\end{equation}
This leads to the following expression for $\ave{\at}$:
\begin{align}
	\ave{\at}  &= \begin{cases}
	               \makebox[33mm][l]{$N(1-\overline p_t)$}
		                          & \text{if } h_t=-1 \\
 	               N\overline p_t     & \text{if } h_t=+1
	             \end{cases}  \; . \\
\intertext{\noindent
Taking the time average in exactly the same way as in Sec.
\ref{sec:perftrad-3a}, in equilibrium where \pbar\ is approximately constant,
yields expressions for $\ave{\at}_t$ and $\ave{(\at)^2}_t$:
}									
\label{eqn:ourresultone-3a}
	\ave{\at}_t\!&= \!\!\begin{cases}
			\makebox[39.5mm][l]{$\frac{N}{4}(3-2\ave{\pbar}_t)$} 
			    & \text{if } l\! <\! 0.5 \\
	                \frac{N}{4}(1 + 2\ave{\pbar}_t)	
			    & \text{if } l\!>\! 0.5 
	              \end{cases}  \\
	\ave{\left[\at\right]^2}_t\! &= \!\!\begin{cases} 
	  \makebox[39.5mm][l]{$
	  \frac{N^2}{4} \!\!\left(
	    4\ave{\pbar}_t^2\!-\!6\ave{\pbar}_t\!+\!3 \right)\!\!+\!\sigma^2
	     $} & \text{if } l\!<\!0.5\\ 
	    \frac{N^2}{4} \!\!\left( 
	    4\ave{\pbar}_t^2\!-\!2\ave{\pbar}_t\!+\!1 \right)\!\!+\!\sigma^2  & 
	       \text{if } l\!>\!0.5
	                             \end{cases}
\end{align}
where $\sigma$ is the standard deviation of \ind{\Pi} introduced in Sec.
\ref{sec:tradintro-3a} which will be of order unity.
If we assume that the agents adopt the optimal distribution in Eq.
\eqref{eqn:idealP-3a} then we can take $\sigma=0$. We now obtain an expression
for $\std{\at}t$ as follows:
\begin{align}
\label{eqn:ourresulttwo_a-3a}
	\left[\std{\at}t\right]^2
	           &= \ave{\left[\at\right]^2}_t - 
	              \left[\ave{\at}_t\right]^2 \notag\\
	           &= \frac{3N^2}{4}\left(\ave{\pbar}_t-\frac{1}{2}\right)^2
		      + \sigma^2 \\
\intertext{Taking $\sigma=0$ gives:}
\label{eqn:ourresulttwo-3a}
	\std{\at}t &= \frac{N\sqrt{3}}{2}\left|\ave{\pbar}_t-\frac{1}{2}\right|
	\; .
\end{align}

Ref. \cite{evol-freeze} used a mean-field approximation to derive
expressions for \at\ and $\std{\at}t$ in terms of $\ave{h_t}_t$ and
$\ave{\pbar}_t$.
In particular:
\begin{align}
\label{eqn:jatone-3a}
	\ave{\at}_t &= 
	  \begin{cases}
  	    \frac{N}{4} ( 3 -2\ave{\pbar}_t ) & \text{if } l<0.5 \\
  	    \frac{N}{4} ( 1 +2\ave{\pbar}_t ) & \text{if } l>0.5
	  \end{cases}
	 \\
\label{eqn:jattwo-3a}
	\std{\at}t&=\frac{N\sqrt{3}}{2}\left|\ave{\pbar}_t-\frac{1}{2}\right|\;.
\end{align}
Note that we have substituted for $\ave{h_t}_t$ in the
expressions of Ref. \cite{evol-freeze}, with the values of $\ave{h_t}_t$ that obtain
for
$l_{c1}<l<l_{c2}$ ($\ave{h_t}_t=0.75$ for $l<0.5$ and $\ave{h_t}_t=0.25$ for
$l>0.5$ \cite{evol-freeze}). Thus, we can see that the expressions that we
derived in Eqs. \eqref{eqn:ourresultone-3a} and \eqref{eqn:ourresulttwo-3a} are
consistent with those obtained in Ref. \cite{evol-freeze}.

\subsection{Numerical Results}
\label{sec:numerics-3a}

\bigfloatfig[\floatplace]{fig:attencomp-3a}{
$\ave{\at}_t$ and $\std{\at}t$ as a function of the resource level $l$ in the
original and memoryless Genetic Models.
}{
a, Numerical results for $\ave{\at}_t$ and $\std{\at}t$ as a function of the
resource level $l$ in the original Genetic Model. We also include lines which
represent: $\opt[\at]=Nl$ from Eq. \eqref{eqn:optattrad-3a} and the analytical
expressions for $\ave{\at}_t$ and \std{\at}t presented in Eq.
\eqref{eqn:analytic-3a}.
b, $\ave{\at}_t$ and $\std{\at}t$ as a function of the resource level $l$ in the
memoryless Genetic Model. 
}{
\begin{center}
\resizebox{0.476\textwidth}{!}{\small 
\begingroup%
  \makeatletter%
  \newcommand{\GNUPLOTspecial}{%
    \catcode`\%=14\relax\special}%
  \setlength{\unitlength}{0.1bp}%
\begin{picture}(3600,3240)(0,0)%
\includegraphics{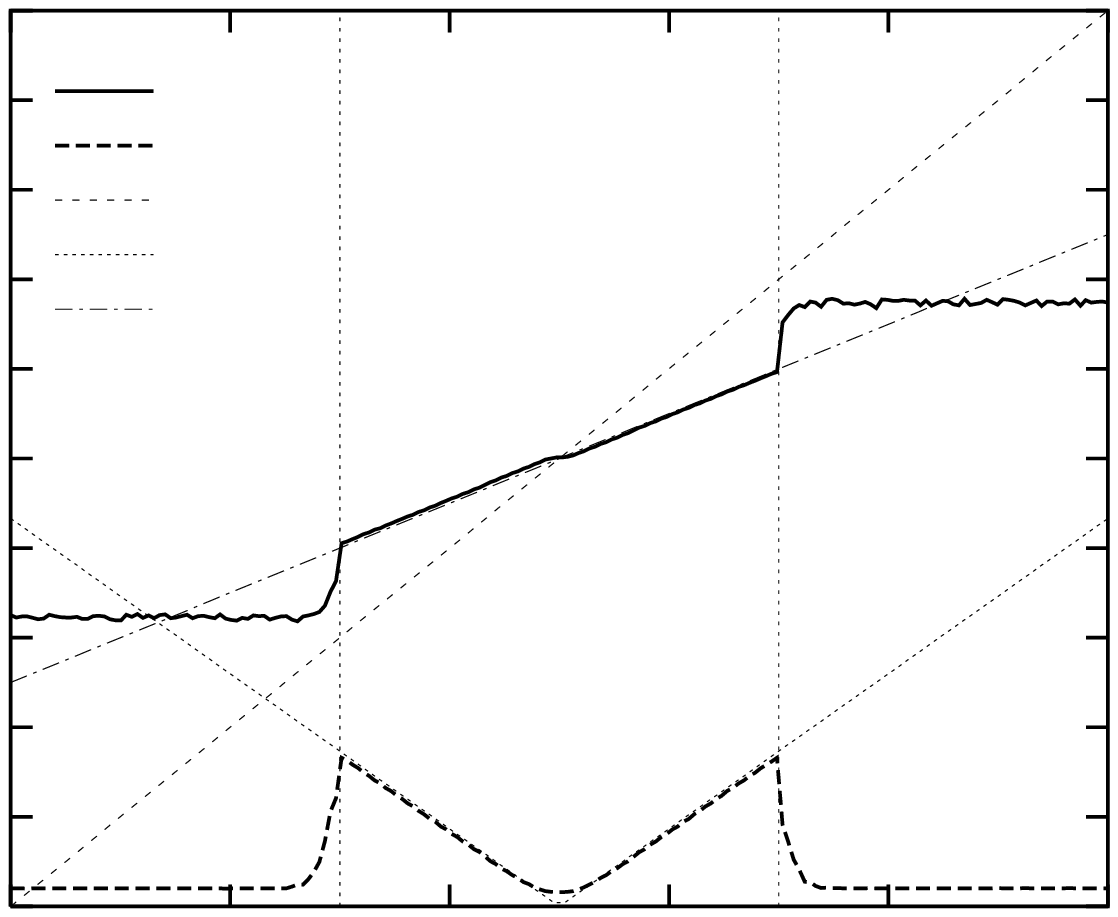}
\put(851,2050){\makebox(0,0)[l]{$\ave{\at}_t=\frac{N}{4}(1+2l)$}}%
\put(851,2207){\makebox(0,0)[l]{$\std{\at}t=\frac{N\sqrt{3}}{2}\left|l-\frac{1}{2}\right|$}}%
\put(851,2364){\makebox(0,0)[l]{$\opt[\at]=Nl$}}%
\put(851,2521){\makebox(0,0)[l]{Numerical: \std{\at}t}}%
\put(851,2678){\makebox(0,0)[l]{Numerical: $\ave{\at}_t$}}%
\put(2597,149){\makebox(0,0){$l_{c2}$}}%
\put(1333,149){\makebox(0,0){$l_{c1}$}}%
\put(1965,3075){\makebox(0,0){\Large a)}}%
\put(1965,55){\makebox(0,0){$l$}}%
\put(55,1620){%
\makebox(0,0)[b]{\shortstack{ }}%
}%
\put(3545,220){\makebox(0,0){1}}%
\put(2913,220){\makebox(0,0){0.8}}%
\put(2281,220){\makebox(0,0){0.6}}%
\put(1649,220){\makebox(0,0){0.4}}%
\put(1017,220){\makebox(0,0){0.2}}%
\put(385,220){\makebox(0,0){0}}%
\put(330,2910){\makebox(0,0)[r]{500}}%
\put(330,2652){\makebox(0,0)[r]{450}}%
\put(330,2394){\makebox(0,0)[r]{400}}%
\put(330,2136){\makebox(0,0)[r]{350}}%
\put(330,1878){\makebox(0,0)[r]{300}}%
\put(330,1620){\makebox(0,0)[r]{250}}%
\put(330,1362){\makebox(0,0)[r]{200}}%
\put(330,1104){\makebox(0,0)[r]{150}}%
\put(330,846){\makebox(0,0)[r]{100}}%
\put(330,588){\makebox(0,0)[r]{50}}%
\put(330,330){\makebox(0,0)[r]{0}}%
\end{picture}%
\endgroup

}
\resizebox{0.476\textwidth}{!}{\small 
\begingroup%
  \makeatletter%
  \newcommand{\GNUPLOTspecial}{%
    \catcode`\%=14\relax\special}%
  \setlength{\unitlength}{0.1bp}%
\begin{picture}(3600,3240)(0,0)%
\includegraphics{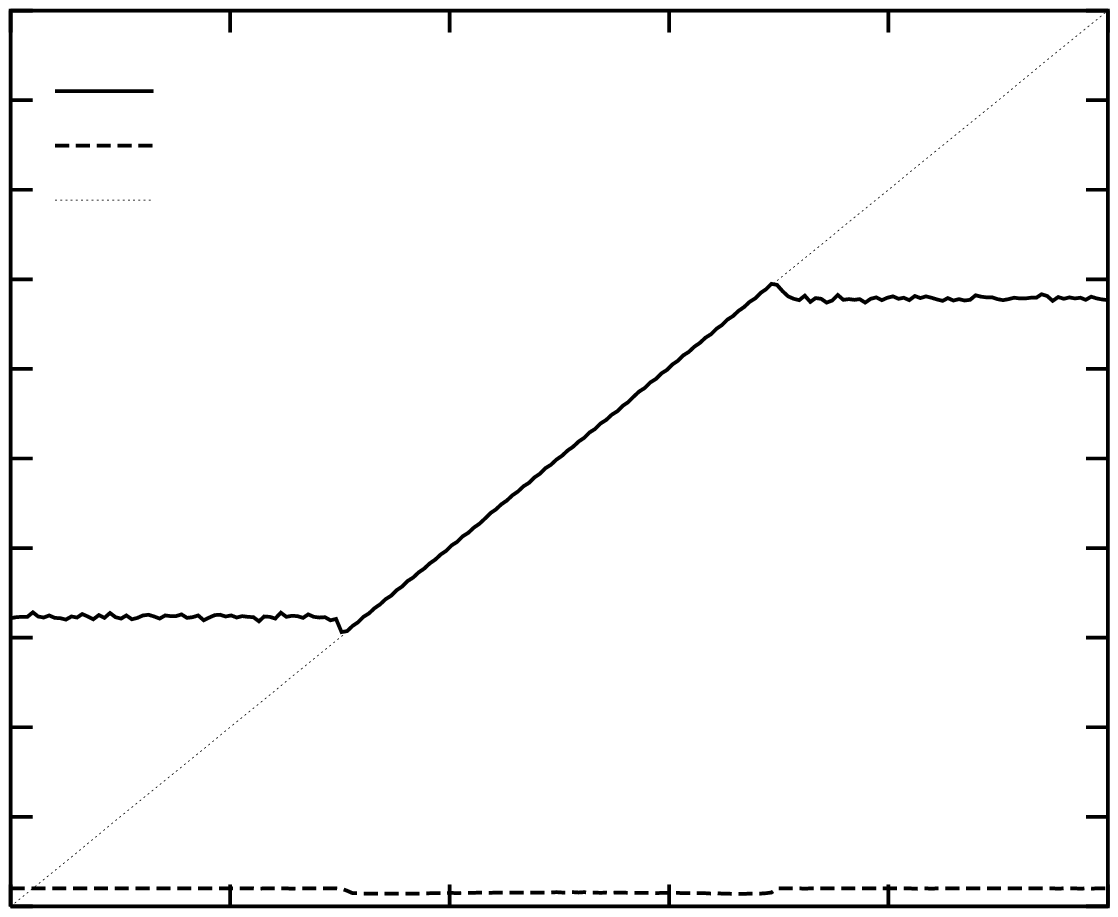}
\put(851,2364){\makebox(0,0)[l]{$\opt[\at]=Nl$}}%
\put(851,2521){\makebox(0,0)[l]{Numerical: \std{\at}t}}%
\put(851,2678){\makebox(0,0)[l]{Numerical: $\ave{\at}_t$}}%
\put(1965,3075){\makebox(0,0){\Large b)}}%
\put(1965,55){\makebox(0,0){$l$}}%
\put(55,1620){%
\makebox(0,0)[b]{\shortstack{ }}%
}%
\put(3545,220){\makebox(0,0){1}}%
\put(2913,220){\makebox(0,0){0.8}}%
\put(2281,220){\makebox(0,0){0.6}}%
\put(1649,220){\makebox(0,0){0.4}}%
\put(1017,220){\makebox(0,0){0.2}}%
\put(385,220){\makebox(0,0){0}}%
\put(330,2910){\makebox(0,0)[r]{500}}%
\put(330,2652){\makebox(0,0)[r]{450}}%
\put(330,2394){\makebox(0,0)[r]{400}}%
\put(330,2136){\makebox(0,0)[r]{350}}%
\put(330,1878){\makebox(0,0)[r]{300}}%
\put(330,1620){\makebox(0,0)[r]{250}}%
\put(330,1362){\makebox(0,0)[r]{200}}%
\put(330,1104){\makebox(0,0)[r]{150}}%
\put(330,846){\makebox(0,0)[r]{100}}%
\put(330,588){\makebox(0,0)[r]{50}}%
\put(330,330){\makebox(0,0)[r]{0}}%
\end{picture}%
\endgroup

}
\end{center}
}
\bigfloatfig[\floatplace]{fig:mphis-3a}{
Numerical results for $\overline p_t$ and $\ave{h_t}_t$ in the original
and memoryless models.
}{
a, Numerical results for $\overline p_t$ as a function of the resource level
$l$ in the original and memoryless models.
b, Numerical results for $\ave{h_t}_t$ as a function of $l$ in the original
model.
}{
\begin{center}
\resizebox{0.476\textwidth}{!}{\small 
\begingroup%
  \makeatletter%
  \newcommand{\GNUPLOTspecial}{%
    \catcode`\%=14\relax\special}%
  \setlength{\unitlength}{0.1bp}%
\begin{picture}(3600,3240)(0,0)%
\includegraphics{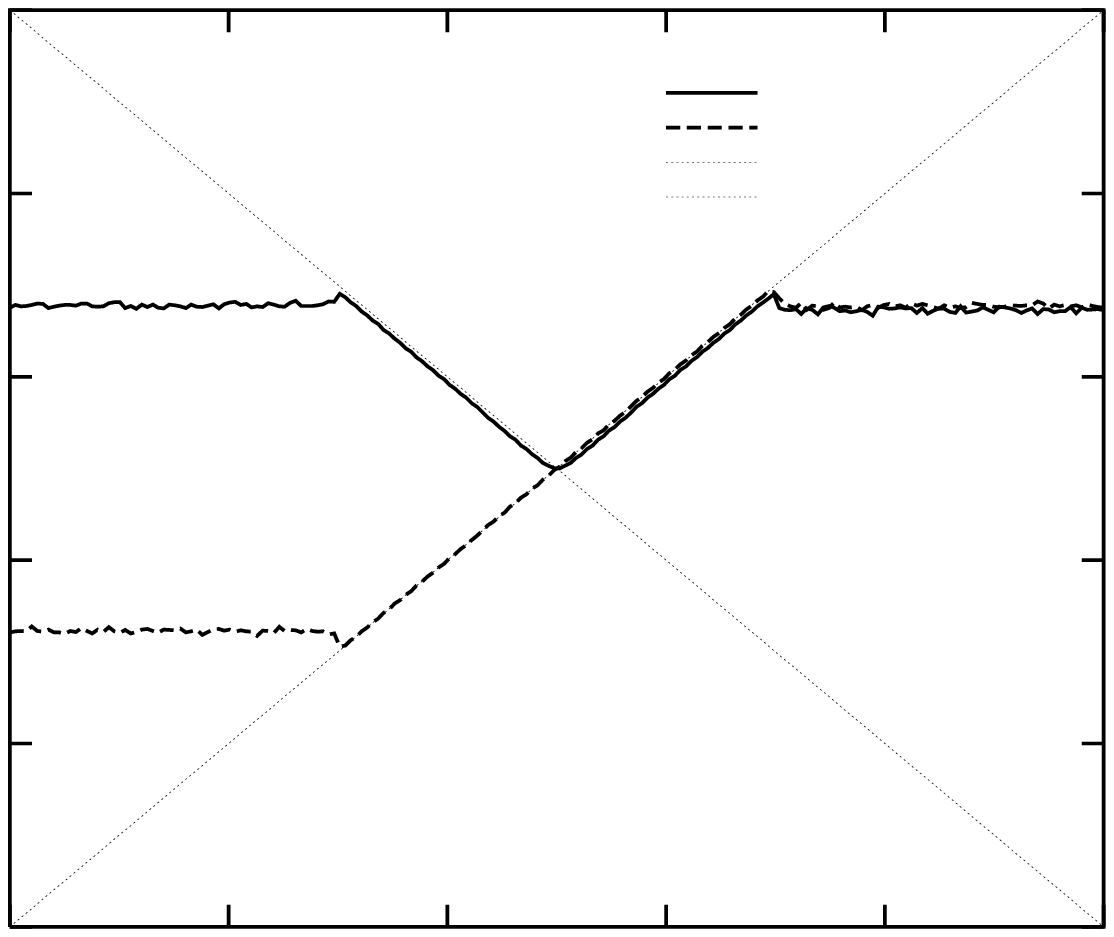}
\put(1740,2402){\makebox(0,0)[l]{$1-l$}}%
\put(1740,2502){\makebox(0,0)[l]{$l$}}%
\put(1740,2602){\makebox(0,0)[l]{Memoryless}}%
\put(1740,2702){\makebox(0,0)[l]{Original}}%
\put(1975,3090){\makebox(0,0){\Large a)}}%
\put(1975,50){\makebox(0,0){$l$}}%
\put(100,1620){%
\makebox(0,0)[b]{\shortstack{$\ave{\overline p_t}_t$}}%
}%
\put(3550,200){\makebox(0,0){1}}%
\put(2920,200){\makebox(0,0){0.8}}%
\put(2290,200){\makebox(0,0){0.6}}%
\put(1660,200){\makebox(0,0){0.4}}%
\put(1030,200){\makebox(0,0){0.2}}%
\put(400,200){\makebox(0,0){0}}%
\put(350,2940){\makebox(0,0)[r]{1}}%
\put(350,2412){\makebox(0,0)[r]{0.8}}%
\put(350,1884){\makebox(0,0)[r]{0.6}}%
\put(350,1356){\makebox(0,0)[r]{0.4}}%
\put(350,828){\makebox(0,0)[r]{0.2}}%
\put(350,300){\makebox(0,0)[r]{0}}%
\end{picture}%
\endgroup

}
\resizebox{0.476\textwidth}{!}{\small 
\begingroup%
  \makeatletter%
  \newcommand{\GNUPLOTspecial}{%
    \catcode`\%=14\relax\special}%
  \setlength{\unitlength}{0.1bp}%
\begin{picture}(3600,3240)(0,0)%
\includegraphics{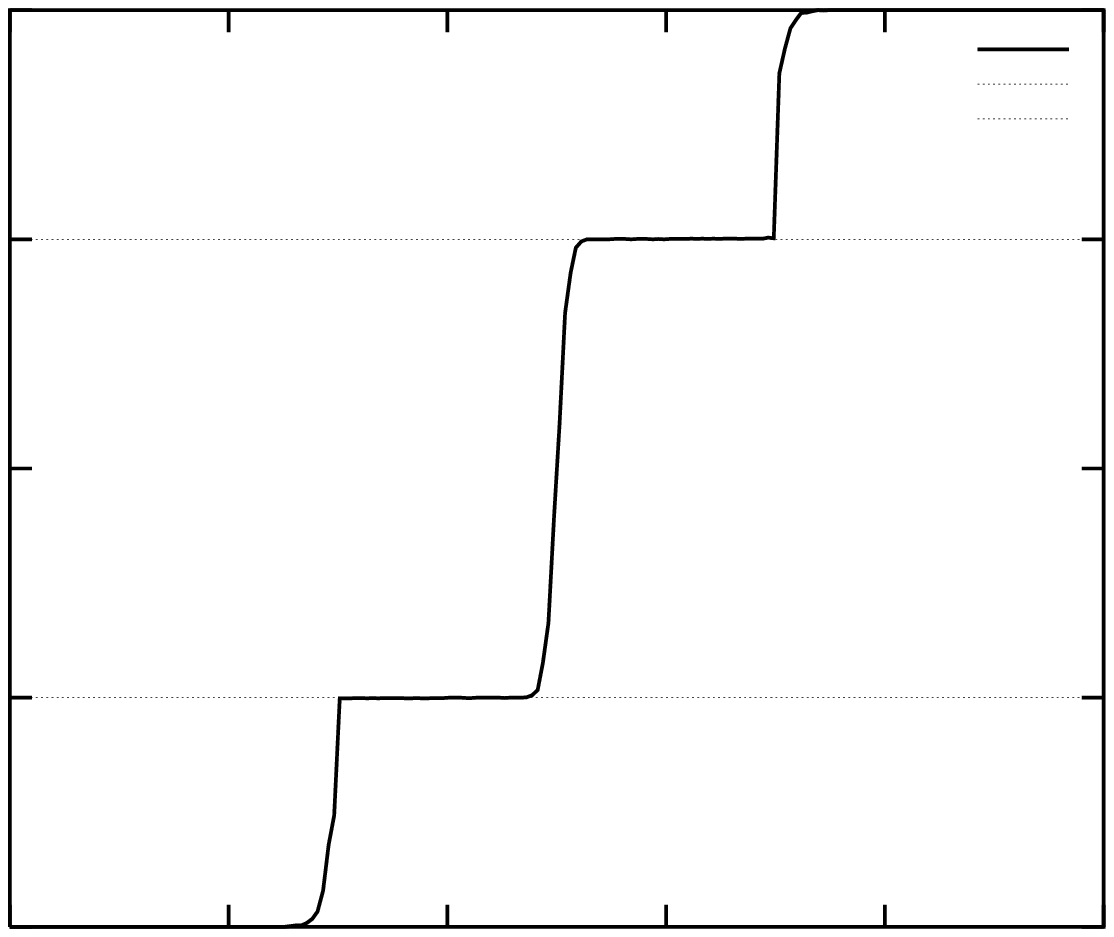}
\put(2787,2627){\makebox(0,0)[l]{$-0.5$}}%
\put(2787,2727){\makebox(0,0)[l]{$+0.5$}}%
\put(2787,2827){\makebox(0,0)[l]{$\ave{h_t}_t$}}%
\put(1975,3090){\makebox(0,0){\Large b)}}%
\put(1975,50){\makebox(0,0){$l$}}%
\put(250,1620){%
\makebox(0,0)[b]{\shortstack{$\ave{h_t}_t$}}%
}%
\put(3550,200){\makebox(0,0){1}}%
\put(2920,200){\makebox(0,0){0.8}}%
\put(2290,200){\makebox(0,0){0.6}}%
\put(1660,200){\makebox(0,0){0.4}}%
\put(1030,200){\makebox(0,0){0.2}}%
\put(400,200){\makebox(0,0){0}}%
\put(350,2940){\makebox(0,0)[r]{1}}%
\put(350,2280){\makebox(0,0)[r]{0.5}}%
\put(350,1620){\makebox(0,0)[r]{0}}%
\put(350,960){\makebox(0,0)[r]{-0.5}}%
\put(350,300){\makebox(0,0)[r]{-1}}%
\end{picture}%
\endgroup

}
\end{center}
}

In this section we present numerical data which supports the analytical results
that we presented in the previous sections.  Ref. \cite{evol-freeze}
investigated the behavior of $\ave{\at}_t$ and \std{\at}t as a function of the
resource level $l$.  Further work was done with regard to the memoryless Genetic
Model and general $l$ by Burgos \ea\ in Refs. \cite{ceva-quench,ceva-thermal}.
Figure \ref{fig:attencomp-3a}a recalls the results of Ref. \cite{evol-freeze}.
We can clearly see the dynamic and frozen regimes for $l_{c1}<l<l_{c2}$ and
$l<l_{c1}$, $l>l_{c_2}$ respectively. Figure \ref{fig:attencomp-3a}b shows
$\ave{\at}_t$ and \std{\at}t in the memoryless variant. Figure
\ref{fig:mphis-3a}a shows $\ave{\pbar}_t$ as a function of $l$ in the original
and memoryless models while Fig. \ref{fig:mphis-3a}b shows $\ave{h_t}_t$ in the
original model.

First of all note that in Fig. \ref{fig:mphis-3a}a $\ave{\pbar}_t$ lies to a
very good approximation on the following lines:
\begin{align}
	\text{memory: } \ave{\pbar}_t \!= \begin{cases}
	                                  1-l & l\!<\!0.5 \\
					  l   & l\!>\!0.5 
					\end{cases}
	&&
	\text{no memory: } \ave{\pbar}_t \!= l 
\end{align}
This confirms that the population of agents is capable of evolving to achieve
the optimal values of $\overline p_t$ given in Eqs. \eqref{eqn:optp-3a} and
\eqref{eqn:optpml-3a}.

In Sec. \ref{sec:consistency-3a} we presented expressions in Eqs.
\eqref{eqn:ourresultone-3a}, \eqref{eqn:ourresulttwo-3a}, \eqref{eqn:jatone-3a}
and \eqref{eqn:jattwo-3a} for $\ave{\at}_t$ and \std{\at}t If we substitute for
$\ave{\pbar}_t$ in these equations with the optimal values from Eq.
\eqref{eqn:optp-3a}, neglecting the term $\frac{1}{N}$, we obtain the following
analytical equations for $\ave{\at}_t$ and \std{\at}t for $l_{c1}<l<l_{c2}$:
\begin{align}
\label{eqn:analytic-3a}
	\ave{\at}_t = \frac{N}{4}(1+2l) &&
	\std{\at}t  = \frac{N\sqrt{3}}{2}\left|l-\half\right| \; .
\end{align}
In Fig. \ref{fig:attencomp-3a}a we show these analytic expressions together with
the numerical data and $\opt[\at]=Nl$ from Eq. \eqref{eqn:optattrad-3a}. We can
see that $\ave{\at}_t$ deviates from the optimal value of $Nl$ for
$l_{c1}<l<0.5$ and $0.5<l<l_{c2}$ as pointed out in Ref.
\cite{evol-freeze}. We now know, from Sec. \ref{sec:directcomp-3a},
that the reason for this is that the population of
agents can only control \fol\ directly and not \at. Thus their performance is
reduced by the action of $h_t$. We can see that for $l_{c1}<l<l_{c2}$
$\ave{\at}_t$ instead lies on the line defined by Eq. \eqref{eqn:analytic-3a}
\footnote{Note that in fact $\ave{\at}_t$ lies slightly closer to $\frac{N}{2}$
than the expression in Eq. \eqref{eqn:analytic-3a}. This is due to the deviation
depicted in Fig. \ref{fig:ndist-3a} which we shall discuss in Secs. \ref{sec:hbar-3a} and \ref{sec:dev-3a}.}.
In Fig. \ref{fig:attencomp-3a}a we also see that the numerical data for
\std{\at}t agrees to a good approximation with the expressions in Eq.
\eqref{eqn:analytic-3a}. 
The analytic expression deviates from the numerical data in the vicinity of
$l=0.5$. The reason for this is that when we derived the expression for
$\std{\at}t$ in Eq. \eqref{eqn:ourresulttwo-3a} we assumed that the agent gene
value distribution is as given by Eq. \eqref{eqn:idealP-3a} and so $\sigma=0$.
In fact $\sigma\ne0$ and near $l=0.5$ the $\sigma^2$ term in Eq.
\eqref{eqn:ourresulttwo_a-3a} dominates. Therefore $\std{\at}t$ does not go to
zero as predicted.


In Fig. \ref{fig:attencomp-3a}b we see that, as predicted by Sec.
\ref{sec:directcomp-3a}, $\ave{\at}_t$ in the memoryless model does lie on the
optimal line defined by $\opt[\at]=Nl$ for $l_{c1}<l<l_{c2}$. 
We can also see that \std{\at}t in the memoryless model is much lower than in
the original model. The large value of \std{\at}t in the original model
results from the fact that \at\ is a function of both \fol, determined via
$\ind{\Pi}$ by the distribution of agent \pval s $P(x)$, and $h_t$. In the
memoryless model \at\ is a function of $P(x)$ only which in equilibrium will be 
approximately constant in form. The small remaining fluctuations are due to the
fact that the agent population does not achieve the ideal distribution of
Eq. \eqref{eqn:idealP-3a}.
We can therefore say that the memoryless model is efficient in
accessing the available resources.

\newcommand{\R}[1]{\ensuremath{\mc R_{#1}}}
\newcommand{\M}{\ensuremath{\mc S}}
\subsection{Generation of the prediction from an exogenous source}
\label{sec:external-3a}

\floatfig[\floatplace]{fig:exog-3a}{
Numerical results for $\ave{\at}_t$ and \std{at}t in the memoryless model using
exogenous sources.
}{
Numerical results for $\ave{\at}_t$ and \std{\at}t in the memoryless model using
exogenous sources \R{0.5} and \R{1.0} for $h_t$. The results for $\ave{\at}_t$
and \std{\at}t in the original model, where $h_t$ is determined by the memory
($\{h_t\}_t=\mc S$), are included for comparison. Each pair of lines shows 
$\ave{\at}_t$ and \std{\at}t for $\{h_t\}_t$ given as indicated.
}{
\begin{center}
\resizebox{0.476\textwidth}{!}{\small 
\begingroup%
  \makeatletter%
  \newcommand{\GNUPLOTspecial}{%
    \catcode`\%=14\relax\special}%
  \setlength{\unitlength}{0.1bp}%
\begin{picture}(3600,2376)(0,0)%
\includegraphics{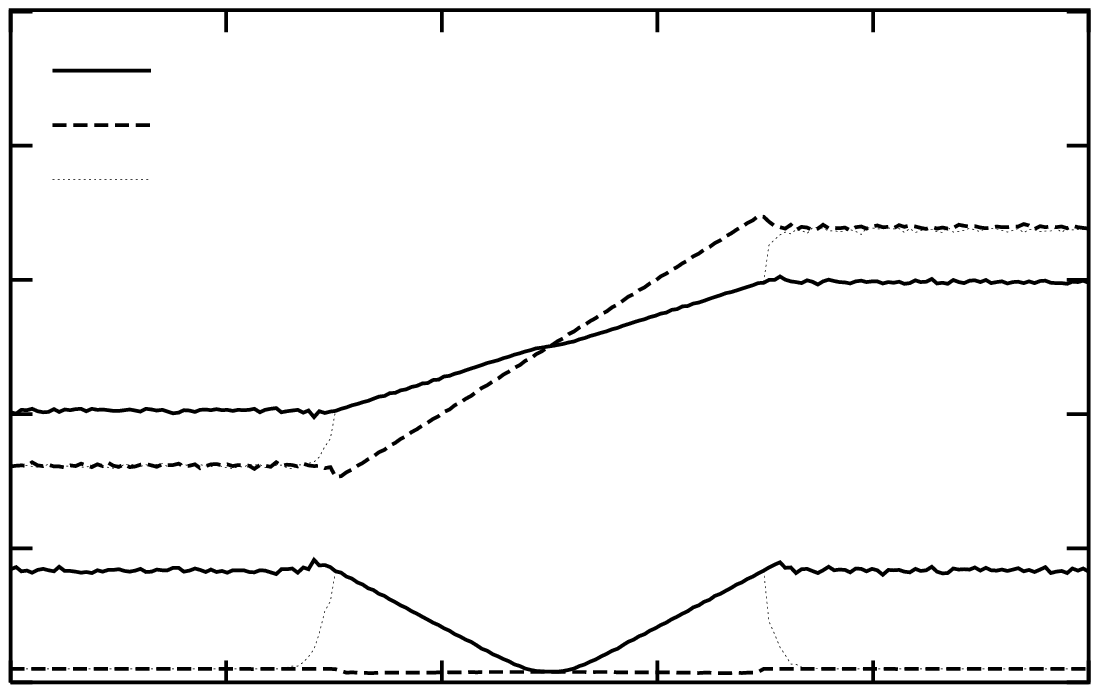}
\put(789,1778){\makebox(0,0)[l]{$\{h_t\}_t=\mc S$}}%
\put(789,1935){\makebox(0,0)[l]{$\{h_t\}_t=\mc R_{1.0}$}}%
\put(789,2092){\makebox(0,0)[l]{$\{h_t\}_t=\mc R_{0.5}$}}%
\put(1882,55){\makebox(0,0){$l$}}%
\put(3435,220){\makebox(0,0){1}}%
\put(2814,220){\makebox(0,0){0.8}}%
\put(2193,220){\makebox(0,0){0.6}}%
\put(1572,220){\makebox(0,0){0.4}}%
\put(951,220){\makebox(0,0){0.2}}%
\put(330,220){\makebox(0,0){0}}%
\put(275,2262){\makebox(0,0)[r]{500}}%
\put(275,1876){\makebox(0,0)[r]{400}}%
\put(275,1489){\makebox(0,0)[r]{300}}%
\put(275,1103){\makebox(0,0)[r]{200}}%
\put(275,716){\makebox(0,0)[r]{100}}%
\put(275,330){\makebox(0,0)[r]{0}}%
\end{picture}%
\endgroup

}
\end{center}
}

In Sec. \ref{sec:comptradml-3a} we showed that the effect of the prediction
$h_t$ is to reduce the agents' performance via its effect on $\at$. This being
the case we should expect that the effect of the prediction on the model would
be no different to that of an exogenous source provided that the value of
$\ave{h_t}_t$ is preserved.
In this section we shall check this by comparing the behavior of the
original model with a different memoryless variant. In this variant the
prediction $h_t$ will be generated by a random source, external to the model,
rather than taking the value $+1\;\forall\;t$. We shall let \R{\alpha} represent
the output of such a random exogenous source which contains only the two
values $-1$ and $+1$ and for which $\alpha$ is the time average, 
$\ave{\R\alpha}_t=\alpha$.
We represent the binary sequence generated by the memory for $h_t$ in the
original Genetic Model by \M. 

Figure \ref{fig:exog-3a} shows numerical results for $\ave{\at}_t$ in the
memoryless model with $h_t$ given by the exogenous sources \R{0.5} and \R{1.0}.
The results for the original model, $\{h_t\}_t=\mc S$, are included for
comparison. The results for $\{h_t\}_t=\R{1.0}$ duplicate those
presented in Fig. \ref{fig:attencomp-3a}b since $\{h_t\}_t=\R{1.0}$
is equivalent to $h_t=+1\;\forall\; t$. In other words taking
$\{h_t\}_t=\R{1.0}$ is exactly equivalent to the memoryless model that we
considered in previous sections.
%
%
In Fig. \ref{fig:exog-3a} the data produced using $\{h_t\}_t=\R{0.5}$ for
$\ave{\at}_t$ and \std{\at}t agrees with that from the original model for
$l_{c1}<l<l_{c2}$. For $l<l_{c1}$ and $l>l_{c2}$ the data from the original
model switches to agree with that from the memoryless model with $\{h_t\}_t=
\R{1.0}$ corresponding to the value of $\ave{h_t}_t$ from the original model
in these regions. Note there is no need to consider \R{-0.5} and \R{-1.0}. The
lack of physical significance attached to the labeling of the states of $h_t$
means that the model behaves equivalently for $\{h_t\}_t = \R{\pm\alpha}$.

These results confirm that the original Genetic Model and the memoryless Genetic Model with $h_t$
taken from an exogenous source can be regarded as equivalent when considering
$\ave{\at}_t$. In contrast we shall see in Sec. \ref{sec:autoc-3a} that
this does not apply when considering higher moments.

\subsection{The values of $\ave{h_t}_t$}
\label{sec:hbar-3a}

So far we have treated the values of $\ave{h_t}_t$ that obtain
for $l_{c1}<l<0.5$ and $0.5<l<l_{c2}$ as values to be derived empirically by
numerical simulation. Now we shall discuss the theoretical reasons for their
observed values.

Lo has presented a theory \cite{lo-thesis} that predicts, using our
conventions, the following values for $\ave{h_t}_t$:
\begin{equation}
\label{eqn:lovalues-3a}
	\ave{h_t}_t = \begin{cases}
	                 -1		& \text{For } l<l_{c1} \\
			 -\frac{1}{3}	& \text{For } l_{c1}<l<0.5 \\
			 +\frac{1}{3}	& \text{For } 0.5<l<l_{c2} \\
			 +1		& \text{For } l_{c2}<l
	              \end{cases}
\end{equation}
However numerical simulation robustly yields values of $\ave{h_t}_t\approx
\pm\frac{1}{2}$ in the dynamic regime. In what follows we summarize Lo's
analysis with the addition of some observations which explain why the numerical
and analytical results differ. Note that as we pointed out in Sec.
\ref{sec:perftrad-3a} the absolute values that obtain for $\ave{h_t}_t$ in the
dynamic regime are not important for the theory that we present here.
As long as $\ave{h_t}_t<0$ for $l_{c1}<l<0.5$ and $\ave{h_t}_t>0$ for
$0.5<l<l_{c2}$ everything that we have said about $\opt[\ind p]$ will remain
unchanged. Only the magnitude of the relative performance of the original and
memoryless Genetic Models depends upon the values taken by $\ave{h_t}_t$.

\floatfig[\floatplace]{fig:intexp-3a}{
Illustration of the summations necessary to calculate $\ave{h_t}_t$
}{
Illustration of the summations $I_1$ and $I_2$ in Eq. \eqref{eqn:avehI-3a}
needed to calculate $\ave{h_t}_t$ in the dynamic region of $l_{c1}<l<l_{c2}$.
}{
\begin{center}
\resizebox{0.48\textwidth}{!}{
\begin{picture}(0,0)%
\includegraphics{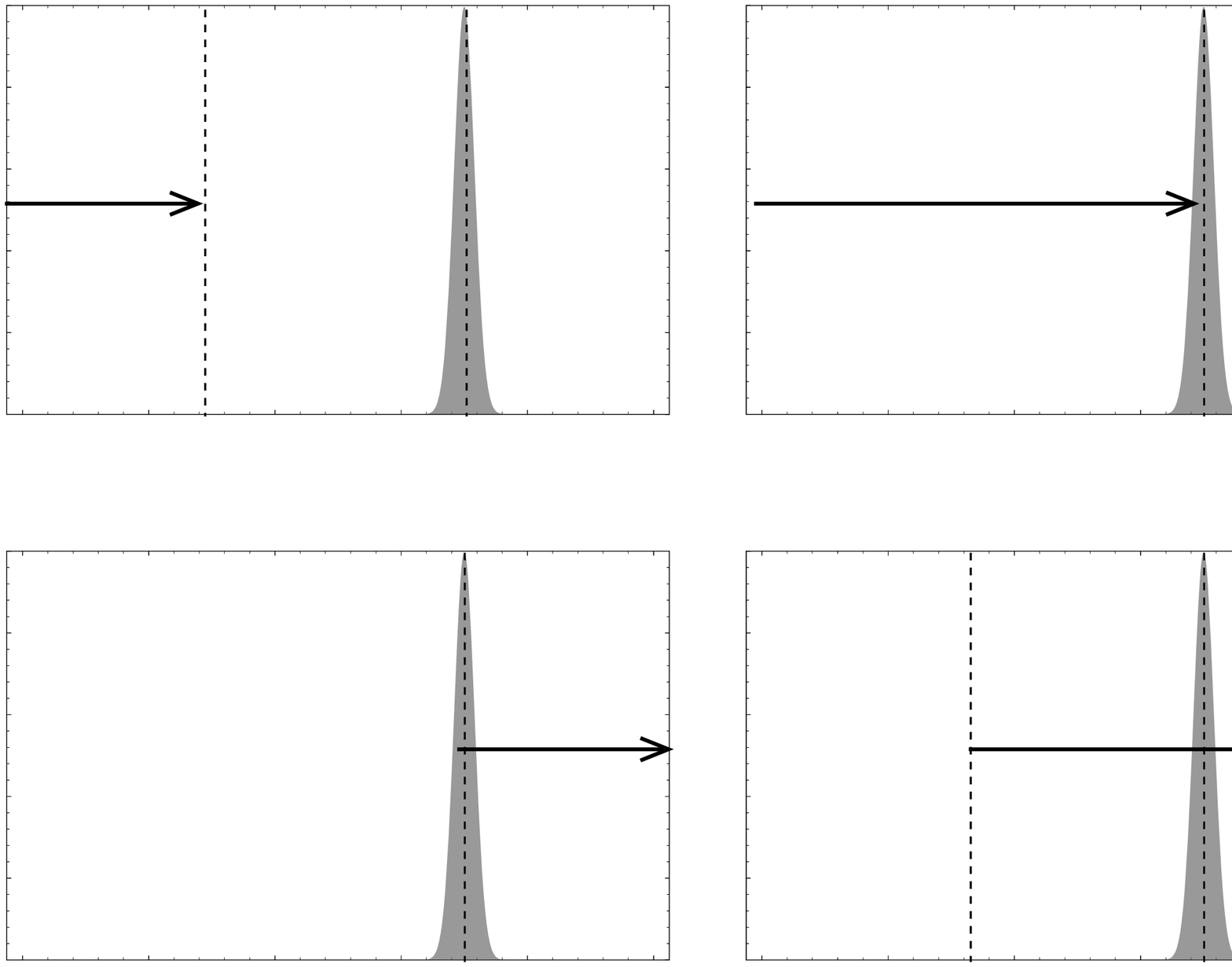}%
\end{picture}%
\setlength{\unitlength}{3947sp}%
\begingroup\makeatletter\ifx\SetFigFont\undefined%
\gdef\SetFigFont#1#2#3#4#5{%
  \fontsize{#1}{#2pt}%
  \fontfamily{#3}\fontseries{#4}\fontshape{#5}%
  \selectfont}%
\fi\endgroup%
\begin{picture}(11717,8544)(-97,-8139)
\put(3035,-3582){\makebox(0,0)[b]{\smash{\SetFigFont{20}{24.0}{\familydefault}{\mddefault}{\updefault}{\color[rgb]{0,0,0}\at}%
}}}
\put(9005,-3582){\makebox(0,0)[b]{\smash{\SetFigFont{20}{24.0}{\familydefault}{\mddefault}{\updefault}{\color[rgb]{0,0,0}\at}%
}}}
\put(682, 58){\makebox(0,0)[b]{\smash{\SetFigFont{20}{24.0}{\familydefault}{\mddefault}{\updefault}{\color[rgb]{0,0,0}a)}%
}}}
\put(682,-4382){\makebox(0,0)[b]{\smash{\SetFigFont{20}{24.0}{\familydefault}{\mddefault}{\updefault}{\color[rgb]{0,0,0}c)}%
}}}
\put(5435,-7742){\makebox(0,0)[b]{\smash{\SetFigFont{20}{24.0}{\familydefault}{\mddefault}{\updefault}{\color[rgb]{0,0,0}$N$}%
}}}
\put(350,-7742){\makebox(0,0)[b]{\smash{\SetFigFont{20}{24.0}{\familydefault}{\mddefault}{\updefault}{\color[rgb]{0,0,0}$0$}%
}}}
\put(5435,-3357){\makebox(0,0)[b]{\smash{\SetFigFont{20}{24.0}{\familydefault}{\mddefault}{\updefault}{\color[rgb]{0,0,0}$N$}%
}}}
\put(350,-3357){\makebox(0,0)[b]{\smash{\SetFigFont{20}{24.0}{\familydefault}{\mddefault}{\updefault}{\color[rgb]{0,0,0}$0$}%
}}}
\put(1865,-3357){\makebox(0,0)[b]{\smash{\SetFigFont{20}{24.0}{\familydefault}{\mddefault}{\updefault}{\color[rgb]{0,0,0}$lN$}%
}}}
\put(172,-5684){\rotatebox{90.0}{\makebox(0,0)[b]{\smash{\SetFigFont{20}{24.0}{\familydefault}{\mddefault}{\updefault}{\color[rgb]{0,0,0}$\Pi_t(n_t^{-1})$}%
}}}}
\put(3935,-3357){\makebox(0,0)[b]{\smash{\SetFigFont{20}{24.0}{\familydefault}{\mddefault}{\updefault}{\color[rgb]{0,0,0}$(1-l)N$}%
}}}
\put(3935,-7742){\makebox(0,0)[b]{\smash{\SetFigFont{20}{24.0}{\familydefault}{\mddefault}{\updefault}{\color[rgb]{0,0,0}$(1-l)N$}%
}}}
\put(6592,-4382){\makebox(0,0)[b]{\smash{\SetFigFont{20}{24.0}{\familydefault}{\mddefault}{\updefault}{\color[rgb]{0,0,0}d)}%
}}}
\put(6592, 58){\makebox(0,0)[b]{\smash{\SetFigFont{20}{24.0}{\familydefault}{\mddefault}{\updefault}{\color[rgb]{0,0,0}b)}%
}}}
\put(11510,-7742){\makebox(0,0)[b]{\smash{\SetFigFont{20}{24.0}{\familydefault}{\mddefault}{\updefault}{\color[rgb]{0,0,0}$lN$}%
}}}
\put(6290,-7742){\makebox(0,0)[b]{\smash{\SetFigFont{20}{24.0}{\familydefault}{\mddefault}{\updefault}{\color[rgb]{0,0,0}$0$}%
}}}
\put(8015,-7742){\makebox(0,0)[b]{\smash{\SetFigFont{20}{24.0}{\familydefault}{\mddefault}{\updefault}{\color[rgb]{0,0,0}$(1-l)N$}%
}}}
\put(11510,-3357){\makebox(0,0)[b]{\smash{\SetFigFont{20}{24.0}{\familydefault}{\mddefault}{\updefault}{\color[rgb]{0,0,0}$lN$}%
}}}
\put(6290,-3357){\makebox(0,0)[b]{\smash{\SetFigFont{20}{24.0}{\familydefault}{\mddefault}{\updefault}{\color[rgb]{0,0,0}$0$}%
}}}
\put(6112,-1184){\rotatebox{90.0}{\makebox(0,0)[b]{\smash{\SetFigFont{20}{24.0}{\familydefault}{\mddefault}{\updefault}{\color[rgb]{0,0,0}$\Pi_t(\at)$}%
}}}}
\put(6112,-5684){\rotatebox{90.0}{\makebox(0,0)[b]{\smash{\SetFigFont{20}{24.0}{\familydefault}{\mddefault}{\updefault}{\color[rgb]{0,0,0}$\Pi_t(n_t^{-1})$}%
}}}}
\put(9890,-3357){\makebox(0,0)[b]{\smash{\SetFigFont{20}{24.0}{\familydefault}{\mddefault}{\updefault}{\color[rgb]{0,0,0}$lN$}%
}}}
\put(9905,-7742){\makebox(0,0)[b]{\smash{\SetFigFont{20}{24.0}{\familydefault}{\mddefault}{\updefault}{\color[rgb]{0,0,0}$lN$}%
}}}
\put(172,-1184){\rotatebox{90.0}{\makebox(0,0)[b]{\smash{\SetFigFont{20}{24.0}{\familydefault}{\mddefault}{\updefault}{\color[rgb]{0,0,0}$\Pi_t(\at)$}%
}}}}
\put(3035,-8139){\makebox(0,0)[b]{\smash{\SetFigFont{20}{24.0}{\familydefault}{\mddefault}{\updefault}{\color[rgb]{0,0,0}$n_t^{-1}$}%
}}}
\put(9005,-8139){\makebox(0,0)[b]{\smash{\SetFigFont{20}{24.0}{\familydefault}{\mddefault}{\updefault}{\color[rgb]{0,0,0}$n_t^{-1}$}%
}}}
\put(3020,-4364){\makebox(0,0)[b]{\smash{\SetFigFont{20}{24.0}{\familydefault}{\mddefault}{\updefault}{\color[rgb]{0,0,0}$l<0.5$}%
}}}
\put(9080,-4364){\makebox(0,0)[b]{\smash{\SetFigFont{20}{24.0}{\familydefault}{\mddefault}{\updefault}{\color[rgb]{0,0,0}$l>0.5$}%
}}}
\put(3020,  1){\makebox(0,0)[b]{\smash{\SetFigFont{20}{24.0}{\familydefault}{\mddefault}{\updefault}{\color[rgb]{0,0,0}$l<0.5$}%
}}}
\put(9080,  1){\makebox(0,0)[b]{\smash{\SetFigFont{20}{24.0}{\familydefault}{\mddefault}{\updefault}{\color[rgb]{0,0,0}$l>0.5$}%
}}}
\put(5122,-17){\makebox(0,0)[b]{\smash{\SetFigFont{25}{30.0}{\familydefault}{\mddefault}{\updefault}{\color[rgb]{0,0,0}$I_1$}%
}}}
\put(5122,-4517){\makebox(0,0)[b]{\smash{\SetFigFont{25}{30.0}{\familydefault}{\mddefault}{\updefault}{\color[rgb]{0,0,0}$I_2$}%
}}}
\put(11032,-4517){\makebox(0,0)[b]{\smash{\SetFigFont{25}{30.0}{\familydefault}{\mddefault}{\updefault}{\color[rgb]{0,0,0}$I_2$}%
}}}
\put(11032,-17){\makebox(0,0)[b]{\smash{\SetFigFont{25}{30.0}{\familydefault}{\mddefault}{\updefault}{\color[rgb]{0,0,0}$I_1$}%
}}}
\end{picture}
}
\end{center}
}
\floatfig[\floatplace]{fig:ndist-3a}{
Numerical data for $\ind{\Pi}$.
}{
Numerical data for $\ind{\Pi}$ illustrating that $\ave{\fol}_t$ deviates from
the value of $\opt[\fol]$ predicted by Eq. \eqref{eqn:optfol-3a}. The numbers
under the curve represent the sum of the values of the data points lying on
either side of the cutoff indicated by the dashed line. Note that this cutoff is
at $\fol=Nl$ if $l>0.5$ and $\fol=N(1-l)$ otherwise.
The model parameters were as given in Tab.
\ref{tab:params-3a} except that $n=10000$ and the value of \fol\ was sampled
over the interval $2000<t<102000$.
}{
\begin{center}
\resizebox{0.48\textwidth}{!}{
\begin{picture}(0,0)%
\includegraphics{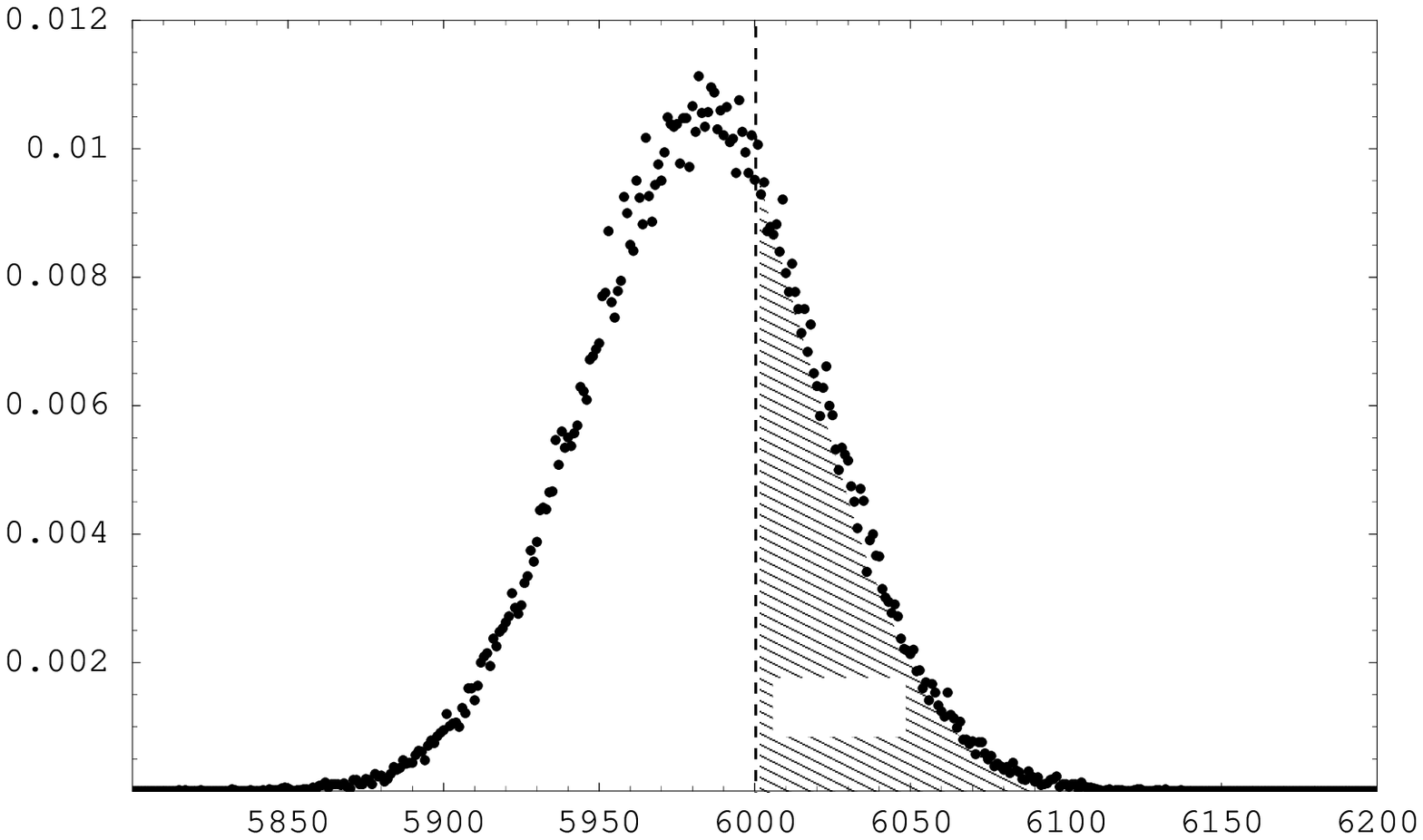}%
\end{picture}%
\setlength{\unitlength}{3947sp}%
\begingroup\makeatletter\ifx\SetFigFont\undefined%
\gdef\SetFigFont#1#2#3#4#5{%
  \fontsize{#1}{#2pt}%
  \fontfamily{#3}\fontseries{#4}\fontshape{#5}%
  \selectfont}%
\fi\endgroup%
\begin{picture}(8254,5085)(1157,-6977)
\put(1517,-4277){\rotatebox{90.0}{\makebox(0,0)[b]{\smash{\SetFigFont{20}{24.0}{\familydefault}{\mddefault}{\updefault}{\color[rgb]{0,0,0}$\ind{\Pi}$}%
}}}}
\put(8987,-6977){\makebox(0,0)[b]{\smash{\SetFigFont{20}{24.0}{\familydefault}{\mddefault}{\updefault}{\color[rgb]{0,0,0}\fol}%
}}}
\put(5696,-6902){\makebox(0,0)[b]{\smash{\SetFigFont{14}{16.8}{\familydefault}{\mddefault}{\updefault}{\color[rgb]{0,0,0}$Nl$}%
}}}
\put(5696,-2117){\makebox(0,0)[b]{\smash{\SetFigFont{14}{16.8}{\familydefault}{\mddefault}{\updefault}{\color[rgb]{0,0,0}$N(1-l)$}%
}}}
\put(3746,-6902){\makebox(0,0)[lb]{\smash{\SetFigFont{14}{16.8}{\familydefault}{\mddefault}{\updefault}{\color[rgb]{0,0,0}$l>0.5$:}%
}}}
\put(3746,-2117){\makebox(0,0)[lb]{\smash{\SetFigFont{14}{16.8}{\familydefault}{\mddefault}{\updefault}{\color[rgb]{0,0,0}$l<0.5$:}%
}}}
\put(5017,-6039){\makebox(0,0)[b]{\smash{\SetFigFont{20}{24.0}{\familydefault}{\mddefault}{\updefault}{\color[rgb]{0,0,0}0.677}%
}}}
\put(6172,-6039){\makebox(0,0)[b]{\smash{\SetFigFont{20}{24.0}{\familydefault}{\mddefault}{\updefault}{\color[rgb]{0,0,0}0.323}%
}}}
\end{picture}
}
\end{center}
}

The analysis hinges on the observation that:
\begin{equation}
\label{eqn:Ahequiv-3a}
	\ave{A_t}_t = \ave{h_t}_t \;.
\end{equation}
From Eq. \ref{eqn:folcond-3a} and the definition of $\ind{\Pi}$ the probability
that $A_t=+1$ is given by:
\begin{equation} 
\label{eqn:probAone-3a}
	P[A_t=+1] = \begin{cases}
	 	      I_2 & \f h_t = -1 \\
		      I_1 & \f h_t = +1 
		    \end{cases}
\end{equation}
where:
\begin{align}
\label{eqn:Idefs-3a}
	I_1 = \sum\limits_{i=0}^{Nl}  \ind{\Pi}        &&
	I_2 = \sum\limits_{i=N(1-l)}^{N} \ind{\Pi} \; .
\end{align}
Therefore, the ensemble average values of the global action $A_t$ when $h_t=+1$
and $-1$ are given by:
\begin{equation}
	\ave{A_t} = \begin{cases}
	  2 I_2 -1 & \text{If } h_t=-1\\
	  2 I_1 -1 & \text{If } h_t=+1
		      \end{cases} \; .
\end{equation}
The time average of the global action is given by the sum of the values
of $\ave{A_t}$ for $h_t=\pm1$ weighted by the probabilities that 
$h_t=\pm1$. Thus:
\begin{equation}
	\ave{A_t}_t = \left(\frac{\ave{h_t}_t+1}{2}\right)(2I_1-1) +
	              \left(\frac{1-\ave{h_t}_t}{2}\right)(2I_2-1) 
\end{equation}
Finally, using Eq. \eqref{eqn:Ahequiv-3a} gives:
\begin{equation}
\label{eqn:avehI-3a}
	\ave{h_t}_t = \frac{I_1+I_2-1}{I_2-I_1+1} \; .
\end{equation}
In Fig. \ref{fig:intexp-3a} we illustrate the summations over $\ind{\Pi}$ in Eq.
\eqref{eqn:Idefs-3a} that define the values of $I_1$ and $I_2$. For the purposes
of calculating these values, we have taken the mean of the distribution to be
given by $\mu=\opt[\fol]$ from Eq.  \eqref{eqn:optfol-3a} and the standard
deviation to be $\sigma=8$. 
We can see from Fig. \ref{fig:intexp-3a} that the values for
$I_1$ and $I_2$ will be given by:
\begin{align}
\label{eqn:Ivals-3a}
	I_1=\begin{cases}0&\text{if }l<0.5\\\half&\text{if }l>0.5
	    \end{cases}  &&
	I_2=\begin{cases}\half&\text{if }l<0.5\\1&\text{if }l>0.5
	    \end{cases}  \;,
\end{align}
since the summations either cover the entire peak of the Gaussian, exactly half
of it or none at all.
From Eqs. \eqref{eqn:avehI-3a} and \eqref{eqn:Ivals-3a} we obtain values for
$\ave{h_t}_t$ in the dynamic regime of:
\begin{equation}
	\ave{h_t}_t = \begin{cases}
	                -\frac{1}{3} & \text{if }l<0.5\\
	                +\frac{1}{3} & \text{if }l>0.5\\
		      \end{cases} \; ,
\end{equation}
which agree exactly with the values found by Lo in Ref. \cite{lo-thesis}.

The values given in Eq. \eqref{eqn:Ivals-3a} depend on the assumption that 
$\ave{\fol}_t$ is given by the optimal value of Eq. \eqref{eqn:optfol-3a}.
In fact, as we see in Fig. \ref{fig:ndist-3a}, numerical simulation using a
large number of agents reveals that this is not the case. In the figure
$\opt[\fol]-\ave{\fol}_t\approx 0.002N$ corresponding to 
$\opt[\overline p_t] - \ave{\overline p_t}_t \approx 0.002$. Careful observation
of Fig.  \ref{fig:mphis-3a}a reveals that $\ave{\overline p_t}_t$ always lies
slightly closer to $0.5$ than the value predicted by Eq. \eqref{eqn:optp-3a} in
the original model. As we can see from Fig. \ref{fig:ndist-3a} this changes
the values of the summations in Eq. \eqref{eqn:Ivals-3a} as follows:
\begin{align}
\label{eqn:Ivalscorrected-3a}
	I_1=\begin{cases}0&\text{if }l<0.5\\0.677&\text{if } l>0.5
	    \end{cases}  &&
	I_2=\begin{cases}0.323&\text{if }l<0.5\\1&\text{if }l>0.5
	    \end{cases}  \;,
\end{align}
which by Eq. \eqref{eqn:avehI-3a} yields the following values for $\ave{h_t}_t$:
\begin{equation}
	\ave{h_t}_t = \begin{cases}
	                -0.511 & \text{if }l<0.5\\
	                +0.511 & \text{if }l>0.5\\
		      \end{cases} \; ,
\end{equation}
These figures agree much more closely with the numerically observed values of
$\ave{h_t}_t = \pm 0.5$ than the values obtained in Eq. \eqref{eqn:Ivals-3a}.

We can therefore see that the values of $\ave{h_t}_t=\pm 0.5$ that obtain in
the dynamic regime result from the fact that $\overline p_t$ deviates slightly
from the optimal value of Eq. \eqref{eqn:optp-3a}. In the absence of this
deviation we would indeed obtain the values given in Eq. \eqref{eqn:lovalues-3a}
for $\ave{h_t}_t$. We can see that the memory does indeed have a non-trivial
role to  play in the Genetic Model. The feedback between the prediction $h_t$
and the global action $A_t$ represented by Eq. \eqref{eqn:Ahequiv-3a} controls
the values of $\ave{h_t}_t$ that obtain in the dynamic regime $l_{c1}<l<l_{c2}$.

\subsection{Deviation of \pbar\ from its optimal value}
\label{sec:dev-3a}

In Sec. \ref{sec:comptradml-3a} when we calculated the optimal values of 
\pbar\ in Eq. \eqref{eqn:optp-3a} we implicitly assumed that the gene value
distribution is given by \eqref{eqn:idealP-3a} and that therefore $\sigma=0$.
If $\sigma$ is non-zero then the values of \at\ will be spread about
$\ave{\at}_t$. An examination of Fig. \ref{fig:utilmeanplusmemless-3a} reveals
that the reduction in \gu\ resulting from a deviation of \at\ towards $0$ is
less than that of a deviation towards $N$. Therefore we might expect that the
agents would evolve such that \pbar\ was less than $\opt[\pbar]$ when $\sigma>0$
in order to compensate for this.
This argument also applies to the memoryless model. We know from
Eqs. \eqref{eqn:guave_a-3a}, \eqref{eqn:guave_b-3a} \eqref{eqn:guml-3a} and Fig.
\ref{fig:utilmeanplusmemless-3a} that the change that occurs in \gu\ at
$\at=Nl$ in the memoryless model is greater than that which occurs in the
original model. Therefore, if the argument presented above were correct, we
would expect a larger deviation in the memoryless case. However, we can see in
Fig.  \ref{fig:mphis-3a} that the deviation of $\overline p_t$ from
$\opt[\overline p_t]$ in the memoryless model is less than that in the
original model since $\ave{\overline p_t}_t$ lies closer to the line
described by $\overline p_t=l$
in the figure. Figure \ref{fig:mlndist-3a} shows numerical data for $\ind{\Pi}$
in the memoryless model. It is the equivalent of Fig. \ref{fig:ndist-3a}. 
Any deviation of the peak of $\ind{\Pi}$ in Fig.
\ref{fig:mlndist-3a} from the optimal value given in Eq. \eqref{eqn:optpml-3a}
is too small to be observable.  Thus we can see that the observed deviation
$\opt[\overline p_t] - \overline p_t$ in the memoryless model is much smaller
than that in the original model.  
Therefore we conclude that a different effect, one most probably related to the
variation of the prediction $h_t$, must be responsible for the deviation
observed in the original model. 

We note in passing that since $\ave{\overline
p_t}_t$ deviates from $\opt[\overline p_t]$ given by Eq. \eqref{eqn:optp-3a}
in the original model while no such deviation is observed in the memoryless
model, the \pval\ distribution of the agents in the two models differs for
general $l$. We would expect this difference to vanish in the case of $l=0.5$,
as found by Burgos and Ceva \cite{ceva-mem}, and for the frozen regimes of
$l<l_{c1}$ and $l>l_{c2}$ in which the original and memoryless models are
equivalent.

\floatfig[\floatplace]{fig:mlndist-3a}{
Numerical data for $\ind{\Pi}$ in the memoryless model.
}{
Numerical data for $\ind{\Pi}$ illustrating that the deviation 
$\opt[\at]-\ave{\at}_t$ in the memoryless model is much less than that in the
original model with memory.
The dashed line represents the value of \at\ above which $A_t=-1$.
The model parameters were as given in Tab.  \ref{tab:params-3a} with $l=0.6$,
except that $n=10000$ and the value of \fol\ was sampled over the interval
$2000<t<102000$.
}{
\begin{center}
\resizebox{0.48\textwidth}{!}{
\begin{picture}(0,0)%
\includegraphics{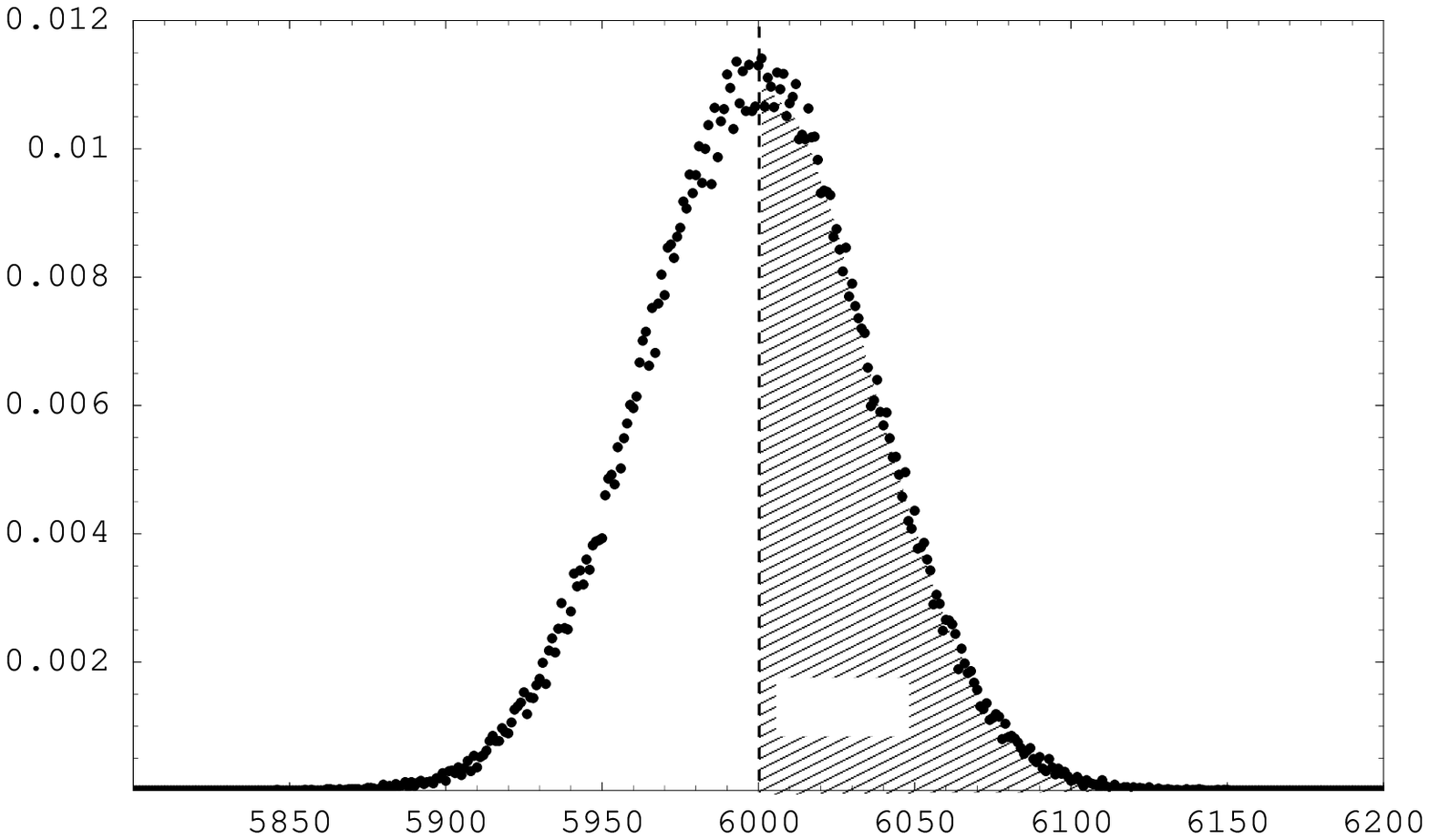}%
\end{picture}%
\setlength{\unitlength}{3947sp}%
\begingroup\makeatletter\ifx\SetFigFont\undefined%
\gdef\SetFigFont#1#2#3#4#5{%
  \fontsize{#1}{#2pt}%
  \fontfamily{#3}\fontseries{#4}\fontshape{#5}%
  \selectfont}%
\fi\endgroup%
\begin{picture}(8321,5045)(1111,-6961)
\put(9031,-6961){\makebox(0,0)[b]{\smash{\SetFigFont{20}{24.0}{\familydefault}{\mddefault}{\updefault}{\color[rgb]{0,0,0}\at}%
}}}
\put(1471,-4276){\rotatebox{90.0}{\makebox(0,0)[b]{\smash{\SetFigFont{20}{24.0}{\familydefault}{\mddefault}{\updefault}{\color[rgb]{0,0,0}$\ind{\Pi}$}%
}}}}
\put(5696,-6902){\makebox(0,0)[b]{\smash{\SetFigFont{14}{16.8}{\familydefault}{\mddefault}{\updefault}{\color[rgb]{0,0,0}$Nl$}%
}}}
\put(5017,-6039){\makebox(0,0)[b]{\smash{\SetFigFont{20}{24.0}{\familydefault}{\mddefault}{\updefault}{\color[rgb]{0,0,0}0.512}%
}}}
\put(6172,-6039){\makebox(0,0)[b]{\smash{\SetFigFont{20}{24.0}{\familydefault}{\mddefault}{\updefault}{\color[rgb]{0,0,0}0.488}%
}}}
\end{picture}
}
\end{center}
}

\subsection{Autocorrelation of \at}
\label{sec:autoc-3a}

In this section we consider another property of the Genetic Model: the
autocorrelation of the time series \at. First of all we shall present
numerical data contrasting the behavior of the autocorrelation in the
original and memoryless models. We shall then consider a Markovian analysis
of the original Genetic Model which explains the behavior of the
autocorrelation of \at\ observed therein in the simplest case of $m=1$.

\subsubsection{Numerical Results}

We define the autocorrelation \autoc{x_t}{\tau}\ of a time series $x_t$ as
follows:
\begin{equation}
\label{eqn:autocdef-3a}
  \autoc{x_t}{\tau} = \frac{\ave{x_tx_{t+\tau}}_t-\ave{x_t}_t^2}
                           {\ave{x_t^2}_t-\ave{x_t}_t^2} \;,
\end{equation}
where $\tau$ gives the lag time.

Figure \ref{fig:autocwithmem-3a} shows the autocorrelation of the attendance
time series \autoc{\at}{\tau} for 
$0\le\tau\le10$ and $1\le m\le5$. We can see from the figure that for $l=0.5$
there is no significant correlation for $\tau>0$. For $l\ne0.5$, in
contrast, \autoc{\at}{\tau} as a function of $\tau$ has a clear structure. The
magnitude of \autoc{\at}{\tau} is non-zero for $\tau=(m+1)i$, where
$i=1,2,3\ldots$. \at\ and \genat{t+\tau} are anticorrelated for odd values of
$i$ and correlated for even values. 
Figure \ref{fig:autocranmem-3a} shows \autoc{\at}{\tau} for the same
values of $\tau$ and $l$ depicted in Fig. \ref{fig:autocwithmem-3a}.
However, this time $h_t$ was derived from a random exogenous source rather
than the memory. It is clear from a comparison of Figs.
\ref{fig:autocwithmem-3a} and \ref{fig:autocranmem-3a} that the structure
observed in \autoc{\at}{\tau} in Fig. \ref{fig:autocwithmem-3a} is only
present in the original model with memory.
Figure \ref{fig:hautocwithmem-3a} shows the autocorrelation of the
prediction time series $h_t$. From this we can see that the structure present
in Fig. \ref{fig:autocwithmem-3a} derives from structure present in the
prediction time series.

From Figs. \ref{fig:autocwithmem-3a}-\ref{fig:hautocwithmem-3a} we
can see that one of the functions of the memory is to introduce non-zero
autocorrelations into the prediction time series $h_t$. These will clearly not
be present in either of the memoryless models in which either
$h_t=+1\;\;\forall\;\;t$ or $h_t=\R{\alpha}$. In Ref. \cite{ceva-mem} Burgos
\ea\ found that for $l=0.5$ the gene value distribution functions in the Genetic
Model and the memoryless Genetic Model are equivalent. Further in Ref.
\cite{ceva-quench} Burgos \ea\ stated that ``a simplified version of the model,
\emph{that makes no use of memory} is indistinguishable from the original
formulation''. In contrast however, the numerical data presented in Figs.
\ref{fig:autocwithmem-3a} and \ref{fig:hautocwithmem-3a} demonstrates that the
two formulations \emph{are} distinguishable. While the memoryless model yields
equivalent results for $\ave{\at}_t$, it does not do so for higher moments. Any
analysis in which possible autocorrelations in the \at\ time series could be a
factor could not be conducted using the memoryless model. However, this effect
would not be noticeable in an investigation restricted to the minority case of
$l=0.5$.

\bigfloatfig[\floatplace]{fig:autocwithmem-3a}{
Autocorrelation of the attendance time series.
}{
Autocorrelation of the \at\ time series for the range of values of $m$ shown.
Model parameters: $n=1000$, a) $l=0.4$. b) $l=0.5$.
}{
\begin{center}
\resizebox{0.48\textwidth}{!}{%
\begingroup%
  \makeatletter%
  \newcommand{\GNUPLOTspecial}{%
    \catcode`\%=14\relax\special}%
  \setlength{\unitlength}{0.1bp}%
\begin{picture}(3600,3240)(0,0)%
\includegraphics{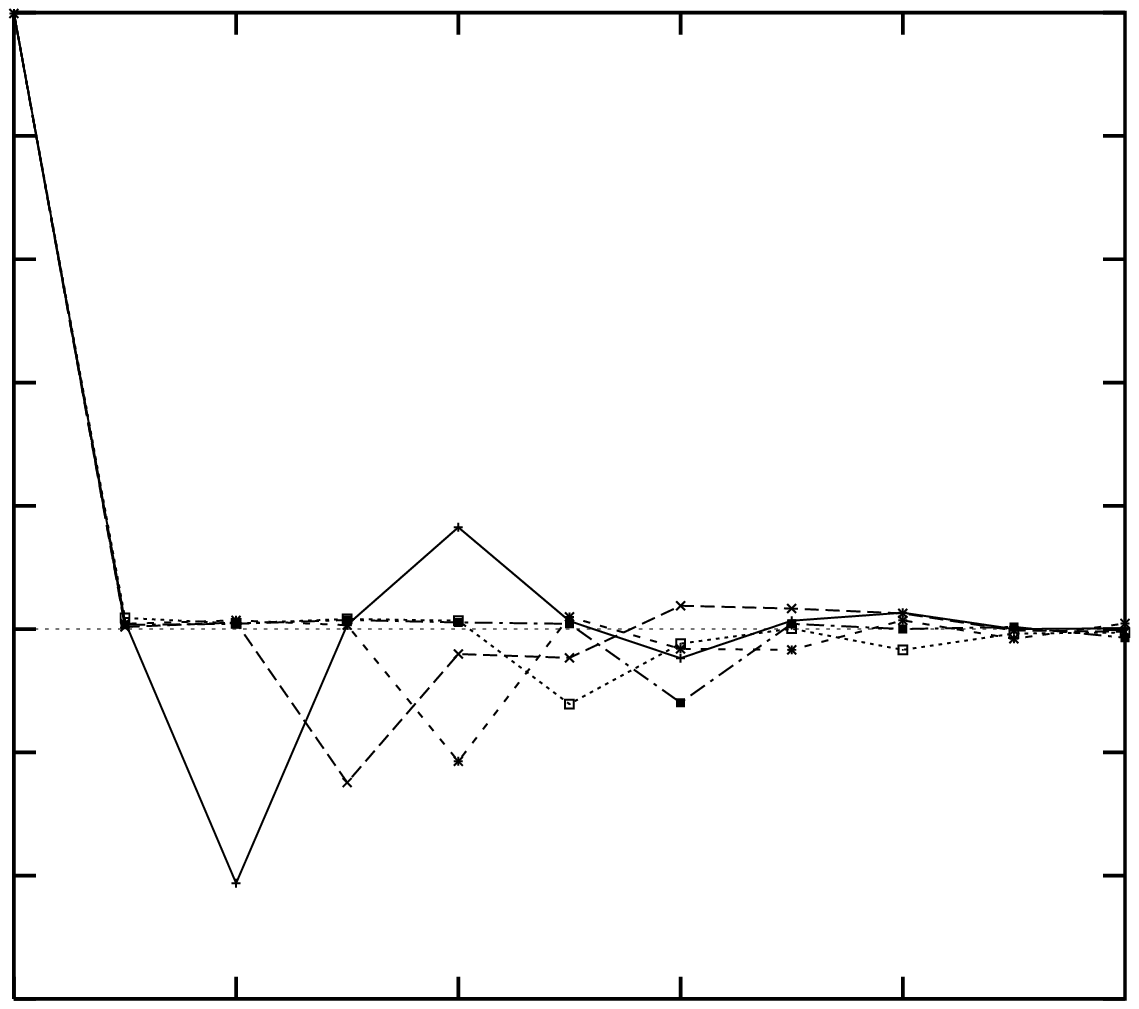}
\put(640,2856){\makebox(0,0)[l]{\Large a)}}%
\put(2000,50){\makebox(0,0){Lag: $\tau$}}%
\put(50,1720){%
\makebox(0,0)[b]{\shortstack{\autoc{\at}{\tau}}}%
}%
\put(2960,200){\makebox(0,0){8}}%
\put(2320,200){\makebox(0,0){6}}%
\put(1680,200){\makebox(0,0){4}}%
\put(1040,200){\makebox(0,0){2}}%
\put(400,200){\makebox(0,0){0}}%
\put(350,3140){\makebox(0,0)[r]{1}}%
\put(350,2785){\makebox(0,0)[r]{0.8}}%
\put(350,2430){\makebox(0,0)[r]{0.6}}%
\put(350,2075){\makebox(0,0)[r]{0.4}}%
\put(350,1720){\makebox(0,0)[r]{0.2}}%
\put(350,1365){\makebox(0,0)[r]{0}}%
\put(350,1010){\makebox(0,0)[r]{-0.2}}%
\put(350,655){\makebox(0,0)[r]{-0.4}}%
\put(350,300){\makebox(0,0)[r]{-0.6}}%
\end{picture}%
\endgroup

}\resizebox{0.48\textwidth}{!}{%
\begingroup%
  \makeatletter%
  \newcommand{\GNUPLOTspecial}{%
    \catcode`\%=14\relax\special}%
  \setlength{\unitlength}{0.1bp}%
\begin{picture}(3600,3240)(0,0)%
\includegraphics{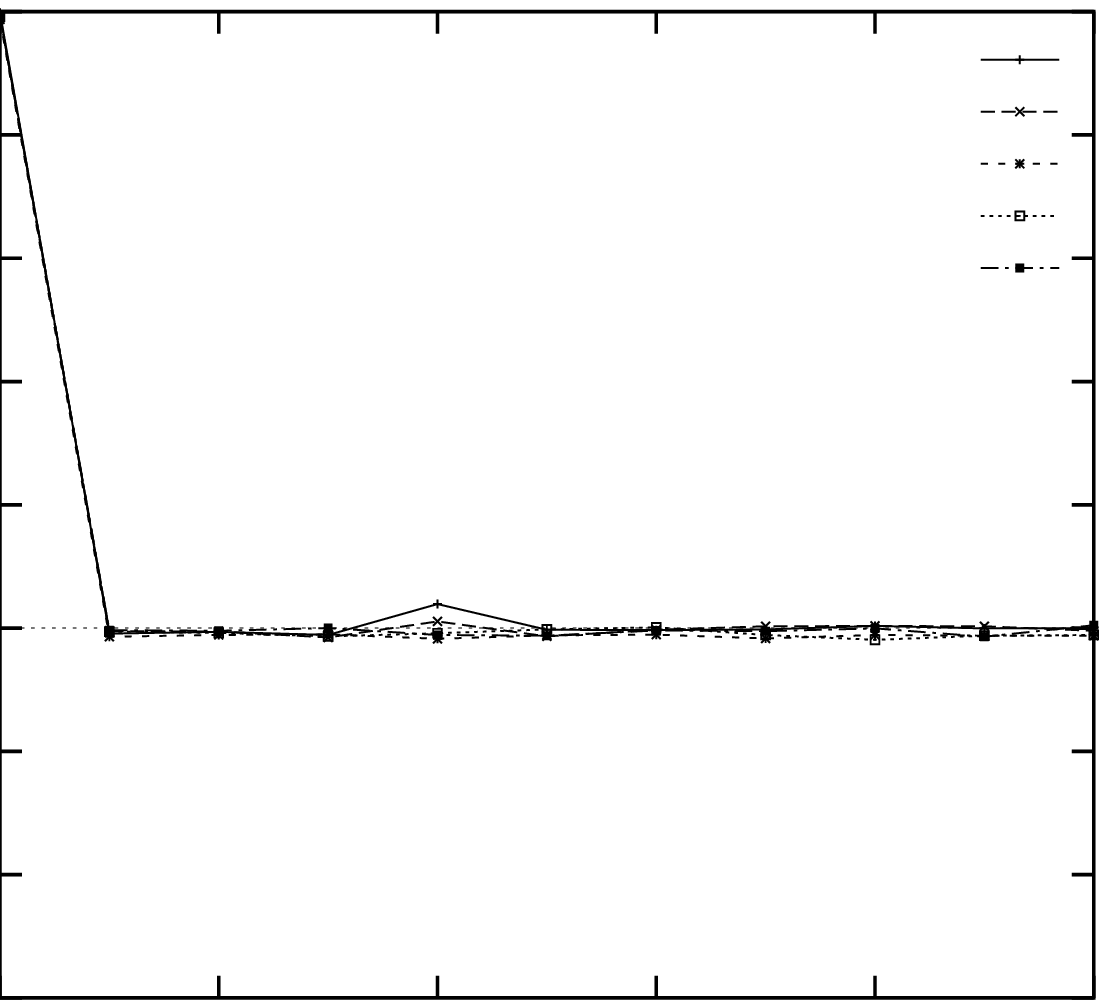}
\put(2775,2402){\makebox(0,0)[r]{$m=5$}}%
\put(2775,2552){\makebox(0,0)[r]{$m=4$}}%
\put(2775,2702){\makebox(0,0)[r]{$m=3$}}%
\put(2775,2852){\makebox(0,0)[r]{$m=2$}}%
\put(2775,3002){\makebox(0,0)[r]{$m=1$}}%
\put(236,2856){\makebox(0,0)[l]{\Large b)}}%
\put(1575,50){\makebox(0,0){Lag: $\tau$}}%
\put(3150,200){\makebox(0,0){10}}%
\put(2520,200){\makebox(0,0){8}}%
\put(1890,200){\makebox(0,0){6}}%
\put(1260,200){\makebox(0,0){4}}%
\put(630,200){\makebox(0,0){2}}%
\put(0,200){\makebox(0,0){0}}%
\end{picture}%
\endgroup

}
\end{center}
}
\bigfloatfig[\floatplace]{fig:autocranmem-3a}{
Autocorrelation of the attendance time series with the prediction given by
a random exogenous source.
}{
Autocorrelation of the \at\ time series with $h_t$ given by a random exogenous
source.
Model parameters: $n=1000$. a) $l=0.4$, $h_t=\R{-0.5}$. b) $l=0.5$,
$h_t=\R{0.0}$.
}{
\begin{center}
\resizebox{0.48\textwidth}{!}{%
\begingroup%
  \makeatletter%
  \newcommand{\GNUPLOTspecial}{%
    \catcode`\%=14\relax\special}%
  \setlength{\unitlength}{0.1bp}%
\begin{picture}(3600,3240)(0,0)%
\includegraphics{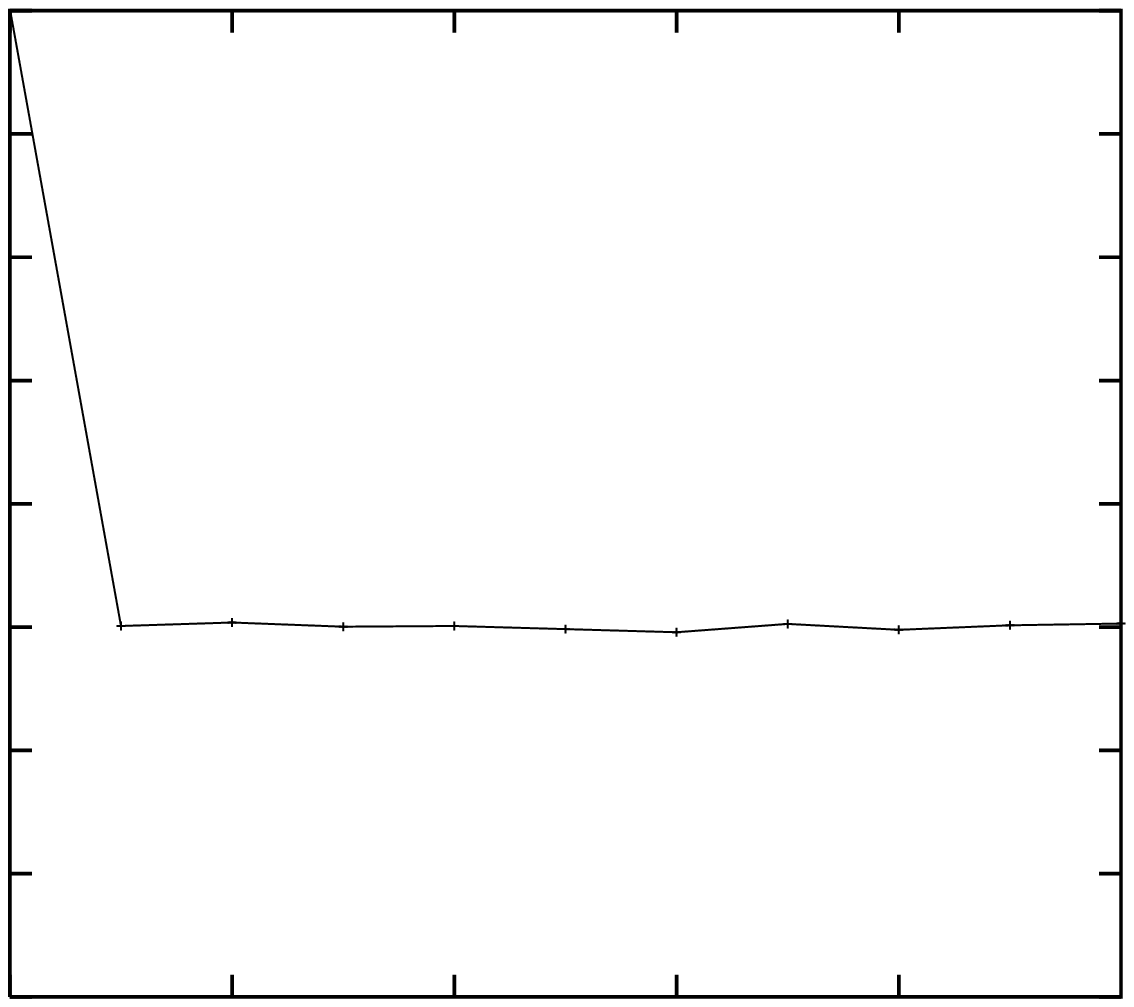}
\put(640,2856){\makebox(0,0)[l]{\Large a)}}%
\put(2000,50){\makebox(0,0){Lag: $\tau$}}%
\put(50,1720){%
\makebox(0,0)[b]{\shortstack{\autoc{\at}{\tau}}}%
}%
\put(2960,200){\makebox(0,0){8}}%
\put(2320,200){\makebox(0,0){6}}%
\put(1680,200){\makebox(0,0){4}}%
\put(1040,200){\makebox(0,0){2}}%
\put(400,200){\makebox(0,0){0}}%
\put(350,3140){\makebox(0,0)[r]{1}}%
\put(350,2785){\makebox(0,0)[r]{0.8}}%
\put(350,2430){\makebox(0,0)[r]{0.6}}%
\put(350,2075){\makebox(0,0)[r]{0.4}}%
\put(350,1720){\makebox(0,0)[r]{0.2}}%
\put(350,1365){\makebox(0,0)[r]{0}}%
\put(350,1010){\makebox(0,0)[r]{-0.2}}%
\put(350,655){\makebox(0,0)[r]{-0.4}}%
\put(350,300){\makebox(0,0)[r]{-0.6}}%
\end{picture}%
\endgroup

}\resizebox{0.48\textwidth}{!}{%
\begingroup%
  \makeatletter%
  \newcommand{\GNUPLOTspecial}{%
    \catcode`\%=14\relax\special}%
  \setlength{\unitlength}{0.1bp}%
\begin{picture}(3600,3240)(0,0)%
\includegraphics{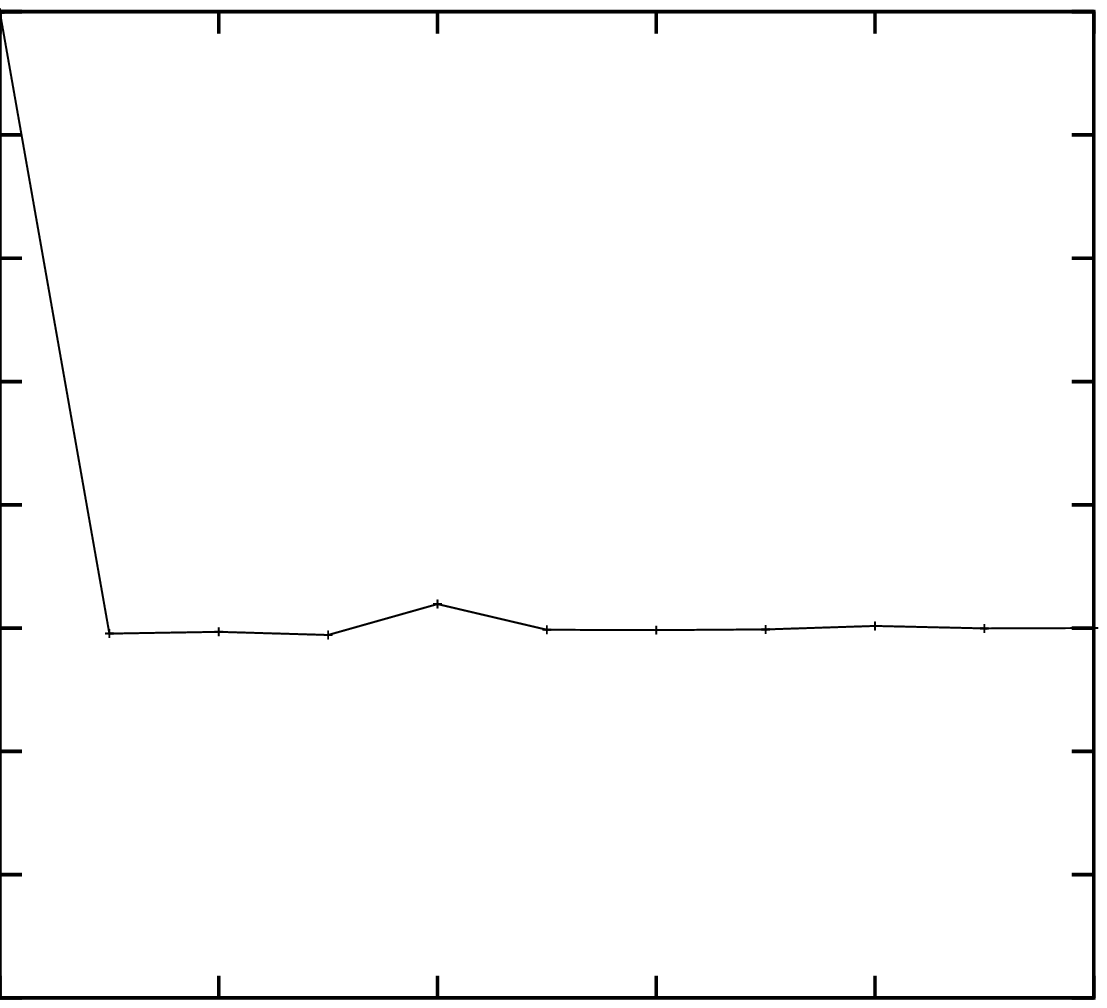}
\put(236,2856){\makebox(0,0)[l]{\Large b)}}%
\put(1575,50){\makebox(0,0){Lag: $\tau$}}%
\put(3150,200){\makebox(0,0){10}}%
\put(2520,200){\makebox(0,0){8}}%
\put(1890,200){\makebox(0,0){6}}%
\put(1260,200){\makebox(0,0){4}}%
\put(630,200){\makebox(0,0){2}}%
\put(0,200){\makebox(0,0){0}}%
\end{picture}%
\endgroup

}
\end{center}
}
\bigfloatfig[\floatplace]{fig:hautocwithmem-3a}{
Autocorrelation of the prediction time series.
}{
Autocorrelation of the $h_t$ time series for the range of values of $m$ shown.
Model parameters: $n=1000$, a) $l=0.4$. b) $l=0.5$.
}{
\begin{center}
\resizebox{0.48\textwidth}{!}{%
\begingroup%
  \makeatletter%
  \newcommand{\GNUPLOTspecial}{%
    \catcode`\%=14\relax\special}%
  \setlength{\unitlength}{0.1bp}%
\begin{picture}(3600,3240)(0,0)%
\includegraphics{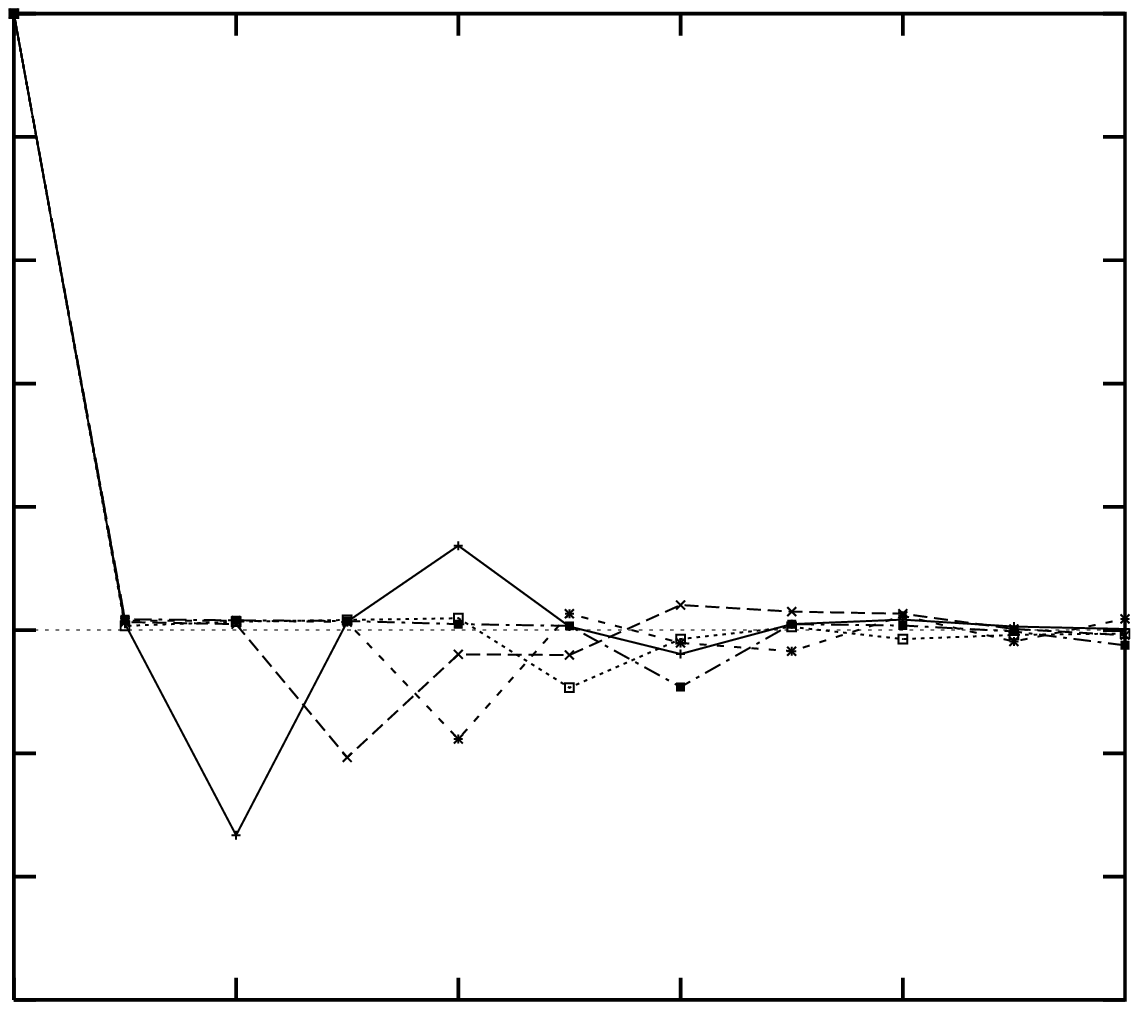}
\put(640,2856){\makebox(0,0)[l]{\Large a)}}%
\put(2000,50){\makebox(0,0){Lag: $\tau$}}%
\put(50,1720){%
\makebox(0,0)[b]{\shortstack{\autoc{h_t}{\tau}}}%
}%
\put(2960,200){\makebox(0,0){8}}%
\put(2320,200){\makebox(0,0){6}}%
\put(1680,200){\makebox(0,0){4}}%
\put(1040,200){\makebox(0,0){2}}%
\put(400,200){\makebox(0,0){0}}%
\put(350,3140){\makebox(0,0)[r]{1}}%
\put(350,2785){\makebox(0,0)[r]{0.8}}%
\put(350,2430){\makebox(0,0)[r]{0.6}}%
\put(350,2075){\makebox(0,0)[r]{0.4}}%
\put(350,1720){\makebox(0,0)[r]{0.2}}%
\put(350,1365){\makebox(0,0)[r]{0}}%
\put(350,1010){\makebox(0,0)[r]{-0.2}}%
\put(350,655){\makebox(0,0)[r]{-0.4}}%
\put(350,300){\makebox(0,0)[r]{-0.6}}%
\end{picture}%
\endgroup

}\resizebox{0.48\textwidth}{!}{%
\begingroup%
  \makeatletter%
  \newcommand{\GNUPLOTspecial}{%
    \catcode`\%=14\relax\special}%
  \setlength{\unitlength}{0.1bp}%
\begin{picture}(3600,3240)(0,0)%
\includegraphics{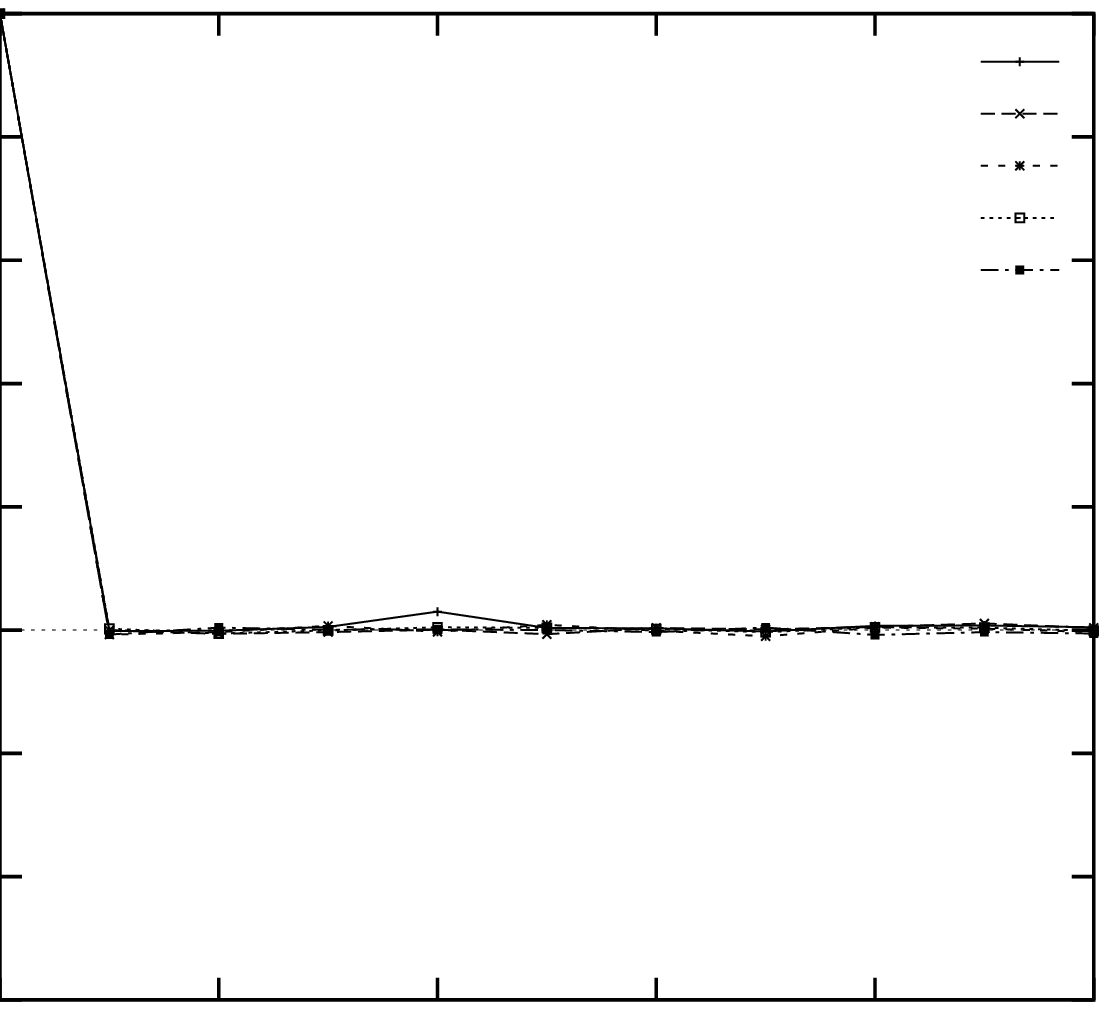}
\put(2775,2402){\makebox(0,0)[r]{$m=5$}}%
\put(2775,2552){\makebox(0,0)[r]{$m=4$}}%
\put(2775,2702){\makebox(0,0)[r]{$m=3$}}%
\put(2775,2852){\makebox(0,0)[r]{$m=2$}}%
\put(2775,3002){\makebox(0,0)[r]{$m=1$}}%
\put(236,2856){\makebox(0,0)[l]{\Large b)}}%
\put(1575,50){\makebox(0,0){Lag: $\tau$}}%
\put(3150,200){\makebox(0,0){10}}%
\put(2520,200){\makebox(0,0){8}}%
\put(1890,200){\makebox(0,0){6}}%
\put(1260,200){\makebox(0,0){4}}%
\put(630,200){\makebox(0,0){2}}%
\put(0,200){\makebox(0,0){0}}%
\end{picture}%
\endgroup

}
\end{center}
}

\subsubsection{Markovian analysis}
\label{sec:acmarkov-3a}
\newcommand{\mmin}{\ensuremath{m^{-1}_t}} 
\newcommand{\mplu}{\ensuremath{m^{+1}_t}}

In this section we will present a Markovian analysis of the action of the memory
in the Genetic Model.  This analysis must be performed separately for each value of $m$ of
interest since each leads to a distinct state space. Here we shall present the
analysis for $m=1$ only. Treatment of higher values of $m$ is possible,
although cumbersome since they lead to state spaces which are too large to be
treated conveniently by hand. 

The first stage of our analysis is to
define a convention for labeling the states of the memory.
Each state label must define the values of $h_t$, $A_t$ and the state of
the memory. This is the minimum set of information needed to calculate the
state transition probabilities. For $m=1$ the memory will contain 
$2^{m=1}=2$ entries corresponding to the two possible histories $A_{t-1}=-1$ and
$A_{t-1}=+1$. We shall label these entries $m_t^{-1}$ and $m_t^{+1}$
respectively. Each state therefore comprises four attributes each of which
can take values $\pm1$. There are therefore sixteen possible states which
we denote using the shorthand notation described in Fig.
\ref{fig:allstates-3a}b. We label these states by analogy with the binary 
number system as shown in Fig. \ref{fig:allstates-3a}a.

\floatfig[\floatplace]{fig:allstates-3a}{
State labels.
}{
a) The labels used to index the sixteen possible states. The states that are
crossed out are those in which $h_t=+1$ and $A_t=+1$. These can be omitted due
to Eq. \eqref{eqn:Aprobs-3a}. b) A key for the graphical representation of each
state.
}{
\newcommand{\tmpstk}{$\begin{cases}h_t=-1\\A_t=+1\\m_t^{-1}=+1\\m_t^{+1}=-1\end{cases}$}
\begin{center}
\rule[-2.8cm]{0pt}{5.6cm}\raisebox{-0.5\height}{
	\resizebox{0.476\textwidth}{!}{
\begin{picture}(0,0)%
\includegraphics{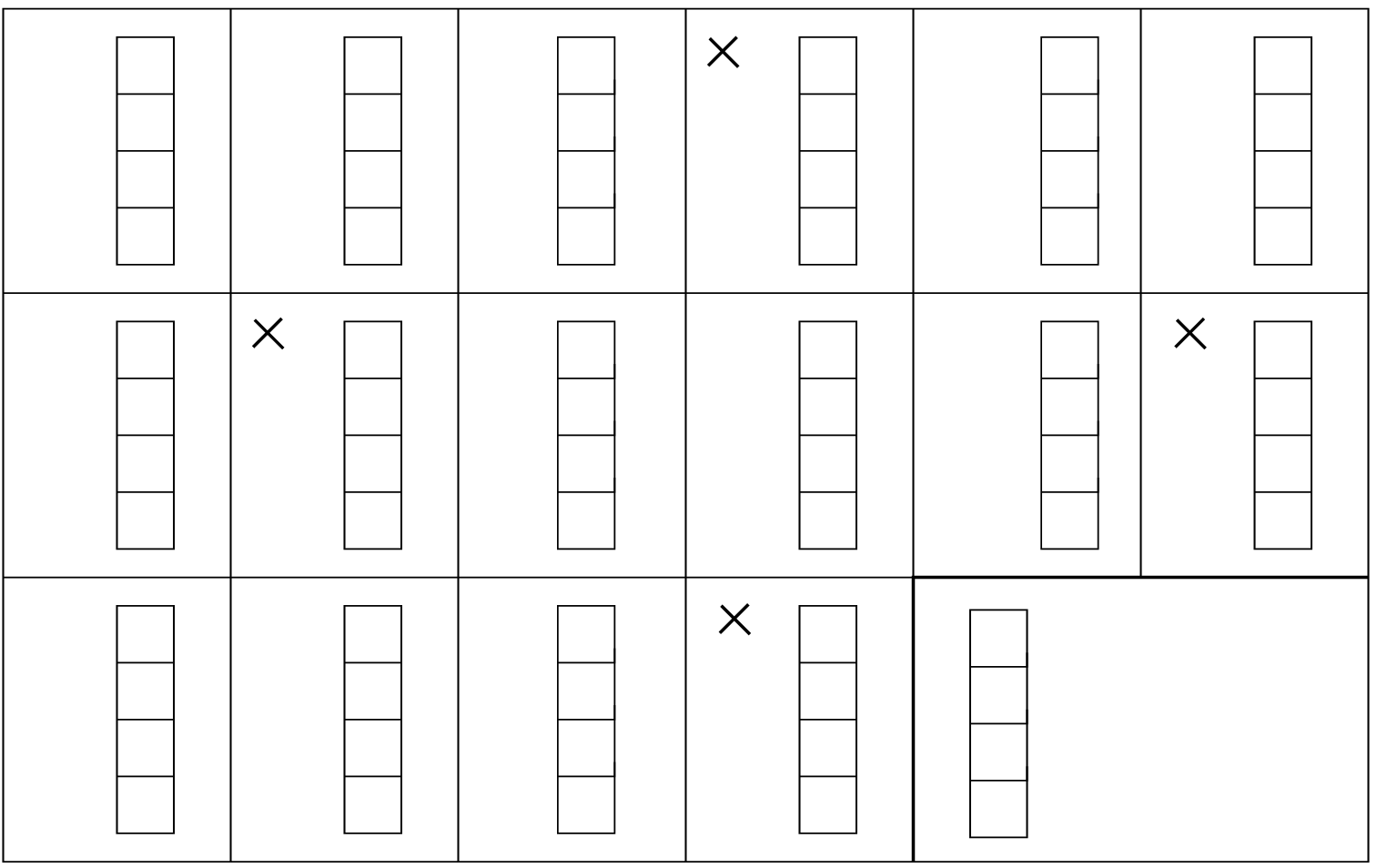}%
\end{picture}%
\setlength{\unitlength}{3947sp}%
\begingroup\makeatletter\ifx\SetFigFont\undefined%
\gdef\SetFigFont#1#2#3#4#5{%
  \fontsize{#1}{#2pt}%
  \fontfamily{#3}\fontseries{#4}\fontshape{#5}%
  \selectfont}%
\fi\endgroup%
\begin{picture}(7234,4534)(-11,-4283)
\put(1951,-3136){\makebox(0,0)[b]{\smash{\SetFigFont{20}{24.0}{\familydefault}{\mddefault}{\updefault}{\color[rgb]{0,0,0}$+$}%
}}}
\put(1951,-3436){\makebox(0,0)[b]{\smash{\SetFigFont{20}{24.0}{\familydefault}{\mddefault}{\updefault}{\color[rgb]{0,0,0}$-$}%
}}}
\put(1951,-3736){\makebox(0,0)[b]{\smash{\SetFigFont{20}{24.0}{\familydefault}{\mddefault}{\updefault}{\color[rgb]{0,0,0}$+$}%
}}}
\put(1951,-4036){\makebox(0,0)[b]{\smash{\SetFigFont{20}{24.0}{\familydefault}{\mddefault}{\updefault}{\color[rgb]{0,0,0}$+$}%
}}}
\put(3076,-3136){\makebox(0,0)[b]{\smash{\SetFigFont{20}{24.0}{\familydefault}{\mddefault}{\updefault}{\color[rgb]{0,0,0}$-$}%
}}}
\put(3076,-3436){\makebox(0,0)[b]{\smash{\SetFigFont{20}{24.0}{\familydefault}{\mddefault}{\updefault}{\color[rgb]{0,0,0}$+$}%
}}}
\put(3076,-3736){\makebox(0,0)[b]{\smash{\SetFigFont{20}{24.0}{\familydefault}{\mddefault}{\updefault}{\color[rgb]{0,0,0}$+$}%
}}}
\put(3076,-4036){\makebox(0,0)[b]{\smash{\SetFigFont{20}{24.0}{\familydefault}{\mddefault}{\updefault}{\color[rgb]{0,0,0}$+$}%
}}}
\put(4351,-3136){\makebox(0,0)[b]{\smash{\SetFigFont{20}{24.0}{\familydefault}{\mddefault}{\updefault}{\color[rgb]{0,0,0}$+$}%
}}}
\put(4351,-3436){\makebox(0,0)[b]{\smash{\SetFigFont{20}{24.0}{\familydefault}{\mddefault}{\updefault}{\color[rgb]{0,0,0}$+$}%
}}}
\put(4351,-3736){\makebox(0,0)[b]{\smash{\SetFigFont{20}{24.0}{\familydefault}{\mddefault}{\updefault}{\color[rgb]{0,0,0}$+$}%
}}}
\put(4351,-4036){\makebox(0,0)[b]{\smash{\SetFigFont{20}{24.0}{\familydefault}{\mddefault}{\updefault}{\color[rgb]{0,0,0}$+$}%
}}}
\put(5251,-3158){\makebox(0,0)[b]{\smash{\SetFigFont{20}{24.0}{\familydefault}{\mddefault}{\updefault}{\color[rgb]{0,0,0}$-$}%
}}}
\put(5251,-3458){\makebox(0,0)[b]{\smash{\SetFigFont{20}{24.0}{\familydefault}{\mddefault}{\updefault}{\color[rgb]{0,0,0}$+$}%
}}}
\put(5251,-3758){\makebox(0,0)[b]{\smash{\SetFigFont{20}{24.0}{\familydefault}{\mddefault}{\updefault}{\color[rgb]{0,0,0}$+$}%
}}}
\put(5251,-4058){\makebox(0,0)[b]{\smash{\SetFigFont{20}{24.0}{\familydefault}{\mddefault}{\updefault}{\color[rgb]{0,0,0}$-$}%
}}}
\put(5950,-3586){\makebox(0,0)[lb]{\smash{\SetFigFont{12}{14.4}{\familydefault}{\mddefault}{\updefault}{\color[rgb]{0,0,0}\tmpstk}%
}}}
\put(5711,-3616){\makebox(0,0)[b]{\smash{\SetFigFont{20}{24.0}{\familydefault}{\mddefault}{\updefault}{\color[rgb]{0,0,0}$=$}%
}}}
\put(5626,-136){\makebox(0,0)[b]{\smash{\SetFigFont{20}{24.0}{\familydefault}{\mddefault}{\updefault}{\color[rgb]{0,0,0}$-$}%
}}}
\put(5626,-436){\makebox(0,0)[b]{\smash{\SetFigFont{20}{24.0}{\familydefault}{\mddefault}{\updefault}{\color[rgb]{0,0,0}$-$}%
}}}
\put(5626,-736){\makebox(0,0)[b]{\smash{\SetFigFont{20}{24.0}{\familydefault}{\mddefault}{\updefault}{\color[rgb]{0,0,0}$+$}%
}}}
\put(5626,-1036){\makebox(0,0)[b]{\smash{\SetFigFont{20}{24.0}{\familydefault}{\mddefault}{\updefault}{\color[rgb]{0,0,0}$-$}%
}}}
\put(6751,-136){\makebox(0,0)[b]{\smash{\SetFigFont{20}{24.0}{\familydefault}{\mddefault}{\updefault}{\color[rgb]{0,0,0}$+$}%
}}}
\put(6751,-436){\makebox(0,0)[b]{\smash{\SetFigFont{20}{24.0}{\familydefault}{\mddefault}{\updefault}{\color[rgb]{0,0,0}$-$}%
}}}
\put(6751,-736){\makebox(0,0)[b]{\smash{\SetFigFont{20}{24.0}{\familydefault}{\mddefault}{\updefault}{\color[rgb]{0,0,0}$+$}%
}}}
\put(6751,-1036){\makebox(0,0)[b]{\smash{\SetFigFont{20}{24.0}{\familydefault}{\mddefault}{\updefault}{\color[rgb]{0,0,0}$-$}%
}}}
\put(1951,-136){\makebox(0,0)[b]{\smash{\SetFigFont{20}{24.0}{\familydefault}{\mddefault}{\updefault}{\color[rgb]{0,0,0}$+$}%
}}}
\put(1951,-436){\makebox(0,0)[b]{\smash{\SetFigFont{20}{24.0}{\familydefault}{\mddefault}{\updefault}{\color[rgb]{0,0,0}$-$}%
}}}
\put(1951,-736){\makebox(0,0)[b]{\smash{\SetFigFont{20}{24.0}{\familydefault}{\mddefault}{\updefault}{\color[rgb]{0,0,0}$-$}%
}}}
\put(1951,-1036){\makebox(0,0)[b]{\smash{\SetFigFont{20}{24.0}{\familydefault}{\mddefault}{\updefault}{\color[rgb]{0,0,0}$-$}%
}}}
\put(3076,-136){\makebox(0,0)[b]{\smash{\SetFigFont{20}{24.0}{\familydefault}{\mddefault}{\updefault}{\color[rgb]{0,0,0}$-$}%
}}}
\put(3076,-436){\makebox(0,0)[b]{\smash{\SetFigFont{20}{24.0}{\familydefault}{\mddefault}{\updefault}{\color[rgb]{0,0,0}$+$}%
}}}
\put(3076,-736){\makebox(0,0)[b]{\smash{\SetFigFont{20}{24.0}{\familydefault}{\mddefault}{\updefault}{\color[rgb]{0,0,0}$-$}%
}}}
\put(3076,-1036){\makebox(0,0)[b]{\smash{\SetFigFont{20}{24.0}{\familydefault}{\mddefault}{\updefault}{\color[rgb]{0,0,0}$-$}%
}}}
\put(4351,-136){\makebox(0,0)[b]{\smash{\SetFigFont{20}{24.0}{\familydefault}{\mddefault}{\updefault}{\color[rgb]{0,0,0}$+$}%
}}}
\put(4351,-436){\makebox(0,0)[b]{\smash{\SetFigFont{20}{24.0}{\familydefault}{\mddefault}{\updefault}{\color[rgb]{0,0,0}$+$}%
}}}
\put(4351,-736){\makebox(0,0)[b]{\smash{\SetFigFont{20}{24.0}{\familydefault}{\mddefault}{\updefault}{\color[rgb]{0,0,0}$-$}%
}}}
\put(4351,-1036){\makebox(0,0)[b]{\smash{\SetFigFont{20}{24.0}{\familydefault}{\mddefault}{\updefault}{\color[rgb]{0,0,0}$-$}%
}}}
\put(1951,-1636){\makebox(0,0)[b]{\smash{\SetFigFont{20}{24.0}{\familydefault}{\mddefault}{\updefault}{\color[rgb]{0,0,0}$+$}%
}}}
\put(1951,-1936){\makebox(0,0)[b]{\smash{\SetFigFont{20}{24.0}{\familydefault}{\mddefault}{\updefault}{\color[rgb]{0,0,0}$+$}%
}}}
\put(1951,-2236){\makebox(0,0)[b]{\smash{\SetFigFont{20}{24.0}{\familydefault}{\mddefault}{\updefault}{\color[rgb]{0,0,0}$+$}%
}}}
\put(1951,-2536){\makebox(0,0)[b]{\smash{\SetFigFont{20}{24.0}{\familydefault}{\mddefault}{\updefault}{\color[rgb]{0,0,0}$-$}%
}}}
\put(3076,-1636){\makebox(0,0)[b]{\smash{\SetFigFont{20}{24.0}{\familydefault}{\mddefault}{\updefault}{\color[rgb]{0,0,0}$-$}%
}}}
\put(3076,-1936){\makebox(0,0)[b]{\smash{\SetFigFont{20}{24.0}{\familydefault}{\mddefault}{\updefault}{\color[rgb]{0,0,0}$-$}%
}}}
\put(3076,-2236){\makebox(0,0)[b]{\smash{\SetFigFont{20}{24.0}{\familydefault}{\mddefault}{\updefault}{\color[rgb]{0,0,0}$-$}%
}}}
\put(3076,-2536){\makebox(0,0)[b]{\smash{\SetFigFont{20}{24.0}{\familydefault}{\mddefault}{\updefault}{\color[rgb]{0,0,0}$+$}%
}}}
\put(4351,-1636){\makebox(0,0)[b]{\smash{\SetFigFont{20}{24.0}{\familydefault}{\mddefault}{\updefault}{\color[rgb]{0,0,0}$+$}%
}}}
\put(4351,-1936){\makebox(0,0)[b]{\smash{\SetFigFont{20}{24.0}{\familydefault}{\mddefault}{\updefault}{\color[rgb]{0,0,0}$-$}%
}}}
\put(4351,-2236){\makebox(0,0)[b]{\smash{\SetFigFont{20}{24.0}{\familydefault}{\mddefault}{\updefault}{\color[rgb]{0,0,0}$-$}%
}}}
\put(4351,-2536){\makebox(0,0)[b]{\smash{\SetFigFont{20}{24.0}{\familydefault}{\mddefault}{\updefault}{\color[rgb]{0,0,0}$+$}%
}}}
\put(5626,-1636){\makebox(0,0)[b]{\smash{\SetFigFont{20}{24.0}{\familydefault}{\mddefault}{\updefault}{\color[rgb]{0,0,0}$-$}%
}}}
\put(5626,-1936){\makebox(0,0)[b]{\smash{\SetFigFont{20}{24.0}{\familydefault}{\mddefault}{\updefault}{\color[rgb]{0,0,0}$+$}%
}}}
\put(5626,-2236){\makebox(0,0)[b]{\smash{\SetFigFont{20}{24.0}{\familydefault}{\mddefault}{\updefault}{\color[rgb]{0,0,0}$-$}%
}}}
\put(5626,-2536){\makebox(0,0)[b]{\smash{\SetFigFont{20}{24.0}{\familydefault}{\mddefault}{\updefault}{\color[rgb]{0,0,0}$+$}%
}}}
\put(6751,-1636){\makebox(0,0)[b]{\smash{\SetFigFont{20}{24.0}{\familydefault}{\mddefault}{\updefault}{\color[rgb]{0,0,0}$+$}%
}}}
\put(6751,-1936){\makebox(0,0)[b]{\smash{\SetFigFont{20}{24.0}{\familydefault}{\mddefault}{\updefault}{\color[rgb]{0,0,0}$+$}%
}}}
\put(6751,-2236){\makebox(0,0)[b]{\smash{\SetFigFont{20}{24.0}{\familydefault}{\mddefault}{\updefault}{\color[rgb]{0,0,0}$-$}%
}}}
\put(6751,-2536){\makebox(0,0)[b]{\smash{\SetFigFont{20}{24.0}{\familydefault}{\mddefault}{\updefault}{\color[rgb]{0,0,0}$+$}%
}}}
\put(751,-136){\makebox(0,0)[b]{\smash{\SetFigFont{20}{24.0}{\familydefault}{\mddefault}{\updefault}{\color[rgb]{0,0,0}$-$}%
}}}
\put(751,-436){\makebox(0,0)[b]{\smash{\SetFigFont{20}{24.0}{\familydefault}{\mddefault}{\updefault}{\color[rgb]{0,0,0}$-$}%
}}}
\put(751,-736){\makebox(0,0)[b]{\smash{\SetFigFont{20}{24.0}{\familydefault}{\mddefault}{\updefault}{\color[rgb]{0,0,0}$-$}%
}}}
\put(751,-1036){\makebox(0,0)[b]{\smash{\SetFigFont{20}{24.0}{\familydefault}{\mddefault}{\updefault}{\color[rgb]{0,0,0}$-$}%
}}}
\put(751,-1636){\makebox(0,0)[b]{\smash{\SetFigFont{20}{24.0}{\familydefault}{\mddefault}{\updefault}{\color[rgb]{0,0,0}$-$}%
}}}
\put(751,-1936){\makebox(0,0)[b]{\smash{\SetFigFont{20}{24.0}{\familydefault}{\mddefault}{\updefault}{\color[rgb]{0,0,0}$+$}%
}}}
\put(751,-2236){\makebox(0,0)[b]{\smash{\SetFigFont{20}{24.0}{\familydefault}{\mddefault}{\updefault}{\color[rgb]{0,0,0}$+$}%
}}}
\put(751,-2536){\makebox(0,0)[b]{\smash{\SetFigFont{20}{24.0}{\familydefault}{\mddefault}{\updefault}{\color[rgb]{0,0,0}$-$}%
}}}
\put(751,-3136){\makebox(0,0)[b]{\smash{\SetFigFont{20}{24.0}{\familydefault}{\mddefault}{\updefault}{\color[rgb]{0,0,0}$-$}%
}}}
\put(751,-3436){\makebox(0,0)[b]{\smash{\SetFigFont{20}{24.0}{\familydefault}{\mddefault}{\updefault}{\color[rgb]{0,0,0}$-$}%
}}}
\put(751,-3736){\makebox(0,0)[b]{\smash{\SetFigFont{20}{24.0}{\familydefault}{\mddefault}{\updefault}{\color[rgb]{0,0,0}$+$}%
}}}
\put(751,-4036){\makebox(0,0)[b]{\smash{\SetFigFont{20}{24.0}{\familydefault}{\mddefault}{\updefault}{\color[rgb]{0,0,0}$+$}%
}}}
\put(5026,-61){\makebox(0,0)[lb]{\smash{\SetFigFont{14}{16.8}{\familydefault}{\mddefault}{\updefault}{\color[rgb]{0,0,0}4}%
}}}
\put(6151,-61){\makebox(0,0)[lb]{\smash{\SetFigFont{14}{16.8}{\familydefault}{\mddefault}{\updefault}{\color[rgb]{0,0,0}5}%
}}}
\put(2476,-61){\makebox(0,0)[lb]{\smash{\SetFigFont{14}{16.8}{\familydefault}{\mddefault}{\updefault}{\color[rgb]{0,0,0}2}%
}}}
\put(3751,-61){\makebox(0,0)[lb]{\smash{\SetFigFont{14}{16.8}{\familydefault}{\mddefault}{\updefault}{\color[rgb]{0,0,0}3}%
}}}
\put(151,-3061){\makebox(0,0)[lb]{\smash{\SetFigFont{14}{16.8}{\familydefault}{\mddefault}{\updefault}{\color[rgb]{0,0,0}12}%
}}}
\put(1351,-3061){\makebox(0,0)[lb]{\smash{\SetFigFont{14}{16.8}{\familydefault}{\mddefault}{\updefault}{\color[rgb]{0,0,0}13}%
}}}
\put(2476,-3061){\makebox(0,0)[lb]{\smash{\SetFigFont{14}{16.8}{\familydefault}{\mddefault}{\updefault}{\color[rgb]{0,0,0}14}%
}}}
\put(3751,-3061){\makebox(0,0)[lb]{\smash{\SetFigFont{14}{16.8}{\familydefault}{\mddefault}{\updefault}{\color[rgb]{0,0,0}15}%
}}}
\put(5026,-1561){\makebox(0,0)[lb]{\smash{\SetFigFont{14}{16.8}{\familydefault}{\mddefault}{\updefault}{\color[rgb]{0,0,0}10}%
}}}
\put(6151,-1561){\makebox(0,0)[lb]{\smash{\SetFigFont{14}{16.8}{\familydefault}{\mddefault}{\updefault}{\color[rgb]{0,0,0}11}%
}}}
\put(1351,-1561){\makebox(0,0)[lb]{\smash{\SetFigFont{14}{16.8}{\familydefault}{\mddefault}{\updefault}{\color[rgb]{0,0,0}7}%
}}}
\put(2476,-1561){\makebox(0,0)[lb]{\smash{\SetFigFont{14}{16.8}{\familydefault}{\mddefault}{\updefault}{\color[rgb]{0,0,0}8}%
}}}
\put(3751,-1561){\makebox(0,0)[lb]{\smash{\SetFigFont{14}{16.8}{\familydefault}{\mddefault}{\updefault}{\color[rgb]{0,0,0}9}%
}}}
\put(151,-1561){\makebox(0,0)[lb]{\smash{\SetFigFont{14}{16.8}{\familydefault}{\mddefault}{\updefault}{\color[rgb]{0,0,0}6}%
}}}
\put(4846,-2933){\makebox(0,0)[lb]{\smash{\SetFigFont{12}{14.4}{\familydefault}{\mddefault}{\updefault}{\color[rgb]{0,0,0}b)}%
}}}
\put(226,-61){\makebox(0,0)[lb]{\smash{\SetFigFont{14}{16.8}{\familydefault}{\mddefault}{\updefault}{\color[rgb]{0,0,0}0}%
}}}
\put( 39, 66){\makebox(0,0)[lb]{\smash{\SetFigFont{12}{14.4}{\familydefault}{\mddefault}{\updefault}{\color[rgb]{0,0,0}a)}%
}}}
\put(1351,-61){\makebox(0,0)[lb]{\smash{\SetFigFont{14}{16.8}{\familydefault}{\mddefault}{\updefault}{\color[rgb]{0,0,0}1}%
}}}
\end{picture}
	}
}
\end{center}
}

For the sake of clarity we shall consider the case of $l<0.5$. By
symmetry the results that we derive will also apply to $l>0.5$.  
From Eqs. \eqref{eqn:probAone-3a} and \eqref{eqn:Ivalscorrected-3a},
we have the following expression for the probability that $A_t=+1$:
%
%
%
\begin{align}
\label{eqn:Aprobs-3a}
	P[A_t=+1] &= \begin{cases}
		      I_2 = \alpha & \f h_t=-1 \\
		      I_1 = 0      & \f h_t=+1
		    \end{cases}  \; ,
\end{align}
where $\alpha$ represents the numerical value of $I_2$ in Eq. 
\eqref{eqn:Ivalscorrected-3a}.
The result of this is that if the prediction $h_t=+1$ then the global action
$A_t=-1$ with probability $P=1$. We can therefore discard states 3, 7, 11
and 15 from consideration. In each of these states $h_t=+1$ and $A_t=+1$
and thus they will never be visited by the model.

Using Eq. \eqref{eqn:Aprobs-3a} we derive the state transitions shown in Fig.
\ref{fig:transitions-3a}.
We can see that each 
state either makes a transition to one other state with probability $P=1$
or to one of two possible states with probability $P=\alpha$ and $P=1-\alpha$
respectively. This information can be used to further simplify the state
space. We need not consider states which have no inward transitions or states
which only have inwards transitions from states that we have removed. 
This allows us to remove states 2, 8, 9, 10, 12 or 13 from
consideration.  The model will not visit these states once initial transients
have died away.  This reduces our state space from sixteen states to only six:
0, 1, 4, 5, 6 and 14.

\floatfig[\floatplace]{fig:transitions-3a}{
One-step transitions.
}{
List of the one-step transitions that the model can make from each
of the states listed in Fig. \ref{fig:allstates-3a}.
}{
\begin{center}
\rule[-2.8cm]{0pt}{5.6cm}\raisebox{-0.5\height}{
  \resizebox{0.476\textwidth}{!}{
\begin{picture}(0,0)%
\includegraphics{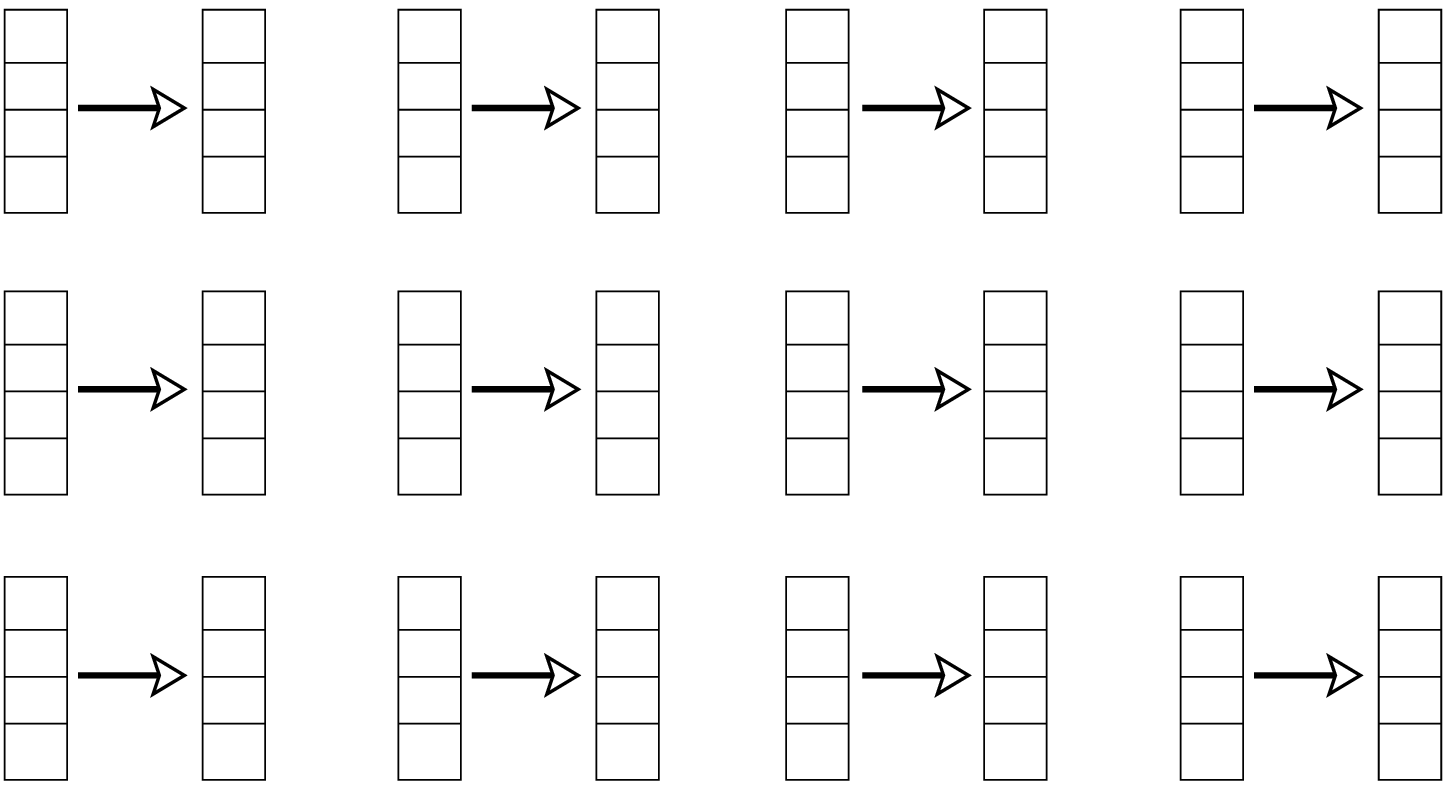}%
\end{picture}%
\setlength{\unitlength}{3947sp}%
\begingroup\makeatletter\ifx\SetFigFont\undefined%
\gdef\SetFigFont#1#2#3#4#5{%
  \fontsize{#1}{#2pt}%
  \fontfamily{#3}\fontseries{#4}\fontshape{#5}%
  \selectfont}%
\fi\endgroup%
\begin{picture}(6920,4162)(11,-4103)
\put(173,-563){\makebox(0,0)[b]{\smash{\SetFigFont{14}{16.8}{\familydefault}{\mddefault}{\updefault}{\color[rgb]{0,0,0}$-$}%
}}}
\put(173,-796){\makebox(0,0)[b]{\smash{\SetFigFont{14}{16.8}{\familydefault}{\mddefault}{\updefault}{\color[rgb]{0,0,0}$-$}%
}}}
\put(173,-1021){\makebox(0,0)[b]{\smash{\SetFigFont{14}{16.8}{\familydefault}{\mddefault}{\updefault}{\color[rgb]{0,0,0}$-$}%
}}}
\put(173,-1246){\makebox(0,0)[b]{\smash{\SetFigFont{14}{16.8}{\familydefault}{\mddefault}{\updefault}{\color[rgb]{0,0,0}$-$}%
}}}
\put(174,-301){\makebox(0,0)[b]{\smash{\SetFigFont{14}{16.8}{\familydefault}{\mddefault}{\updefault}{\color[rgb]{0,0,0}0}%
}}}
\put(173,-1915){\makebox(0,0)[b]{\smash{\SetFigFont{14}{16.8}{\familydefault}{\mddefault}{\updefault}{\color[rgb]{0,0,0}$+$}%
}}}
\put(173,-2148){\makebox(0,0)[b]{\smash{\SetFigFont{14}{16.8}{\familydefault}{\mddefault}{\updefault}{\color[rgb]{0,0,0}$-$}%
}}}
\put(173,-2373){\makebox(0,0)[b]{\smash{\SetFigFont{14}{16.8}{\familydefault}{\mddefault}{\updefault}{\color[rgb]{0,0,0}$-$}%
}}}
\put(173,-2598){\makebox(0,0)[b]{\smash{\SetFigFont{14}{16.8}{\familydefault}{\mddefault}{\updefault}{\color[rgb]{0,0,0}$-$}%
}}}
\put(174,-1653){\makebox(0,0)[b]{\smash{\SetFigFont{14}{16.8}{\familydefault}{\mddefault}{\updefault}{\color[rgb]{0,0,0}1}%
}}}
\put(173,-3285){\makebox(0,0)[b]{\smash{\SetFigFont{14}{16.8}{\familydefault}{\mddefault}{\updefault}{\color[rgb]{0,0,0}$-$}%
}}}
\put(173,-3518){\makebox(0,0)[b]{\smash{\SetFigFont{14}{16.8}{\familydefault}{\mddefault}{\updefault}{\color[rgb]{0,0,0}$+$}%
}}}
\put(173,-3743){\makebox(0,0)[b]{\smash{\SetFigFont{14}{16.8}{\familydefault}{\mddefault}{\updefault}{\color[rgb]{0,0,0}$-$}%
}}}
\put(173,-3968){\makebox(0,0)[b]{\smash{\SetFigFont{14}{16.8}{\familydefault}{\mddefault}{\updefault}{\color[rgb]{0,0,0}$-$}%
}}}
\put(174,-3023){\makebox(0,0)[b]{\smash{\SetFigFont{14}{16.8}{\familydefault}{\mddefault}{\updefault}{\color[rgb]{0,0,0}2}%
}}}
\put(1124,-3285){\makebox(0,0)[b]{\smash{\SetFigFont{14}{16.8}{\familydefault}{\mddefault}{\updefault}{\color[rgb]{0,0,0}$-$}%
}}}
\put(1124,-3518){\makebox(0,0)[b]{\smash{\SetFigFont{14}{16.8}{\familydefault}{\mddefault}{\updefault}{\color[rgb]{0,0,0}$\pm$}%
}}}
\put(1124,-3743){\makebox(0,0)[b]{\smash{\SetFigFont{14}{16.8}{\familydefault}{\mddefault}{\updefault}{\color[rgb]{0,0,0}$-$}%
}}}
\put(1124,-3968){\makebox(0,0)[b]{\smash{\SetFigFont{14}{16.8}{\familydefault}{\mddefault}{\updefault}{\color[rgb]{0,0,0}$\pm$}%
}}}
\put(1125,-3023){\makebox(0,0)[b]{\smash{\SetFigFont{14}{16.8}{\familydefault}{\mddefault}{\updefault}{\color[rgb]{0,0,0}10/0}%
}}}
\put(1124,-563){\makebox(0,0)[b]{\smash{\SetFigFont{14}{16.8}{\familydefault}{\mddefault}{\updefault}{\color[rgb]{0,0,0}$-$}%
}}}
\put(1124,-796){\makebox(0,0)[b]{\smash{\SetFigFont{14}{16.8}{\familydefault}{\mddefault}{\updefault}{\color[rgb]{0,0,0}$\pm$}%
}}}
\put(1124,-1021){\makebox(0,0)[b]{\smash{\SetFigFont{14}{16.8}{\familydefault}{\mddefault}{\updefault}{\color[rgb]{0,0,0}$\pm$}%
}}}
\put(1124,-1246){\makebox(0,0)[b]{\smash{\SetFigFont{14}{16.8}{\familydefault}{\mddefault}{\updefault}{\color[rgb]{0,0,0}$-$}%
}}}
\put(1125,-301){\makebox(0,0)[b]{\smash{\SetFigFont{14}{16.8}{\familydefault}{\mddefault}{\updefault}{\color[rgb]{0,0,0}6/0}%
}}}
\put(1124,-1915){\makebox(0,0)[b]{\smash{\SetFigFont{14}{16.8}{\familydefault}{\mddefault}{\updefault}{\color[rgb]{0,0,0}$-$}%
}}}
\put(1124,-2148){\makebox(0,0)[b]{\smash{\SetFigFont{14}{16.8}{\familydefault}{\mddefault}{\updefault}{\color[rgb]{0,0,0}$\pm$}%
}}}
\put(1124,-2373){\makebox(0,0)[b]{\smash{\SetFigFont{14}{16.8}{\familydefault}{\mddefault}{\updefault}{\color[rgb]{0,0,0}$\pm$}%
}}}
\put(1124,-2598){\makebox(0,0)[b]{\smash{\SetFigFont{14}{16.8}{\familydefault}{\mddefault}{\updefault}{\color[rgb]{0,0,0}$-$}%
}}}
\put(1125,-1653){\makebox(0,0)[b]{\smash{\SetFigFont{14}{16.8}{\familydefault}{\mddefault}{\updefault}{\color[rgb]{0,0,0}6/0}%
}}}
\put(2063,-563){\makebox(0,0)[b]{\smash{\SetFigFont{14}{16.8}{\familydefault}{\mddefault}{\updefault}{\color[rgb]{0,0,0}$-$}%
}}}
\put(2063,-796){\makebox(0,0)[b]{\smash{\SetFigFont{14}{16.8}{\familydefault}{\mddefault}{\updefault}{\color[rgb]{0,0,0}$-$}%
}}}
\put(2063,-1021){\makebox(0,0)[b]{\smash{\SetFigFont{14}{16.8}{\familydefault}{\mddefault}{\updefault}{\color[rgb]{0,0,0}$+$}%
}}}
\put(2063,-1246){\makebox(0,0)[b]{\smash{\SetFigFont{14}{16.8}{\familydefault}{\mddefault}{\updefault}{\color[rgb]{0,0,0}$-$}%
}}}
\put(2064,-301){\makebox(0,0)[b]{\smash{\SetFigFont{14}{16.8}{\familydefault}{\mddefault}{\updefault}{\color[rgb]{0,0,0}4}%
}}}
\put(2063,-3285){\makebox(0,0)[b]{\smash{\SetFigFont{14}{16.8}{\familydefault}{\mddefault}{\updefault}{\color[rgb]{0,0,0}$-$}%
}}}
\put(2063,-3518){\makebox(0,0)[b]{\smash{\SetFigFont{14}{16.8}{\familydefault}{\mddefault}{\updefault}{\color[rgb]{0,0,0}$+$}%
}}}
\put(2063,-3743){\makebox(0,0)[b]{\smash{\SetFigFont{14}{16.8}{\familydefault}{\mddefault}{\updefault}{\color[rgb]{0,0,0}$+$}%
}}}
\put(2063,-3968){\makebox(0,0)[b]{\smash{\SetFigFont{14}{16.8}{\familydefault}{\mddefault}{\updefault}{\color[rgb]{0,0,0}$-$}%
}}}
\put(2064,-3023){\makebox(0,0)[b]{\smash{\SetFigFont{14}{16.8}{\familydefault}{\mddefault}{\updefault}{\color[rgb]{0,0,0}6}%
}}}
\put(3014,-3285){\makebox(0,0)[b]{\smash{\SetFigFont{14}{16.8}{\familydefault}{\mddefault}{\updefault}{\color[rgb]{0,0,0}$-$}%
}}}
\put(3014,-3518){\makebox(0,0)[b]{\smash{\SetFigFont{14}{16.8}{\familydefault}{\mddefault}{\updefault}{\color[rgb]{0,0,0}$\pm$}%
}}}
\put(3014,-3743){\makebox(0,0)[b]{\smash{\SetFigFont{14}{16.8}{\familydefault}{\mddefault}{\updefault}{\color[rgb]{0,0,0}$+$}%
}}}
\put(3014,-3968){\makebox(0,0)[b]{\smash{\SetFigFont{14}{16.8}{\familydefault}{\mddefault}{\updefault}{\color[rgb]{0,0,0}$\pm$}%
}}}
\put(3015,-3023){\makebox(0,0)[b]{\smash{\SetFigFont{14}{16.8}{\familydefault}{\mddefault}{\updefault}{\color[rgb]{0,0,0}14/4}%
}}}
\put(3014,-563){\makebox(0,0)[b]{\smash{\SetFigFont{14}{16.8}{\familydefault}{\mddefault}{\updefault}{\color[rgb]{0,0,0}$+$}%
}}}
\put(3014,-796){\makebox(0,0)[b]{\smash{\SetFigFont{14}{16.8}{\familydefault}{\mddefault}{\updefault}{\color[rgb]{0,0,0}$-$}%
}}}
\put(3014,-1021){\makebox(0,0)[b]{\smash{\SetFigFont{14}{16.8}{\familydefault}{\mddefault}{\updefault}{\color[rgb]{0,0,0}$-$}%
}}}
\put(3014,-1246){\makebox(0,0)[b]{\smash{\SetFigFont{14}{16.8}{\familydefault}{\mddefault}{\updefault}{\color[rgb]{0,0,0}$-$}%
}}}
\put(3015,-301){\makebox(0,0)[b]{\smash{\SetFigFont{14}{16.8}{\familydefault}{\mddefault}{\updefault}{\color[rgb]{0,0,0}1}%
}}}
\put(3014,-1915){\makebox(0,0)[b]{\smash{\SetFigFont{14}{16.8}{\familydefault}{\mddefault}{\updefault}{\color[rgb]{0,0,0}$+$}%
}}}
\put(3014,-2148){\makebox(0,0)[b]{\smash{\SetFigFont{14}{16.8}{\familydefault}{\mddefault}{\updefault}{\color[rgb]{0,0,0}$-$}%
}}}
\put(3014,-2373){\makebox(0,0)[b]{\smash{\SetFigFont{14}{16.8}{\familydefault}{\mddefault}{\updefault}{\color[rgb]{0,0,0}$-$}%
}}}
\put(3014,-2598){\makebox(0,0)[b]{\smash{\SetFigFont{14}{16.8}{\familydefault}{\mddefault}{\updefault}{\color[rgb]{0,0,0}$-$}%
}}}
\put(3015,-1653){\makebox(0,0)[b]{\smash{\SetFigFont{14}{16.8}{\familydefault}{\mddefault}{\updefault}{\color[rgb]{0,0,0}1}%
}}}
\put(2063,-1915){\makebox(0,0)[b]{\smash{\SetFigFont{14}{16.8}{\familydefault}{\mddefault}{\updefault}{\color[rgb]{0,0,0}$+$}%
}}}
\put(2063,-2148){\makebox(0,0)[b]{\smash{\SetFigFont{14}{16.8}{\familydefault}{\mddefault}{\updefault}{\color[rgb]{0,0,0}$-$}%
}}}
\put(2063,-2373){\makebox(0,0)[b]{\smash{\SetFigFont{14}{16.8}{\familydefault}{\mddefault}{\updefault}{\color[rgb]{0,0,0}$+$}%
}}}
\put(2063,-2598){\makebox(0,0)[b]{\smash{\SetFigFont{14}{16.8}{\familydefault}{\mddefault}{\updefault}{\color[rgb]{0,0,0}$-$}%
}}}
\put(2064,-1653){\makebox(0,0)[b]{\smash{\SetFigFont{14}{16.8}{\familydefault}{\mddefault}{\updefault}{\color[rgb]{0,0,0}5}%
}}}
\put(3924,-563){\makebox(0,0)[b]{\smash{\SetFigFont{14}{16.8}{\familydefault}{\mddefault}{\updefault}{\color[rgb]{0,0,0}$-$}%
}}}
\put(3924,-796){\makebox(0,0)[b]{\smash{\SetFigFont{14}{16.8}{\familydefault}{\mddefault}{\updefault}{\color[rgb]{0,0,0}$-$}%
}}}
\put(3924,-1021){\makebox(0,0)[b]{\smash{\SetFigFont{14}{16.8}{\familydefault}{\mddefault}{\updefault}{\color[rgb]{0,0,0}$-$}%
}}}
\put(3924,-1246){\makebox(0,0)[b]{\smash{\SetFigFont{14}{16.8}{\familydefault}{\mddefault}{\updefault}{\color[rgb]{0,0,0}$+$}%
}}}
\put(3925,-301){\makebox(0,0)[b]{\smash{\SetFigFont{14}{16.8}{\familydefault}{\mddefault}{\updefault}{\color[rgb]{0,0,0}8}%
}}}
\put(3924,-1915){\makebox(0,0)[b]{\smash{\SetFigFont{14}{16.8}{\familydefault}{\mddefault}{\updefault}{\color[rgb]{0,0,0}$+$}%
}}}
\put(3924,-2148){\makebox(0,0)[b]{\smash{\SetFigFont{14}{16.8}{\familydefault}{\mddefault}{\updefault}{\color[rgb]{0,0,0}$-$}%
}}}
\put(3924,-2373){\makebox(0,0)[b]{\smash{\SetFigFont{14}{16.8}{\familydefault}{\mddefault}{\updefault}{\color[rgb]{0,0,0}$-$}%
}}}
\put(3924,-2598){\makebox(0,0)[b]{\smash{\SetFigFont{14}{16.8}{\familydefault}{\mddefault}{\updefault}{\color[rgb]{0,0,0}$+$}%
}}}
\put(3925,-1653){\makebox(0,0)[b]{\smash{\SetFigFont{14}{16.8}{\familydefault}{\mddefault}{\updefault}{\color[rgb]{0,0,0}9}%
}}}
\put(3924,-3285){\makebox(0,0)[b]{\smash{\SetFigFont{14}{16.8}{\familydefault}{\mddefault}{\updefault}{\color[rgb]{0,0,0}$-$}%
}}}
\put(3924,-3518){\makebox(0,0)[b]{\smash{\SetFigFont{14}{16.8}{\familydefault}{\mddefault}{\updefault}{\color[rgb]{0,0,0}$+$}%
}}}
\put(3924,-3743){\makebox(0,0)[b]{\smash{\SetFigFont{14}{16.8}{\familydefault}{\mddefault}{\updefault}{\color[rgb]{0,0,0}$-$}%
}}}
\put(3924,-3968){\makebox(0,0)[b]{\smash{\SetFigFont{14}{16.8}{\familydefault}{\mddefault}{\updefault}{\color[rgb]{0,0,0}$+$}%
}}}
\put(3925,-3023){\makebox(0,0)[b]{\smash{\SetFigFont{14}{16.8}{\familydefault}{\mddefault}{\updefault}{\color[rgb]{0,0,0}10}%
}}}
\put(4875,-3285){\makebox(0,0)[b]{\smash{\SetFigFont{14}{16.8}{\familydefault}{\mddefault}{\updefault}{\color[rgb]{0,0,0}$+$}%
}}}
\put(4875,-3518){\makebox(0,0)[b]{\smash{\SetFigFont{14}{16.8}{\familydefault}{\mddefault}{\updefault}{\color[rgb]{0,0,0}$-$}%
}}}
\put(4875,-3743){\makebox(0,0)[b]{\smash{\SetFigFont{14}{16.8}{\familydefault}{\mddefault}{\updefault}{\color[rgb]{0,0,0}$-$}%
}}}
\put(4875,-3968){\makebox(0,0)[b]{\smash{\SetFigFont{14}{16.8}{\familydefault}{\mddefault}{\updefault}{\color[rgb]{0,0,0}$-$}%
}}}
\put(4876,-3023){\makebox(0,0)[b]{\smash{\SetFigFont{14}{16.8}{\familydefault}{\mddefault}{\updefault}{\color[rgb]{0,0,0}1}%
}}}
\put(4875,-563){\makebox(0,0)[b]{\smash{\SetFigFont{14}{16.8}{\familydefault}{\mddefault}{\updefault}{\color[rgb]{0,0,0}$-$}%
}}}
\put(4875,-796){\makebox(0,0)[b]{\smash{\SetFigFont{14}{16.8}{\familydefault}{\mddefault}{\updefault}{\color[rgb]{0,0,0}$\pm$}%
}}}
\put(4875,-1021){\makebox(0,0)[b]{\smash{\SetFigFont{14}{16.8}{\familydefault}{\mddefault}{\updefault}{\color[rgb]{0,0,0}$\pm$}%
}}}
\put(4875,-1246){\makebox(0,0)[b]{\smash{\SetFigFont{14}{16.8}{\familydefault}{\mddefault}{\updefault}{\color[rgb]{0,0,0}$+$}%
}}}
\put(4876,-301){\makebox(0,0)[b]{\smash{\SetFigFont{14}{16.8}{\familydefault}{\mddefault}{\updefault}{\color[rgb]{0,0,0}14/8}%
}}}
\put(4875,-1915){\makebox(0,0)[b]{\smash{\SetFigFont{14}{16.8}{\familydefault}{\mddefault}{\updefault}{\color[rgb]{0,0,0}$-$}%
}}}
\put(4875,-2148){\makebox(0,0)[b]{\smash{\SetFigFont{14}{16.8}{\familydefault}{\mddefault}{\updefault}{\color[rgb]{0,0,0}$\pm$}%
}}}
\put(4875,-2373){\makebox(0,0)[b]{\smash{\SetFigFont{14}{16.8}{\familydefault}{\mddefault}{\updefault}{\color[rgb]{0,0,0}$\pm$}%
}}}
\put(4875,-2598){\makebox(0,0)[b]{\smash{\SetFigFont{14}{16.8}{\familydefault}{\mddefault}{\updefault}{\color[rgb]{0,0,0}$+$}%
}}}
\put(4876,-1653){\makebox(0,0)[b]{\smash{\SetFigFont{14}{16.8}{\familydefault}{\mddefault}{\updefault}{\color[rgb]{0,0,0}14/8}%
}}}
\put(5818,-563){\makebox(0,0)[b]{\smash{\SetFigFont{14}{16.8}{\familydefault}{\mddefault}{\updefault}{\color[rgb]{0,0,0}$-$}%
}}}
\put(5818,-796){\makebox(0,0)[b]{\smash{\SetFigFont{14}{16.8}{\familydefault}{\mddefault}{\updefault}{\color[rgb]{0,0,0}$-$}%
}}}
\put(5818,-1021){\makebox(0,0)[b]{\smash{\SetFigFont{14}{16.8}{\familydefault}{\mddefault}{\updefault}{\color[rgb]{0,0,0}$+$}%
}}}
\put(5818,-1246){\makebox(0,0)[b]{\smash{\SetFigFont{14}{16.8}{\familydefault}{\mddefault}{\updefault}{\color[rgb]{0,0,0}$+$}%
}}}
\put(5819,-301){\makebox(0,0)[b]{\smash{\SetFigFont{14}{16.8}{\familydefault}{\mddefault}{\updefault}{\color[rgb]{0,0,0}12}%
}}}
\put(5818,-1915){\makebox(0,0)[b]{\smash{\SetFigFont{14}{16.8}{\familydefault}{\mddefault}{\updefault}{\color[rgb]{0,0,0}$+$}%
}}}
\put(5818,-2148){\makebox(0,0)[b]{\smash{\SetFigFont{14}{16.8}{\familydefault}{\mddefault}{\updefault}{\color[rgb]{0,0,0}$-$}%
}}}
\put(5818,-2373){\makebox(0,0)[b]{\smash{\SetFigFont{14}{16.8}{\familydefault}{\mddefault}{\updefault}{\color[rgb]{0,0,0}$+$}%
}}}
\put(5818,-2598){\makebox(0,0)[b]{\smash{\SetFigFont{14}{16.8}{\familydefault}{\mddefault}{\updefault}{\color[rgb]{0,0,0}$+$}%
}}}
\put(5819,-1653){\makebox(0,0)[b]{\smash{\SetFigFont{14}{16.8}{\familydefault}{\mddefault}{\updefault}{\color[rgb]{0,0,0}13}%
}}}
\put(5818,-3285){\makebox(0,0)[b]{\smash{\SetFigFont{14}{16.8}{\familydefault}{\mddefault}{\updefault}{\color[rgb]{0,0,0}$-$}%
}}}
\put(5818,-3518){\makebox(0,0)[b]{\smash{\SetFigFont{14}{16.8}{\familydefault}{\mddefault}{\updefault}{\color[rgb]{0,0,0}$+$}%
}}}
\put(5818,-3743){\makebox(0,0)[b]{\smash{\SetFigFont{14}{16.8}{\familydefault}{\mddefault}{\updefault}{\color[rgb]{0,0,0}$+$}%
}}}
\put(5818,-3968){\makebox(0,0)[b]{\smash{\SetFigFont{14}{16.8}{\familydefault}{\mddefault}{\updefault}{\color[rgb]{0,0,0}$+$}%
}}}
\put(5819,-3023){\makebox(0,0)[b]{\smash{\SetFigFont{14}{16.8}{\familydefault}{\mddefault}{\updefault}{\color[rgb]{0,0,0}14}%
}}}
\put(6769,-3285){\makebox(0,0)[b]{\smash{\SetFigFont{14}{16.8}{\familydefault}{\mddefault}{\updefault}{\color[rgb]{0,0,0}$+$}%
}}}
\put(6769,-3518){\makebox(0,0)[b]{\smash{\SetFigFont{14}{16.8}{\familydefault}{\mddefault}{\updefault}{\color[rgb]{0,0,0}$-$}%
}}}
\put(6769,-3743){\makebox(0,0)[b]{\smash{\SetFigFont{14}{16.8}{\familydefault}{\mddefault}{\updefault}{\color[rgb]{0,0,0}$+$}%
}}}
\put(6769,-3968){\makebox(0,0)[b]{\smash{\SetFigFont{14}{16.8}{\familydefault}{\mddefault}{\updefault}{\color[rgb]{0,0,0}$-$}%
}}}
\put(6770,-3023){\makebox(0,0)[b]{\smash{\SetFigFont{14}{16.8}{\familydefault}{\mddefault}{\updefault}{\color[rgb]{0,0,0}5}%
}}}
\put(6769,-563){\makebox(0,0)[b]{\smash{\SetFigFont{14}{16.8}{\familydefault}{\mddefault}{\updefault}{\color[rgb]{0,0,0}$+$}%
}}}
\put(6769,-796){\makebox(0,0)[b]{\smash{\SetFigFont{14}{16.8}{\familydefault}{\mddefault}{\updefault}{\color[rgb]{0,0,0}$-$}%
}}}
\put(6769,-1021){\makebox(0,0)[b]{\smash{\SetFigFont{14}{16.8}{\familydefault}{\mddefault}{\updefault}{\color[rgb]{0,0,0}$-$}%
}}}
\put(6769,-1246){\makebox(0,0)[b]{\smash{\SetFigFont{14}{16.8}{\familydefault}{\mddefault}{\updefault}{\color[rgb]{0,0,0}$+$}%
}}}
\put(6770,-301){\makebox(0,0)[b]{\smash{\SetFigFont{14}{16.8}{\familydefault}{\mddefault}{\updefault}{\color[rgb]{0,0,0}9}%
}}}
\put(6769,-1915){\makebox(0,0)[b]{\smash{\SetFigFont{14}{16.8}{\familydefault}{\mddefault}{\updefault}{\color[rgb]{0,0,0}$+$}%
}}}
\put(6769,-2148){\makebox(0,0)[b]{\smash{\SetFigFont{14}{16.8}{\familydefault}{\mddefault}{\updefault}{\color[rgb]{0,0,0}$-$}%
}}}
\put(6769,-2373){\makebox(0,0)[b]{\smash{\SetFigFont{14}{16.8}{\familydefault}{\mddefault}{\updefault}{\color[rgb]{0,0,0}$-$}%
}}}
\put(6769,-2598){\makebox(0,0)[b]{\smash{\SetFigFont{14}{16.8}{\familydefault}{\mddefault}{\updefault}{\color[rgb]{0,0,0}$+$}%
}}}
\put(6770,-1653){\makebox(0,0)[b]{\smash{\SetFigFont{14}{16.8}{\familydefault}{\mddefault}{\updefault}{\color[rgb]{0,0,0}9}%
}}}
\end{picture}
  }
}
\end{center}
}
\floatfig[\floatplace]{fig:transdiag-3a}{
State transition diagram.
}{
State transition diagram corresponding to the transitions depicted in
Fig. \ref{fig:transitions-3a} in the simplified state space.
The arrow labels give the transition probabilities.
}{
\begin{center}
\resizebox{0.35\textwidth}{!}{
\begin{picture}(0,0)%
\includegraphics{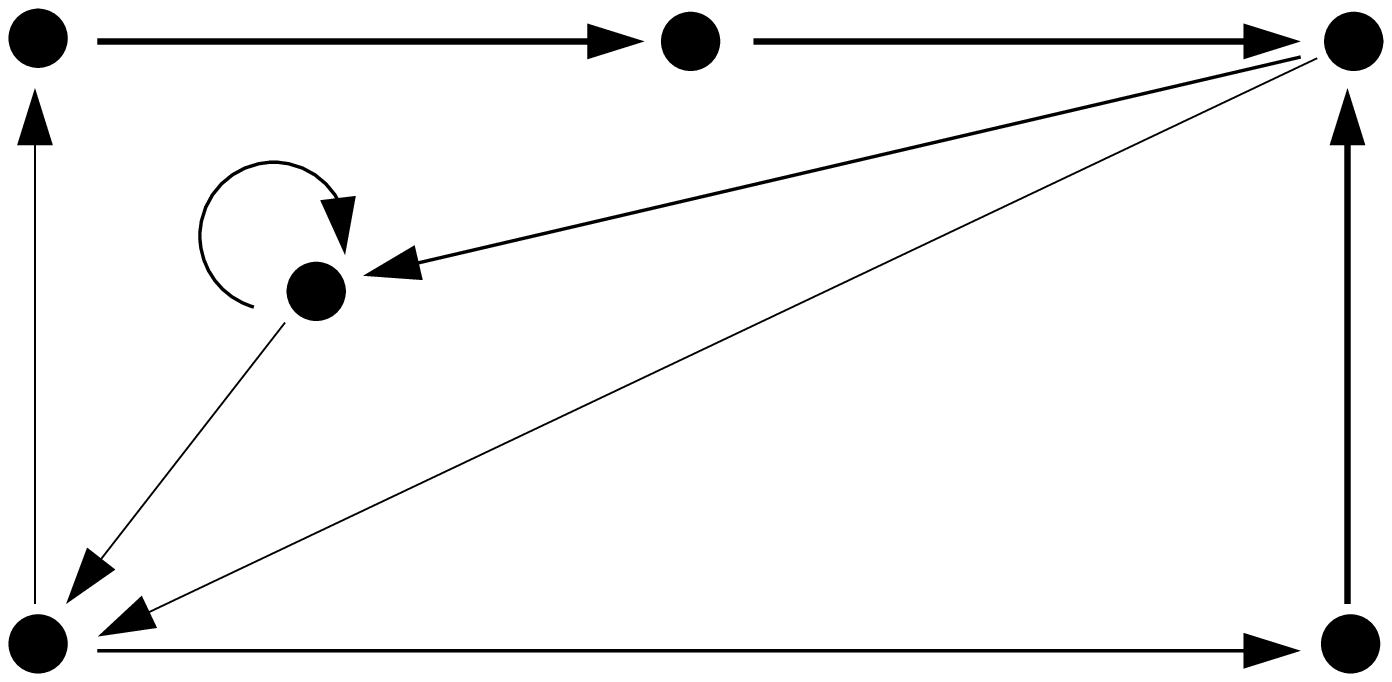}%
\end{picture}%
\setlength{\unitlength}{3947sp}%
\begingroup\makeatletter\ifx\SetFigFont\undefined%
\gdef\SetFigFont#1#2#3#4#5{%
  \fontsize{#1}{#2pt}%
  \fontfamily{#3}\fontseries{#4}\fontshape{#5}%
  \selectfont}%
\fi\endgroup%
\begin{picture}(6900,3645)(1351,-4786)
\put(5026,-2911){\rotatebox{28.0}{\makebox(0,0)[b]{\smash{\SetFigFont{14}{16.8}{\familydefault}{\mddefault}{\updefault}{\color[rgb]{0,0,0}$\alpha$}%
}}}}
\put(4801,-4411){\makebox(0,0)[b]{\smash{\SetFigFont{14}{16.8}{\familydefault}{\mddefault}{\updefault}{\color[rgb]{0,0,0}$1-\alpha$}%
}}}
\put(1576,-3061){\rotatebox{90.0}{\makebox(0,0)[b]{\smash{\SetFigFont{14}{16.8}{\familydefault}{\mddefault}{\updefault}{\color[rgb]{0,0,0}$\alpha$}%
}}}}
\put(8176,-3211){\makebox(0,0)[b]{\smash{\SetFigFont{14}{16.8}{\familydefault}{\mddefault}{\updefault}{\color[rgb]{0,0,0}$1$}%
}}}
\put(2851,-2611){\makebox(0,0)[b]{\smash{\SetFigFont{20}{24.0}{\familydefault}{\mddefault}{\updefault}{\color[rgb]{0,0,0}0}%
}}}
\put(2701,-2086){\makebox(0,0)[b]{\smash{\SetFigFont{14}{16.8}{\familydefault}{\mddefault}{\updefault}{\color[rgb]{0,0,0}$1-\alpha$}%
}}}
\put(2326,-3361){\rotatebox{51.0}{\makebox(0,0)[b]{\smash{\SetFigFont{14}{16.8}{\familydefault}{\mddefault}{\updefault}{\color[rgb]{0,0,0}$\alpha$}%
}}}}
\put(5251,-2086){\rotatebox{12.0}{\makebox(0,0)[b]{\smash{\SetFigFont{14}{16.8}{\familydefault}{\mddefault}{\updefault}{\color[rgb]{0,0,0}$1-\alpha$}%
}}}}
\put(6301,-1486){\makebox(0,0)[b]{\smash{\SetFigFont{14}{16.8}{\familydefault}{\mddefault}{\updefault}{\color[rgb]{0,0,0}$1$}%
}}}
\put(3076,-1486){\makebox(0,0)[b]{\smash{\SetFigFont{14}{16.8}{\familydefault}{\mddefault}{\updefault}{\color[rgb]{0,0,0}$1$}%
}}}
\put(8176,-4786){\makebox(0,0)[b]{\smash{\SetFigFont{20}{24.0}{\familydefault}{\mddefault}{\updefault}{\color[rgb]{0,0,0}4}%
}}}
\put(1426,-4786){\makebox(0,0)[b]{\smash{\SetFigFont{20}{24.0}{\familydefault}{\mddefault}{\updefault}{\color[rgb]{0,0,0}6}%
}}}
\put(4801,-1336){\makebox(0,0)[b]{\smash{\SetFigFont{20}{24.0}{\familydefault}{\mddefault}{\updefault}{\color[rgb]{0,0,0}5}%
}}}
\put(1351,-1336){\makebox(0,0)[b]{\smash{\SetFigFont{20}{24.0}{\familydefault}{\mddefault}{\updefault}{\color[rgb]{0,0,0}14}%
}}}
\put(8251,-1411){\makebox(0,0)[b]{\smash{\SetFigFont{20}{24.0}{\familydefault}{\mddefault}{\updefault}{\color[rgb]{0,0,0}1}%
}}}
\end{picture}
}
\end{center}
}

Figure \ref{fig:transdiag-3a} shows the state transition diagram corresponding
to Fig. \ref{fig:transitions-3a} in the simplified state space described above.
From this we can form the following Markov transition matrix, in which the
remaining states are arranged in numerical order:
\begin{equation}
\label{eqn:matrix-3a}
  \mv{T} = 
  \begin{pmatrix}
    1-\alpha & 1-\alpha & 0 & 0 & 0        & 0 \\
    0        & 0        & 1 & 1 & 0        & 0 \\
    0        & 0        & 0 & 0 & 1-\alpha & 0 \\
    0        & 0        & 0 & 0 & 0        & 1 \\
    \alpha   & \alpha   & 0 & 0 & 0        & 0 \\
    0        & 0        & 0 & 0 & \alpha   & 0 
  \end{pmatrix} \; ,
\end{equation}
which defines the stationary Markov chain.

From Eq. \ref{eqn:matrix-3a} we can derive the $n$-step autocorrelation
functions \autoc{h_t}{n}. Recall the oscillatory dependence of \autoc{h_t}{\tau}
on $\tau$ in Fig. \ref{fig:hautocwithmem-3a}a. If $\autoc{h_t}{1}=0$ and 
$\autoc{h_t}{2}=-\epsilon$ (where $\epsilon$ is of order $0.1$) then we would
expect to observe the dependence depicted in the figure. If $h_t$ and
$h_{t+2}$ are anticorrelated and $|\autoc{h_t}{2}|<1$ then we should expect
$h_t$ and $h_{t+4}$ to be correlated and $|\autoc{h_t}{4}|<|\autoc{h_t}{2}|$ and
so on. Thus the form of \autoc{h_t}{\tau} depicted in Fig.
\ref{fig:hautocwithmem-3a}a for $m=1$ depends only on the values of
$\autoc{h_t}{1}$ and $\autoc{h_t}{2}$. 

First we calculate the stationary state $\mv{s}$ of the Markov chain defined by
Eq. \ref{eqn:matrix-3a}. This gives: 
\begin{equation}
  \mv{s} = \frac{1}{(1+\alpha)^2} 
  \begin{pmatrix}
    1-\alpha \\ \alpha \\ \alpha(1-\alpha) \\ \alpha^2 \\ \alpha \\ \alpha^2
  \end{pmatrix}
\end{equation}
From Eq. \eqref{eqn:autocdef-3a}, the one-step autocorrelation \autoc{h_t}{1} is
given by:
\begin{equation}
\label{eqn:onestep-3a}
  \autoc{h_t}{1} = \frac{\ave{h_th_{t+1}}_t-\ave{h_t}_t^2}
                           {\ave{h_t^2}_t-\ave{h_t}_t^2} \;.
\end{equation}
Therefore we need to calculate values for $\ave{h_t^2}_t$, $\ave{h_t}_t$ and
$\ave{h_th_{t+1}}_t$. Since $h_t=\pm1$, $\ave{h_t^2}_t=1$. In
order to calculate values for the other two averages it is necessary to form
the vectors $\mv{h_0}$ and $\mv{h_1}$ which respectively give the values of
$h_t$ and $h_th_{t+1}$ for each of the states in the simplified state space.
The elements of $\mv{h_0}$ can be read directly from Fig. 
\ref{fig:allstates-3a}. This gives:
\begin{equation}
\label{eqn:hzero-3a}
	\mv{h_0} = 
	\begin{pmatrix}
	  -1 & +1 & -1 & +1 & -1 & -1 \\
	\end{pmatrix} \; .
\end{equation}
Taking the scalar product of $\mv{h_0}$ with $\mv{s}$ gives:
\begin{equation}
\label{eqn:have-3a}
	\ave{h_t}_t = \mv{h_0}\mv{.s} = \frac{\alpha-1}{\alpha+1} \;.
\end{equation}
In Sec. \ref{sec:hbar-3a} we used an alternative method, not restricted to
$m=1$, to derive the expression for $\ave{h_t}_t$ in Eq.  \eqref{eqn:avehI-3a}.
Substituting $I_1=0$ and $I_2=\alpha$ in Eq. \eqref{eqn:avehI-3a} yields Eq.
\eqref{eqn:have-3a} above. Thus, the analysis presented here and in Sec.
\ref{sec:hbar-3a} are consistent. 

We can derive $\mv{h_1}$ by multiplying the values of $h_t$ and
$h_{t+1}$ in Fig. \ref{fig:transitions-3a}. This gives:
\begin{equation}
\label{eqn:hone-3a}
	\mv{h_1} = 
	\begin{pmatrix}
	  +1 & -1 & -1 & +1 & +1 & -1 \\
	\end{pmatrix} \; .
\end{equation}
Once again taking the scalar product with $\mv{s}$ gives:
\begin{equation}
\label{eqn:honeave-3a}
	\ave{h_th_{t+1}}_t=\mv{h_1.s}=\left(\frac{\alpha-1}{\alpha+1}\right)^2
	\; .
\end{equation}
Therefore, from Eqs. \ref{eqn:onestep-3a},  \ref{eqn:have-3a} and
\ref{eqn:honeave-3a} we have:
\begin{equation}
\label{eqn:onestepresult-3a}
	\autoc{h_t}{1} = 0 \;.
\end{equation}
Thus, we expect $h_t$ and $h_{t+1}$ to be uncorrelated which is in agreement
with Fig. \ref{fig:hautocwithmem-3a}a.

\floatfig[\floatplace]{fig:twostep-3a}{
Example of derivation of values of $h_th_{t+2}$.
}{
Example of how to derive the values of $h_th_{t+2}$ for each of the states
in the simplified state space.
}{
\newcommand{\figfooone}[1]{$\therefore h_th_{t+2}=#1$}
\begin{center}
\resizebox{0.38\textwidth}{!}{
\begin{picture}(0,0)%
\includegraphics{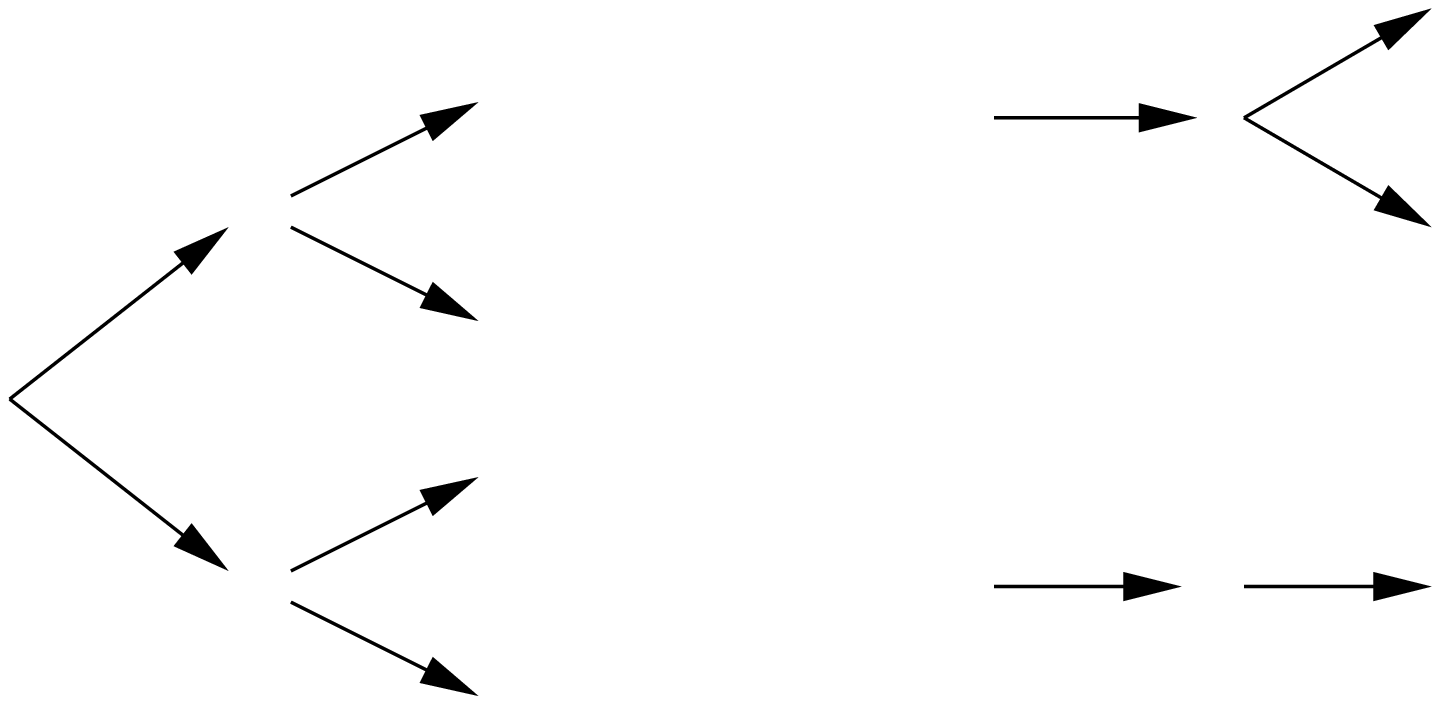}%
\end{picture}%
\setlength{\unitlength}{3947sp}%
\begingroup\makeatletter\ifx\SetFigFont\undefined%
\gdef\SetFigFont#1#2#3#4#5{%
  \fontsize{#1}{#2pt}%
  \fontfamily{#3}\fontseries{#4}\fontshape{#5}%
  \selectfont}%
\fi\endgroup%
\begin{picture}(7125,3898)(1651,-4289)
\put(6376,-3436){\makebox(0,0)[b]{\smash{\SetFigFont{20}{24.0}{\familydefault}{\mddefault}{\updefault}{\color[rgb]{0,0,0}14}%
}}}
\put(7576,-3436){\makebox(0,0)[b]{\smash{\SetFigFont{20}{24.0}{\familydefault}{\mddefault}{\updefault}{\color[rgb]{0,0,0}5}%
}}}
\put(8776,-3436){\makebox(0,0)[b]{\smash{\SetFigFont{20}{24.0}{\familydefault}{\mddefault}{\updefault}{\color[rgb]{0,0,0}1}%
}}}
\put(6376,-3661){\makebox(0,0)[b]{\smash{\SetFigFont{14}{16.8}{\familydefault}{\mddefault}{\updefault}{\color[rgb]{0,0,0}$h_t=-1$}%
}}}
\put(8776,-3661){\makebox(0,0)[b]{\smash{\SetFigFont{14}{16.8}{\familydefault}{\mddefault}{\updefault}{\color[rgb]{0,0,0}$h_{t+2}=+1$}%
}}}
\put(7951,-4186){\makebox(0,0)[rb]{\smash{\SetFigFont{25}{30.0}{\familydefault}{\mddefault}{\updefault}{\color[rgb]{0,0,0}\figfooone{-1}}%
}}}
\put(3001,-1636){\makebox(0,0)[b]{\smash{\SetFigFont{20}{24.0}{\familydefault}{\mddefault}{\updefault}{\color[rgb]{0,0,0}6}%
}}}
\put(4201,-2236){\makebox(0,0)[b]{\smash{\SetFigFont{20}{24.0}{\familydefault}{\mddefault}{\updefault}{\color[rgb]{0,0,0}4}%
}}}
\put(4201,-4036){\makebox(0,0)[b]{\smash{\SetFigFont{20}{24.0}{\familydefault}{\mddefault}{\updefault}{\color[rgb]{0,0,0}0}%
}}}
\put(1651,-2536){\makebox(0,0)[b]{\smash{\SetFigFont{20}{24.0}{\familydefault}{\mddefault}{\updefault}{\color[rgb]{0,0,0}0}%
}}}
\put(3001,-3436){\makebox(0,0)[b]{\smash{\SetFigFont{20}{24.0}{\familydefault}{\mddefault}{\updefault}{\color[rgb]{0,0,0}0}%
}}}
\put(4201,-1036){\makebox(0,0)[b]{\smash{\SetFigFont{20}{24.0}{\familydefault}{\mddefault}{\updefault}{\color[rgb]{0,0,0}14}%
}}}
\put(4201,-2836){\makebox(0,0)[b]{\smash{\SetFigFont{20}{24.0}{\familydefault}{\mddefault}{\updefault}{\color[rgb]{0,0,0}6}%
}}}
\put(4201,-3211){\makebox(0,0)[b]{\smash{\SetFigFont{14}{16.8}{\familydefault}{\mddefault}{\updefault}{\color[rgb]{0,0,0}$h_{t+2}=-1$}%
}}}
\put(4201,-4261){\makebox(0,0)[b]{\smash{\SetFigFont{14}{16.8}{\familydefault}{\mddefault}{\updefault}{\color[rgb]{0,0,0}$h_{t+2}=-1$}%
}}}
\put(4201,-2386){\makebox(0,0)[b]{\smash{\SetFigFont{14}{16.8}{\familydefault}{\mddefault}{\updefault}{\color[rgb]{0,0,0}$h_{t+2}=-1$}%
}}}
\put(4201,-1411){\makebox(0,0)[b]{\smash{\SetFigFont{14}{16.8}{\familydefault}{\mddefault}{\updefault}{\color[rgb]{0,0,0}$h_{t+2}=-1$}%
}}}
\put(3301,-4186){\makebox(0,0)[rb]{\smash{\SetFigFont{25}{30.0}{\familydefault}{\mddefault}{\updefault}{\color[rgb]{0,0,0}\figfooone{+1}}%
}}}
\put(1651,-2836){\makebox(0,0)[b]{\smash{\SetFigFont{14}{16.8}{\familydefault}{\mddefault}{\updefault}{\color[rgb]{0,0,0}$h_t=-1$}%
}}}
\put(6376,-1186){\makebox(0,0)[b]{\smash{\SetFigFont{20}{24.0}{\familydefault}{\mddefault}{\updefault}{\color[rgb]{0,0,0}5}%
}}}
\put(7576,-1186){\makebox(0,0)[b]{\smash{\SetFigFont{20}{24.0}{\familydefault}{\mddefault}{\updefault}{\color[rgb]{0,0,0}1}%
}}}
\put(8776,-586){\makebox(0,0)[b]{\smash{\SetFigFont{20}{24.0}{\familydefault}{\mddefault}{\updefault}{\color[rgb]{0,0,0}6}%
}}}
\put(8776,-1786){\makebox(0,0)[b]{\smash{\SetFigFont{20}{24.0}{\familydefault}{\mddefault}{\updefault}{\color[rgb]{0,0,0}0}%
}}}
\put(6376,-1411){\makebox(0,0)[b]{\smash{\SetFigFont{14}{16.8}{\familydefault}{\mddefault}{\updefault}{\color[rgb]{0,0,0}$h_t=+1$}%
}}}
\put(8776,-2086){\makebox(0,0)[b]{\smash{\SetFigFont{14}{16.8}{\familydefault}{\mddefault}{\updefault}{\color[rgb]{0,0,0}$h_{t+2}=-1$}%
}}}
\put(7951,-2086){\makebox(0,0)[rb]{\smash{\SetFigFont{25}{30.0}{\familydefault}{\mddefault}{\updefault}{\color[rgb]{0,0,0}\figfooone{-1}}%
}}}
\put(8776,-961){\makebox(0,0)[b]{\smash{\SetFigFont{14}{16.8}{\familydefault}{\mddefault}{\updefault}{\color[rgb]{0,0,0}$h_{t+2}=-1$}%
}}}
\end{picture}
}
\end{center}
}

\medskip\noindent
The two-step autocorrelation \autoc{h_t}{2\,} is given by:
\begin{equation}
\label{eqn:twostep-3a}
  \autoc{h_t}{2\,} = \frac{\ave{h_th_{t+2}}_t-\ave{h_t}_t^2}
                           {\ave{h_t^2}_t-\ave{h_t}_t^2} \;.
\end{equation}
Therefore in order to calculate the two-step autocorrelation function
\autoc{h_t}{2\,} we must calculate $\ave{h_th_{t+2}}_t$. As before we must form
the vector $\mv{h_2}$ corresponding to the values of $h_th_{t+2}$ for each of
the states in the simplified space. This can be done following the method
described in Fig. \ref{fig:twostep-3a} which gives:
\begin{equation}
\label{eqn:htwo-3a}
	\mv{h_2} = 
	\begin{pmatrix}
	  +1 & -1 & +1 & -1 & -1 & -1 \\
	\end{pmatrix} \; .
\end{equation}
Taking the scalar product with $\mv{s}$ gives:
\begin{equation}
\label{eqn:htwoave-3a}
	\ave{h_th_{t+2}}_t=\mv{h_2.s}=\frac{(1-3\alpha)(1+\alpha)}{(1+\alpha)^2}
	\; .
\end{equation}
From Eqs. \ref{eqn:have-3a}, \ref{eqn:twostep-3a} and \ref{eqn:htwoave-3a} we
have:
\begin{equation}
\label{eqn:twostepresult-3a}
	\autoc{h_t}{2\,} = -\alpha.
\end{equation}
Therefore, from Eqs. \ref{eqn:onestepresult-3a} and \ref{eqn:twostepresult-3a}
we have that:
\begin{align}
\label{eqn:acseq-3a}
  	\autoc{h_t}{\tau}&=1,0,-\alpha,0,\alpha^2,0,-\alpha^3,\ldots \\
  	\intertext{for:} 
	\tau&=0,1,2,3,4,5,6,\ldots \notag
\end{align}
Note that as a result of this
the autocorrelation of $h_t$ at $m=1$ gives a direct measurement of
the value of $\alpha$. This provides a much more convenient measure of
$\alpha$ than computing the summation for $I_2$ in Eq. \eqref{eqn:Idefs-3a}, as
was done in Sec. \ref{sec:hbar-3a}. 

Figure \ref{fig:autoctheory-3a} shows a comparison of the theoretical values
of \autoc{h_t}{\tau}, obtained with the numerical value of $\alpha=0.323$ from
Sec. \ref{sec:hbar-3a}, with the numerical values given in Fig.
\ref{fig:hautocwithmem-3a}. We can see from Fig. \ref{fig:autoctheory-3a}
that Eq. \eqref{eqn:acseq-3a} correctly predicts the form of \autoc{h_t}{\tau}
although it does underestimate the correlation for $\tau=4$. One possible
explanation of this slight deviation is that in the analysis above we have
assumed that $\alpha$ is a constant. However, the value of $\alpha$ depends
very sensitively on the form of $\ind{\Pi}$. Therefore fluctuations in the
gene value distribution of the agents could cause significant fluctuations in
$\alpha$.
Markovian analysis of the $l=0.5$ case confirms that $\autoc{h_t}{\tau}=0$
for $\tau>1$, in agreement with the numerical results of Fig.
\ref{fig:hautocwithmem-3a}.

\floatfig[\floatplace]{fig:autoctheory-3a}{
Comparison of theoretical and numerical results for the autocorrelation
of $h_t$.
}{
Comparison of the theoretical result of Eqs. \ref{eqn:onestepresult-3a} and
\ref{eqn:twostepresult-3a} with numerical results. The error bars show one
standard deviation on the mean over an ensemble of five separate data sets.
}{
\begin{center}
\resizebox{0.45\textwidth}{!}{
\begingroup%
  \makeatletter%
  \newcommand{\GNUPLOTspecial}{%
    \catcode`\%=14\relax\special}%
  \setlength{\unitlength}{0.1bp}%
\begin{picture}(3600,1620)(0,0)%
\includegraphics{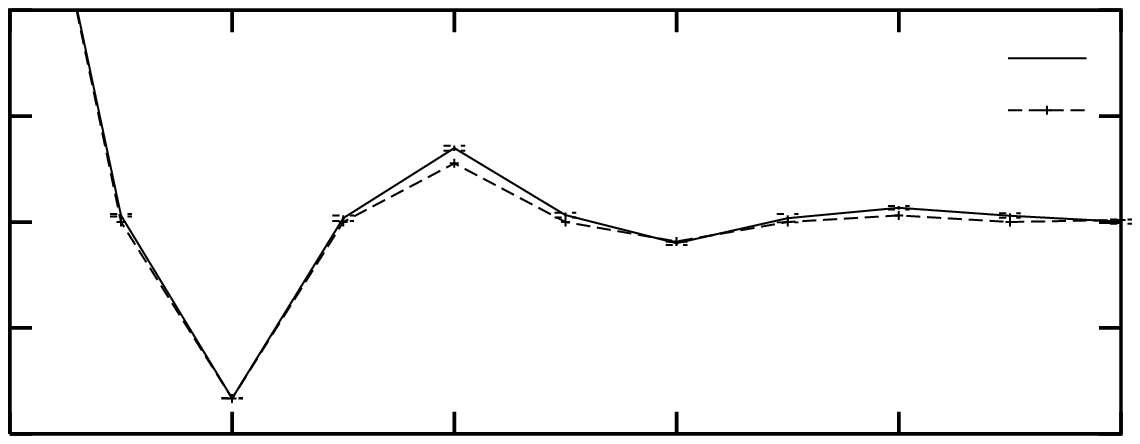}
\put(3225,1232){\makebox(0,0)[r]{Theoretical}}%
\put(3225,1382){\makebox(0,0)[r]{Numerical}}%
\put(2000,50){\makebox(0,0){Lag: $\tau$}}%
\put(50,910){%
\makebox(0,0)[b]{\shortstack{\autoc{h_t}{\tau}}}%
}%
\put(3600,200){\makebox(0,0){10}}%
\put(2960,200){\makebox(0,0){8}}%
\put(2320,200){\makebox(0,0){6}}%
\put(1680,200){\makebox(0,0){4}}%
\put(1040,200){\makebox(0,0){2}}%
\put(400,200){\makebox(0,0){0}}%
\put(350,1520){\makebox(0,0)[r]{0.4}}%
\put(350,1215){\makebox(0,0)[r]{0.2}}%
\put(350,910){\makebox(0,0)[r]{0}}%
\put(350,605){\makebox(0,0)[r]{-0.2}}%
\put(350,300){\makebox(0,0)[r]{-0.4}}%
\end{picture}%
\endgroup

}
\end{center}
}
\subsection{Summary of Memory Characteristics}

We conclude from the results presented in this section that the Genetic Model
performs better in the absence of memory. By this we mean that the average
total number of points scored per time step when $h_t=+1\;\forall\;t$
is twice that when $h_t$ is determined by the global memory. This was
established semi-analytically in Sec. \ref{sec:comptradml-3a} and supporting
numerical data was presented in Sec. \ref{sec:numerics-3a}.
We showed in Sec. \ref{sec:directcomp-3a} that the reason for this reduction in
the performance in the presence of memory is that the agents cannot directly
control the distribution of values for \at\, as they can in the memoryless
model, because \at\ is now also a function of $h_t$. In Sec.
\ref{sec:external-3a} we presented numerical data to show that the 
values of $\ave{\at}_t$ and $\std{\at}t$ can be reproduced
if the prediction $h_t$ is taken from an exogenous source
providing that the value of $\ave{h_t}_t$ is preserved. Thus, the feedback
between the global action of the agents and $h_t$ is of no benefit to the
population of agents. The only function of this feedback is to regulate the
value of $\ave{h_t}_t$.

We also investigated the values observed for the time average of the
prediction $\ave{h_t}_t$ as a function of the resource level $l$. We showed that
the values of $\ave{h_t}_t$ of $\pm 0.5$ that obtain in the dynamic regime
of $l_{c1}<l<l_{c2}$ are due to the deviation observed between the mean of
the agent \pval\ distribution $P(x)$ and the optimal value predicted by Eq. 
\eqref{eqn:optpml-3a}.

In Sec. \ref{sec:autoc-3a} we showed that the form of the autocorrelation
function of the \at\ time series at $l\ne0.5$ occurs as a result of the cycles
in state space performed by the memory. Finally we demonstrated that the
two-step autocorrelation of the prediction \autoc{h_t}{2} can be used to provide
a direct measurement of the deviation described above.

\newcommand{\gcgm}{Grand Canonical Genetic Model}
\newcommand{\act}{\ensuremath{n_t^\text{act}}}
\newcommand{\inact}{\ensuremath{n_t^\text{inact}}}

\newcommand{\expgrphheight}{6.2cm}

\section{Self-induced Shocks in the Grand Canonical Genetic Model}
\label{ch:large-changes}

In this section we move on to consider the important practical property of
self-induce shocks, otherwise known as endogenous large changes (ELC). Such
large changes are arguably a defining characteristic of complex systems, yet
there is no rigorous quantitative description of such events in real-world
realizations of complex systems  (for examples see Refs.
\cite{bubbles,stylized-facts,herd,crash-avoid,anatomy,johnson-review,
trader-dynamics,pred}).

\subsection{Introduction}
\label{sec:intro-3b}

As we show here, the Genetic Model can be generalized in a straightforward way
to produce a model system which demonstrates such large changes. The extent to
which these large changes are insensitive to the memory, then provides a useful
tool for analyzing the microscopic causes underlying these large changes.  In
particular, we introduce an extension of the Genetic Model in Sec.
\ref{ch:gm-memory}, in which the number (or `volume') of active agents is a time
dependent quantity. We shall refer to this variant of the Genetic Model as the
Grand Canonical Genetic Model (GCGM).

One particular application might be to financial markets, where large changes
are called crashes or drawdowns. However the Genetic Model does not directly
yield a price time series. Therefore, if we want to consider the effect of
endogenous large changes on price, we must derive one from fundamental
observables such as $n^{\pm1}_t$. By definition the threshold value of $\at=Nl$
corresponds to the state in which the volume of the item that is being traded
which is available for sale is equal to the demand. Therefore, the more general
case of $l\ne0.5$ represents a system in which the volumes in which an item are
bought and sold are not equal.  The excess demand is then given by:
\begin{equation}
	\Delta = \left(\frac{l}{1-l}\right) N_\text{buy} - N_\text{sell}
\end{equation}
If we let the action of an agent $\ind a=-1,+1$ represent choosing to 
\emph{buy} or \emph{sell} respectively then we obtain the following expression
for the price $\pi_{t+1}$ at time $t+1$ in terms of the price at $t$:
\begin{equation}
\label{eqn:price-3b}
	\pi_{t+1} = \pi_t + \frac{1}{\lambda}\left[
	  \left(\frac{l}{1-l}\right)n^{-1}_t - \at
	  \right]
	  \;,
\end{equation}
where $\lambda$ is known as the \emph{market depth} and determines the
magintude of the change in price caused by a unit change in $\Delta$.
Different expressions for the price $\pi_t$ in terms of the excess
demand $\Delta$ have been discussed in the literature (see, for example, Refs.
\cite{langevin-approach} and \cite{marginally-efficient}). The linear expression
in Eq. \eqref{eqn:price-3b} represents the simplest of these and is not as
realistic as expressions in higher powers of $\Delta$. Nevertheless it is
more than adequate for the illustrative purposes for which we shall need it.
Since in all that follows the units of the price $\pi_t$ are arbitrary we will
take $\lambda=1$.

An example of a large volume change observed in the GCGM is given in
Fig. \ref{fig:crexp2-3b}. We can see that large changes occur in the volume
accompanied by large price movements. However, the behavior of the volume in
the figure is qualitatively different to that observed in the GCMG: Ref.
\cite{trader-dynamics} reported two distinct types of behavior. For
traders with a long memory the volume was observed to be continuously
fluctuating with occasional particularly large fluctuations which were not
instantaneous; much like the `drawdowns' and `drawups' discussed by Sornette
in Ref. \cite{drawdowns}. For traders with short memories the volume was
frequently zero with occasional large instantaneous spikes and corresponding
instantaneous price movements. In contrast, in Fig. \ref{fig:crexp2-3b} the
volume exhibits small fluctuations and occasional instantaneous changes which
are accompanied by periods of large fluctuations in the price. We shall see that
the behavior that we observe in these figures is typical of the behavior of
the volume in the GCGM.

\floatfig[\floatplace]{fig:crexp2-3b}{
Example of an endogenous large change of the volume in the GCGM.
}{
Example of an endogenous large change of the volume in the GCGM with
accompanying price time series. Model Parameters: $n=501$, $m=3$, $r=0.2$,
$l=0.5$ and $T=12$.
}{
\begin{center}
\resizebox{0.48\textwidth}{!}{\footnotesize 
\begingroup%
  \makeatletter%
  \newcommand{\GNUPLOTspecial}{%
    \catcode`\%=14\relax\special}%
  \setlength{\unitlength}{0.1bp}%
\begin{picture}(3600,1296)(0,0)%
\includegraphics{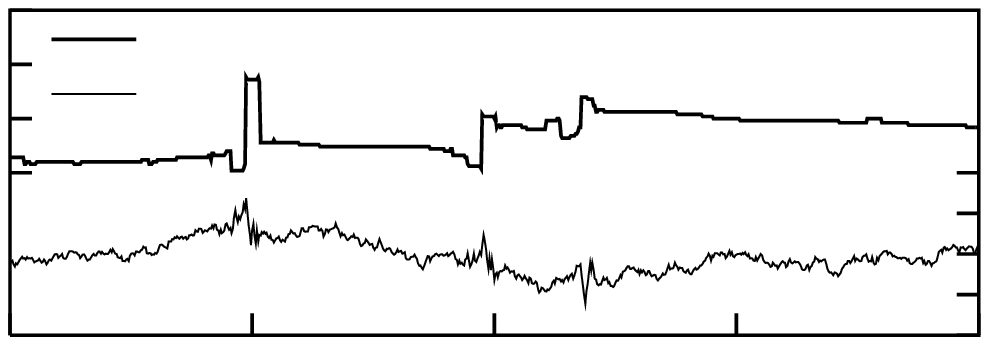}
\put(768,965){\makebox(0,0)[l]{$\pi_t$}}%
\put(768,1122){\makebox(0,0)[l]{$n_t$}}%
\put(1755,45){\makebox(0,0){Time step $t$}}%
\put(3509,738){%
\makebox(0,0)[b]{\shortstack{Price: $\pi_t$}}%
}%
\put(90,738){%
\makebox(0,0)[b]{\shortstack{volume: $n_t$}}%
}%
\put(3195,738){\makebox(0,0)[l]{-400}}%
\put(3195,621){\makebox(0,0)[l]{-600}}%
\put(3195,504){\makebox(0,0)[l]{-800}}%
\put(3195,387){\makebox(0,0)[l]{-1000}}%
\put(3195,270){\makebox(0,0)[l]{-1200}}%
\put(3150,180){\makebox(0,0){29000}}%
\put(2452,180){\makebox(0,0){28800}}%
\put(1755,180){\makebox(0,0){28600}}%
\put(1057,180){\makebox(0,0){28400}}%
\put(360,180){\makebox(0,0){28200}}%
\put(315,1206){\makebox(0,0)[r]{250}}%
\put(315,1050){\makebox(0,0)[r]{225}}%
\put(315,894){\makebox(0,0)[r]{200}}%
\put(315,738){\makebox(0,0)[r]{175}}%
\end{picture}%
\endgroup

}
\end{center}
}

\subsection{The Grand Canonical Genetic Model}
\label{sec:gcgm-3b}

The Genetic Model represents an abstract model of a population competing for a
limited resource and as such it often discussed in the context of financial
markets
\cite{emg-phase,emg-cluster,seg,self-seg,evol-freeze,emg-hetero,emg-theory}.
In such a context, however, it does not seem realistic that the agents trade at
every time step. A real market trader would also have the option of withdrawing
from the market and returning when s/he felt confident of a successful outcome.
In order to model this extra degree of freedom, and in keeping with the work of
\jea\ in Ref. \cite{trader-dynamics}, we shall introduce an extension
of the Genetic Model in which the agents are free to opt in and out of the game.

As in the original Genetic Model described in Sec. \ref{ch:gm-memory} and
Ref. \cite{self-seg} there are $N$ agents participating in the model. However,
unlike the original model they do not all play at each time step. At any time
$t$ there will be two populations of agents, an active population and an
inactive one. Active agents participate in exactly the same way as
they do in the original model. In contrast an inactive agent $i$ will
continue to make its choice as if it were participating, however it will
not be considered when calculating the global action and its score will not be
updated. We can imagine that inactive agents represent traders who do not make a
trade at time $t$.  Instead they make a prediction of whether their best choice
would have been to follow the prediction $a_{i;t}=+h_t$ or to refute it
$a_{i;t}=-h_t$. Since such a trade is \emph{virtual} in that the trader does not
act on it, it has no effect on the market and the trader is protected from
losing, or even making, money. The only effect of such a virtual trade is that
the agent reevaluates its confidence level depending on whether the trade would
have been successful or not. These \emph{virtual trades} are analogous to the
\emph{virtual points} earned by strategies in the Minority Game
\cite{challet-orig-mg}. They allow the agents to keep track of their potential
performance without in any way representing an agent's wealth.

These modifications necessitate some minor changes to the basic equations of the
Genetic Model that we introduced in Sec. \ref{ch:gm-memory}. Let the activity
of an agent $i$ be given by \ind z. If $\ind z=0$ then agent $i$ belongs to the
population of inactive agents and vice-versa for $\ind z=1$. 
Equation \eqref{eqn:Acond-3a} for the global action $A_t$ now becomes:
\begin{align}
\label{eqn:globalact-3b}
	A_t = \begin{cases}
	        +1  &  \at \le ln_t \\
		-1  &  \at >   ln_t
	      \end{cases}   
	      && \text{where: }
	n_t = \sum\limits_i \ind z  \;.
\end{align}
Inactive agents become active and active agents become inactive according to
their performance in the recent past. We defined a new model parameter known as
the \emph{confidence interval}, $T$. An inactive agent will
become active if it would have won for $T$ consecutive time steps. In other
words, an agent $i$ for which $\ind z=0$ will activate, $z_{i;t+1}=1$, if 
$a_{i;\tau}=+A_\tau$ for $t-T<\tau\le t$. In the same way, an active agent will
become inactive if it loses for $T$ consecutive time steps. In order to control
the activation and deactivation of agents we assign each agent a new quantity
with we shall call its virtual points \ind v, in keeping with the virtual points
allocated to strategies in the Minority Game. For an active agent \ind v is
increased each time the agent loses and is reset to zero if it wins. Thus, for
active agents \ind v is the number of consecutive time steps for which the agent
has lost. For an inactive agent \ind v is increased each time that the agent
would have won and is reset to zero each time it would have lost. The
updating rules for \ind v can be summarized as follows.

\noindent
If $\ind z =0$:
\begin{align}
\label{eqn:vrules-3b}
	v_{i;t+1} &= 
	  \begin{cases}
	    \makebox[15mm][l]{0}  & \f\ind a = -A_t \\
	    \ind v +1             & \f\ind a = +A_t 
	  \end{cases}
	\\
	\intertext{If $\ind z =1$:} 
	v_{i;t+1} &= 
	  \begin{cases}
	    \ind v +1             & \f\ind a = -A_t \\
	    \makebox[15mm][l]{0}  & \f\ind a = +A_t 
	  \end{cases} \;.\\
\intertext{The rules for agent activation and deactivation are then,
If $\ind z =0$:}
\label{eqn:zrules-3b}
	z_{i;t+1} &= 
	  \begin{cases}
	    \makebox[15mm][l]{0}  &  \f\ind v < T \\
	    1                     &  \f\ind v = T
	  \end{cases}
	\\
\intertext{If $\ind z=1$:}
	z_{i;t+1} &= 
	  \begin{cases}
	    \makebox[15mm][l]{1}  &  \f\ind v < T \\
	    0                     &  \f\ind v = T
	  \end{cases}  \; .
\end{align}
The expression in Eq. \eqref{eqn:at-3a} for the number of agents for which
$\ind a = +1$ becomes:
\begin{equation}
\label{eqn:at-3b}
	\at = \half \sum\limits_i \ind z \left(a_{i;t} + 1 \right)  \;.
\end{equation}
In this section we shall define \pbar\ to be the mean \pval\ of the 
\emph{active} agent population. Thus, \pbar\ is given by:
\begin{equation}
\label{eqn:pbar-3b}
	\pbar = \frac{1}{n_t} \sum\limits_i\ind z\ind p
\end{equation}
Therefore, the expression in Eq. \eqref{eqn:avefol_en-3a} for the mean number of
agents following the prediction $h_t$ becomes:
\begin{equation}
\label{eqn:avefol-3b}
	\begin{split}
	  \ave{\fol} &= \sum\limits_i \ind z\ind p \\
	               &= n_t\pbar
	\end{split}
\end{equation}

\subsection{Price Time-Series}
\label{sec:pricets-3b}

As we stated in Sec. \ref{sec:intro-3b} the cutoff, now given by $\at=n_tl$, in
the Genetic Model, defined by Eq. \eqref{eqn:globalact-3b}, is by definition the
state in which the excess demand $\Delta=0$. From Eq. \eqref{eqn:price-3b} we
can see that the price change $\Delta\pi=\pi_t-\pi_{t-1}$ is positive and
negative for $\at<n_tl$ and $\at>n_tl$
respectively. By comparison with Eq. \eqref{eqn:globalact-3b} we can see that
the condition that determines the sign of the price change at time $t$ is the
same as that which determines the global action $A_t$; with the exception that
the equality in Eq. \eqref{eqn:globalact-3b} gives a price change of zero.
With reference to Eq. \eqref{eqn:folcond-3a} we have the condition that
$\Delta\pi\ge0$ in terms of \fol:
\begin{align}
	\fol &\ge n_t(1-l)  &  \f h_t&=-1 \notag\\
	\fol &\le n_t l     &  \f h_t&=+1 \;.
\end{align}
From this it follows that the probability $P[\Delta\pi\ge0]$ that the price
rises or remains the same at time $t$ is given by:
\begin{equation}
\label{eqn:priceprob-3b}
	P[\Delta\pi\ge0] = \begin{cases}
	                    I_2 &\f h_t=-1 \\
			    I_1 &\f h_t=+1
			   \end{cases} \; ,
\end{equation}
where $I_1$ and $I_2$ represent the summations defined by Eq.
\eqref{eqn:Idefs-3a} and illustrated by Fig. \ref{fig:intexp-3a}, with the
substitution $N\rightarrow n_t$.
We can see from Eq. \eqref{eqn:priceprob-3b} and Fig. \ref{fig:intexp-3a}
that for $l<0.5$ $P[\Delta\pi\ge0]=0$ for $h_t=+1$ and for $l>0.5$
$P[\Delta\pi\ge0]=1$ for $h_t=-1$. The result of this is that, in the dynamic
regime of the original GCGM with memory, one of the values that $h_t$ can
take will cause the price to rise or fall with probability $P=1$. Note that in
the cases of $l<0.5, h_t=-1$ and $l>0.5, h_t=+1$ we do not expect that
$P[\Delta\pi\ge0]=0.5$ as might be expected from Fig. \ref{fig:intexp-3a}.
Recall that, as we discussed in Sec. \ref{sec:hbar-3a}, \pbar\ deviates from the
optimum values given in Eq. \eqref{eqn:optp-3a} in the Genetic Model. Furthermore, because
the GCGM is frequently perturbed by ELC it does not settle into equilibrium in
the same way as the Genetic Model and so \pbar\ is more variable although, as we shall see,
it does remain close to the values given by Eq. \eqref{eqn:optp-3a} in the
periods between ELC.

\subsection{Endogenous Large Changes (ELC) in the GCGM}
\label{sec:overview-3b}

As we shall see the behavior of the model in this regime is rich and complex.
For this reason we shall initially consider a simplified memoryless variant of
the GCGM which is analogous to the memoryless Genetic Model considered in
Sec.
\ref{ch:gm-memory}. Subsequently we shall consider how the inclusion of memory
affects the behavior of the full GCGM. 

ELC in the GCGM result from a combination of two
factors, both of which must be present if a large change is to occur. We shall
see later that the capacity of the model to undergo an ELC depends sensitively
on the distribution of \pval s $P(p)$. However, a suitable $P(p)$ is not a
sufficient condition for an ELC to occur. It is also necessary for a particular
pattern to occur in the global action time series. We can think of this pattern
as a \emph{trigger} that initiates the ELC, but only if $P(p)$ is in a
susceptible state. Furthermore we shall see that the natural evolution of the
model causes $P(p)$ to evolve towards the most susceptible state while ELCs
move $P(p)$ towards the state in which it is least susceptible.
Thus, rather than settling into equilibrium like the original Genetic Model, the
evolution of the GCGM is characterized by a cyclic behavior: $P(p)$
periodically evolving towards a more susceptible state until its progress is
reversed by an ELC.

\subsubsection
[Susceptibility of the gene value distribution]
{Susceptibility of $P(p)$}
\label{sec:suscep-3b}

\newcommand{\zero}{\emph{zero} }
\newcommand{\one}{\emph{one} }
ELC like the one illustrated in Fig. \ref{fig:crexp2-3b}
are the result of highly correlated behavior of the agents. By this we mean
that a significant number of agents activate or deactivate at the same time
step. This implies highly correlated behavior since in order to do so the
actions of all of the agents involved must be identical for the $T$ preceding 
time steps. It is initially surprising that such a high degree of correlation
could arise in the GCGM because, unlike Minority Game agents, GCGM agents make their
decisions stochastically. The probability of coincidence between the actions of
a large group of agents 
will usually be very small. However, there are 
two groups of agents in the model whose behavior is well correlated.
These two groups are those agents whose gene values lie within a certain small
range $\delta$ of $0$ and $1$. We
shall call agents belonging to these groups \emph{zero agents} and \emph{one
agents} respectively. The degree of correlation of these agents is a decreasing
function of $\delta$, being a maximum for $\delta=0$. It is easy to show that if
we set an upper limit on the fraction $f_d$ of \zero and \one agents whose
actions are not perfectly correlated over a period of $T$ time steps, 
$\delta$ is given by:
\begin{equation}
\label{eqn:d-3b}
	\delta = 1 - (1-f_d)^\frac{1}{T}
\end{equation}
The precise value chosen for $\delta$ is not important since it only serves to
give a measure of the population of \zero and \one agents. Therefore it is
more convenient to choose a fixed value for $\delta$ which gives rise to values
of $f_d$ which lie within an acceptable range, rather than choosing a different
$\delta$ for each value of $T$. In all that follows we shall take $\delta=0.02$
which gives $f_d<0.33$ for $T\le20$.

In short, we see that the probability of highly correlated agent behavior
increases rapidly as the number of \zero and \one agents increases. Thus, gene
values distribution functions $P(p)$ which are biased to favor agents with gene
values near to $p=0.0$ and $p=1.0$ will be the most susceptible to ELC.

Now that we've considered what forms of $P(p)$ are most susceptible we will
discuss how $P(p)$ evolves in the GCGM.  In the periods between the ELC the
number of active agents $n_t$ given by Eq. \eqref{eqn:globalact-3b} is a slowly
varying function of time. As we remarked above, correlations between the
behavior of large numbers agents are expected to be rare and so $n_t$ will
fluctuate slowly with time as individual agents activate and deactivate. Hence
the results of Refs.  \cite{seg,evol-freeze} can be applied to the GCGM in these
periods.
Refs. \cite{seg,evol-freeze} describe how in the Genetic Model the agents
\emph{self-segregate} into two populations having low and high gene values
-- these two populations can be thought of as a `Crowd' and `Anticrowd'.
Thus, although we have yet to consider what the effect of an ELC will be on
$P(p)$ we can see that after such an event $P(p)$ will evolve continuously
towards the extremised distribution described by Ref. \cite{seg}. From
our discussion above we know that it is this type of extremised gene value
distribution which is most susceptible to ELC.

\subsubsection{Triggers in the global action time series}
\label{sec:triggers-3b}
\newcommand{\lp}{\ensuremath{\lambda^+_t}}    
\newcommand{\lm}{\ensuremath{\lambda^-_t}}    
\newcommand{\lpm}{\ensuremath{\lambda^\pm_t}} 

We saw above that it is only the \zero and \one agents which can participate in
the highly correlated behavior necessary for an ELC. Therefore, in order to
think about what patterns in the global action time series might induce an ELC
it is necessary to consider these agents.  If at time step $t$, $A_t=+h_t$ then
each \zero agent will lose while each \one agent will win. If $A_t=+h_t$ for $T$
consecutive time steps then immediately following the $T$th time step a fraction
$1-f_d$ of the active \zero agents will deactivate while the same fraction of
the inactive \one agents will activate. Similarly if $A_t=-h_t$ for $T$ time
steps then a fraction $1-f_d$ of the inactive \zero agents will activate and a
fraction $1-f_d$ of the active \one agents will deactivate.  Thus we can see
that sequences of time steps in which $A_t=+h_t$ or $A_t=-h_t$ for $T$ time
steps will be important for the correlated
agent activations and deactivations that make up an ELC.

Since such sequences are important in the occurrence of ELC it would be useful
to have an expression for the probability that they will occur. The first step
is to derive expressions for the probability that $A_t=\pm h_t$. We shall see
later that these expressions are important in their own right. From Eq.
\eqref{eqn:probAone-3a}, substituting $n_t$ for $N$, we have the following
expressions for the probability that $A_t=\pm h_t$:
\begin{equation}
\begin{split}
	P[A_t=-h_t] &= 
	\begin{cases}
	  \sum\limits_{\makebox[15mm]{\scriptsize $i=(1-l)n_t$}}^{n_t}\ind{\Pi}
	     & \f h_t=-1 \\
	  \sum\limits_{\makebox[15mm]{\scriptsize $i=l n_t+1$}}^{n_t} \ind{\Pi} 
	     & \f h_t=+1 
	\end{cases}
	\\
	P[A_t=+h_t] &= 
	\begin{cases}
	  \sum\limits_{\makebox[15mm]{\scriptsize $i=0$}}^{(1-l)n_t-1}\ind{\Pi} 
	     & \f h_t=-1 \\
	  \sum\limits_{\makebox[15mm]{\scriptsize $i=0$}}^{l n_t} \ind{\Pi}   
	     & \f h_t=+1 
	\end{cases}
\end{split}
\end{equation}
We shall see later that $n_t\gg 1$. Therefore we can use the continuous
approximation of $\Pi_t(x)$ for $\ind{\Pi}$.
As demonstrated in Ref. \cite{emg-theory}, $\ind{\Pi}$
will be a Gaussian. In the GCGM only active agents contribute to the mean $\mu$
and variance $\sigma^2$ of $\Pi_t(x)$. Therefore:
\begin{equation}
\label{eqn:contpigauss-3b}
	\Pi(x) = \frac{1}{\sigma\sqrt{2\pi}} 
                 e^{-\half\left(\frac{x-\mu}{\sigma}\right)^2}  \;,
\end{equation}
where: 
\begin{align}
		\mu = n_t\overline p_t &&
		\sigma^2 = \sum_i \ind z\ind p (1-\ind p)
	\; .
\end{align}
In the continuous approximation we have:
\begin{equation}
\label{eqn:contapp-3b}
\begin{split}
	P[A_t=-h_t] &\approx 
	\begin{cases}
	  \int\limits_{\makebox[18mm]{\scriptsize $(1-l)n_t-\half$}}^{+\infty}
	     \hspace{-6mm}\Pi_t(x)\,dx & \f h_t=-1 \\
	  \int\limits_{\makebox[18mm]{\scriptsize $l n_t+\half$}}^{+\infty} 
	     \hspace{-6mm}\Pi_t(x)\,dx & \f h_t=+1 
	\end{cases}
	\\
	P[A_t=+h_t] &\approx 
	\begin{cases}
	  \int\limits_{\makebox[18mm]{\scriptsize $-\infty$}}^{(1-l)n_t-\half}
	     \hspace{-6mm}\Pi_t(x)\,dx & \f h_t=-1 \\
	  \int\limits_{\makebox[18mm]{\scriptsize $-\infty$}}^{l n_t+\half} 
	     \hspace{-6mm}\Pi_t(x)\,dx & \f h_t=+1 
	\end{cases} \; ,
\end{split}
\end{equation}
where we have used the fact that since $\sigma$ is of order unity the integrands
in Eq. \eqref{eqn:contapp-3b} will approximately vanish for $x<0$ and $x>N$.
This has allowed us to replace lower limits of $0$ with $-\infty$ and upper
limits of $N$ with $+\infty$.

Finally we can express Eq. \eqref{eqn:contapp-3b} in terms of erf functions as,
for $h_t=\pm1$:
\begin{equation}
\label{eqn:reffolprobs-3b}
\begin{split}
	P[A_t=-h_t] &\approx
	  \half\left[1\mp\erf\left(\frac{\half\mp n_t\lpm}{\sigma\sqrt{2}}
	     \right)\right]  \\
	P[A_t=+h_t] &\approx 
	  \half\left[1\pm\erf\left(\frac{\half\mp n_t\lpm}{\sigma\sqrt{2}}
	     \right)\right]  \; ,
\end{split}
\end{equation}
where for convenience we have defined \lpm\ as follows:
\begin{align}
\label{eqn:ldefs-3b}
	\lm = \pbar - (1-l) && \lp = \pbar -l \;\;.
\end{align}
The erf function is defined as:
\begin{equation}
\label{eqn:erf-3b}
	\erf(x)=\frac{2}{\sqrt{\pi}} \int\limits_0^x e^{-t^2} dt \;\; .
\end{equation}
	
We can see from Eq. \eqref{eqn:reffolprobs-3b} that the expressions for the
probability that $A_t=\pm h_t$ depend upon the the values taken by $h_t$. For
this reason it will not be possible in general to derive a simple expression for
the probability that $A_t=\pm h_t$ for $T$ consecutive time steps. Such an
expression would depend upon the realization of the prediction time series
during the specific $T$ time steps under consideration. In the next section,
however, we shall see that a simple expression can be derived in the case of the
memoryless model.

Ref. \cite{nasdaq} presented an analysis of crashes in
financial markets such as the one that occurred on the NASDAQ in April 2000. The
authors propose that such crashes result from speculative bubbles in which large
numbers of traders share the same unrealistic expectations of the future
performance of the companies in question. These bubbles eventually burst,
apparently in response to some event which acts as a trigger. We can draw a 
broad qualitative analogy between this and the GCGM. The extremised
\pval\ distribution discussed in Sec. \ref{sec:suscep-3b} corresponds to a
state of the model in which large numbers of agents share the same unrealistic
expectation that the global action will be equal to or the opposite of the
prediction. It is this \emph{speculative} distribution that is most susceptible
to triggers that occur from time to time in the global action time series. 

\subsection{ELC in the memoryless GCGM}

We form the memoryless GCGM from the full model described in Sec.
\ref{sec:gcgm-3b} by taking $h_t=+1\;\forall\;t$ in exactly the same way as we
did in Sec. \ref{ch:gm-memory}. Our discussion of the susceptibility of the gene
value distribution function $P(p)$ in Sec. \ref{sec:suscep-3b} applies equally
to the memoryless and the full GCGM. The equivalent of the patterns of $T$ time
steps in which $A_t=-h_t$ or $A_t=+h_t$ are those in which 
$A_t=-1$ or $A_t=+1$. In the memoryless case, the expressions in Eq.
\eqref{eqn:reffolprobs-3b} become:
\begin{equation}
\label{eqn:reffolprobs_ml-3b}
	P[A_t=\pm1] \approx
	  \half\left[1\pm\erf\left(\frac{\half-
	  n_t\lp}{\sigma\sqrt{2}}\right)\right] 
\;\; .
\end{equation}
This expression is no longer dependent on $h_t$, since $h_t=+1\;\forall\;t$.
Therefore, in the case of the memoryless GCGM, we can derive the following
simple expression for the probability $\Lambda_t$ that $A_t=\pm 1$ for $T$
consecutive time steps:
\begin{multline}
\label{eqn:Lambda-3b}
	\Lambda_t = \frac{1}{2^T} \left[1-\erf\left(
	  \frac{\half-n_t\lp}{\sigma\sqrt{2}}
	  \right)\right]^T
	  + \\
	  \frac{1}{2^T} \left[1+\erf\left(
	  \frac{\half-n_t\lp}{\sigma\sqrt{2}}
	  \right)\right]^T
	\;\; .
\end{multline}
Figure \ref{fig:lambda-3b}a shows $\Lambda_t$ given by Eq. \eqref{eqn:Lambda-3b}
as a function of \lp. However, it is easier to interpret a plot of the 
average waiting time, given by $\Lambda_t^{-1}$. A plot of this is
included in Fig. \ref{fig:lambda-3b}b.

There are a couple of points to notice in Fig. \ref{fig:lambda-3b}. First of
all the minimum of $\Lambda_t$ does not occur at $\lp=0$ but at a value of
$\lp=\frac{1}{2n_t}$. This results from the fact that in the case of 
$\at=ln_t$ our model tie breaks by declaring the global action $A_t=+1$.
The most obvious feature of Fig. \ref{fig:lambda-3b} is
that $\Lambda_t$ increases rapidly with increasing \lp. This means that the
probability of a trigger sequence occurring in the global action time
series $A_t$ increases with the deviation of \pbar\ from $l$. Therefore we
can consider the value of \lp\ to be controlling the probability that a
trigger sequence will occur. Furthermore, note that $\Lambda_t$ never
vanishes and so the average waiting time never goes to infinity. Crucially,
this means that regardless of the value of \pbar\ there is always a non-zero
probability of that a trigger sequence will occur.

\floatfig[\floatplace]{fig:lambda-3b}{
The probability and average waiting time for a \emph{trigger} pattern to occur
in the global action time series.
}{
a) The probability $\Lambda_t$ (defined by Eq. \eqref{eqn:Lambda-3b}) that
$A_t=\pm 1$ for $T$ consecutive time steps as a function of \lp\ defined by
Eq. \eqref{eqn:ldefs-3b}.
b) The average waiting time $\Lambda_t^{-1}$ for such a pattern in the global
action time series $A_t$ to occur. We have taken $n_t=205$ and $\sigma_t=4.9$
which correspond to their mean values over the period described by Figs.
\ref{fig:exampleone-3b} and \ref{fig:exampleone_graphs-3b}.
}{
\begin{center}
\resizebox{0.475\textwidth}{!}{\small 
\begingroup%
  \makeatletter%
  \newcommand{\GNUPLOTspecial}{%
    \catcode`\%=14\relax\special}%
  \setlength{\unitlength}{0.1bp}%
\begin{picture}(3600,1728)(0,0)%
\includegraphics{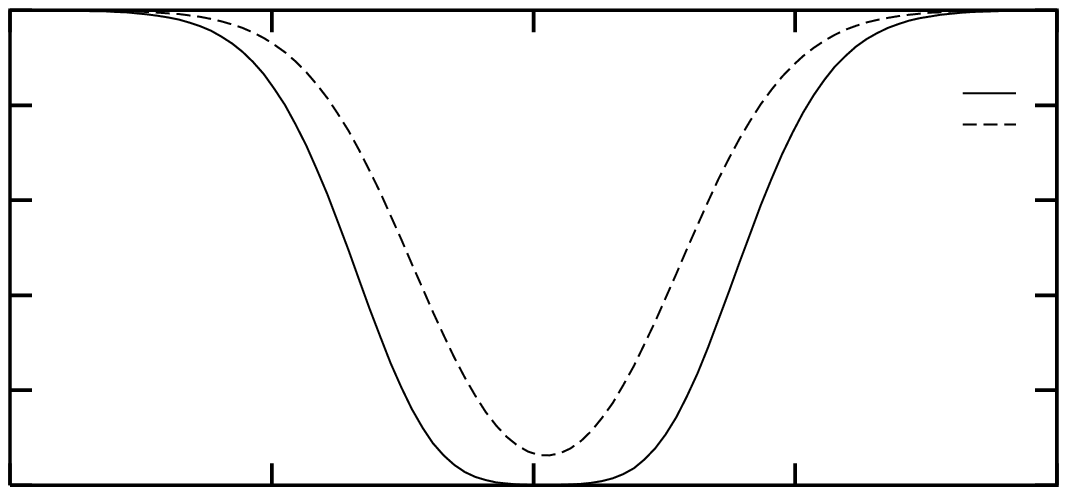}
\put(2744,1309){\makebox(0,0)[l]{$T=5$}}%
\put(2744,1399){\makebox(0,0)[l]{$T=12$}}%
\put(511,1501){\makebox(0,0)[l]{a)}}%
\put(1867,45){\makebox(0,0){$\lp$}}%
\put(90,954){%
\makebox(0,0)[b]{\shortstack{$\Lambda_t$}}%
}%
\put(3375,180){\makebox(0,0){0.1}}%
\put(2621,180){\makebox(0,0){0.05}}%
\put(1868,180){\makebox(0,0){0}}%
\put(1114,180){\makebox(0,0){-0.05}}%
\put(360,180){\makebox(0,0){-0.1}}%
\put(315,1638){\makebox(0,0)[r]{1}}%
\put(315,1364){\makebox(0,0)[r]{0.8}}%
\put(315,1091){\makebox(0,0)[r]{0.6}}%
\put(315,817){\makebox(0,0)[r]{0.4}}%
\put(315,544){\makebox(0,0)[r]{0.2}}%
\put(315,270){\makebox(0,0)[r]{0}}%
\end{picture}%
\endgroup

}
\resizebox{0.475\textwidth}{!}{\small 
\begingroup%
  \makeatletter%
  \newcommand{\GNUPLOTspecial}{%
    \catcode`\%=14\relax\special}%
  \setlength{\unitlength}{0.1bp}%
\begin{picture}(3600,1728)(0,0)%
\includegraphics{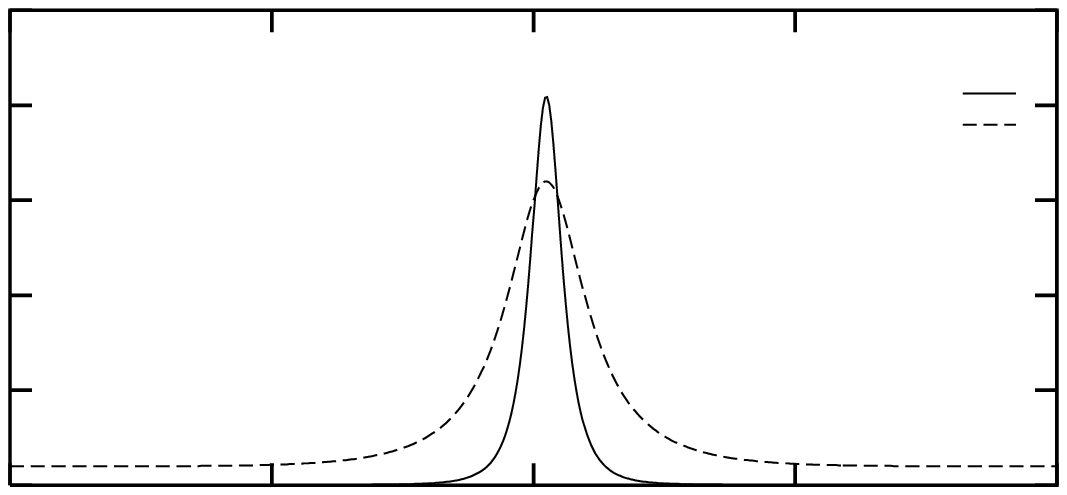}
\put(2744,1308){\makebox(0,0)[l]{$T=5$}}%
\put(2744,1398){\makebox(0,0)[l]{$T=12$}}%
\put(511,1501){\makebox(0,0)[l]{b)}}%
\put(1867,45){\makebox(0,0){$\lp$}}%
\put(3689,954){%
\makebox(0,0)[b]{\shortstack{$\Lambda_t^{-1}$ for $T=5$}}%
}%
\put(45,954){%
\makebox(0,0)[b]{\shortstack{$\Lambda_t^{-1}$ for $T=12$}}%
}%
\put(3420,1638){\makebox(0,0)[l]{25}}%
\put(3420,1364){\makebox(0,0)[l]{20}}%
\put(3420,1091){\makebox(0,0)[l]{15}}%
\put(3420,817){\makebox(0,0)[l]{10}}%
\put(3420,544){\makebox(0,0)[l]{5}}%
\put(3420,270){\makebox(0,0)[l]{0}}%
\put(3375,180){\makebox(0,0){0.1}}%
\put(2621,180){\makebox(0,0){0.05}}%
\put(1868,180){\makebox(0,0){0}}%
\put(1114,180){\makebox(0,0){-0.05}}%
\put(360,180){\makebox(0,0){-0.1}}%
\put(315,1638){\makebox(0,0)[r]{2500}}%
\put(315,1364){\makebox(0,0)[r]{2000}}%
\put(315,1091){\makebox(0,0)[r]{1500}}%
\put(315,817){\makebox(0,0)[r]{1000}}%
\put(315,544){\makebox(0,0)[r]{500}}%
\put(315,270){\makebox(0,0)[r]{0}}%
\end{picture}%
\endgroup

}
\end{center}
}

\subsubsection{Price Time-Series in the memoryless GCGM}
\label{sec:priceml-3b}

One of the advantages of the memoryless GCGM is that the dynamics of the price
time series are particularly simple. We saw in Sec. \ref{sec:pricets-3b}
that the probability of a price fall at time $t$ is given by the same condition
that the global action $A_t=-1$. Therefore, $P[\Delta\pi<0]=P[A_t=-1]$ given
by Eq. \eqref{eqn:reffolprobs_ml-3b}. Thus the price will fall with probability
$P<0.5$ if $\lp<0$ and with probability $P>0.5$
if $\lp>0$.

\subsubsection{Example ELC}
\label{sec:exampleone-3b}
\newcommand{\q}{\equiv}

Now that we have introduced the memoryless GCGM we shall consider a specific
example of an ELC. This will allow us to see how the elements discussed in
Sec. \ref{sec:overview-3b} are involved in ELC in the GCGM. In order to do this,
however, we need to introduce one further quantity which we shall call the
\emph{prediction performance} and denote by $\eta_t$. At time step $t$ the
value of $\eta_t$ gives the number of previous consecutive time steps at which
the prediction was the same as the global action $A_t=+h_t$. In the memoryless
GCGM the value of $\eta_t$ gives the number of consecutive time steps preceding
$t$ at which $A_t=+1$. The reason that $\eta_t$ is useful is that during one of
the so called trigger sequences that we discussed in Sec.
\ref{sec:triggers-3b} $\eta_t$ will become large and, thus, it can be used to
identify these events.

Figure \ref{fig:exampleone-3b} lists the values of the global action $A_t$ and
the prediction performance $\eta_t$ for a period of time in which an ELC occurs
in the memoryless GCGM with $l=0.4$ and $T=12$. In order to show the behavior
of the \one and \zero agents that we discussed in Sec. \ref{sec:suscep-3b} we
have also included the virtual points $\ind v$ of an agent $i$ which is inactive
and has $\ind p=1.0$ at the beginning of the time period shown. 
Since the actions of \zero and
\one agents are anti-correlated (as we saw in Sec. \ref{sec:suscep-3b}) the
virtual points of inactive \one agents and active \zero agents will always be
the same. The same applies to the virtual points of active \one and inactive
\zero agents. For this reason it is only necessary to give $\ind v$ for one of
these four groups in Fig. \ref{sec:suscep-3b} since from this we can infer the
virtual points of the others.
Figure \ref{fig:exampleone_graphs-3b}a shows $\eta_t$ graphically for the same
time period as well as the probability $P[A_t=-1]$ that the global action
$A_t=-1$ at each time step, given in terms of \lp\ by Eq.
\eqref{eqn:reffolprobs_ml-3b}. Figure \ref{fig:exampleone_graphs-3b}b shows the
deviation of \pbar\ from $l$, $\lp$, over the same time period. The quantities
depicted in Figs.  \ref{fig:exampleone-3b} and \ref{fig:exampleone_graphs-3b}
are those that play an important role in the mechanism that causes ELC. Later,
in Fig. \ref{fig:exampleone_obs-3b}, we shall demonstrate what effect the ELC
described here has on external \emph{observables} such as the price and the
volume. In the following paragraphs we describe the significant features in
Figs. \ref{fig:exampleone-3b} and \ref{fig:exampleone_graphs-3b}. The paragraph
labels {\bf a}-{\bf e} correspond to the identically labeled time intervals in
Fig. \ref{fig:exampleone-3b}.

\paragraph*{\bf a} This sequence of $T$ time steps in which $A_t=+1$ provides
the trigger sequence discussed in Sec. \ref{sec:triggers-3b}. At each
time step during this period the virtual points $\ind v$ of the inactive \one
agents and active \zero agents increases. When the model reaches the final time
step in this period $\ind v=T=12$ for both inactive \one agents and active \zero
agents. The inactive \one agents will then activate while the active \zero
agents will deactivate. The virtual points of these agents will then be reset to
$0$. Unless the numbers of \one agents activating and \zero agents
deactivating are approximately equal, this correlated behavior will lead to a
step change in the volume like the one depicted in Fig. \ref{fig:crexp2-3b}. 
Only active agents contribute to \pbar\ (see Eq. \eqref{eqn:pbar-3b}) and so
this instantaneous loss of \zero agents and gain of \one agents causes \pbar,
and therefore \lp, to undergo a step increase. We can see this clearly in Fig.
\ref{fig:exampleone_graphs-3b}.

\paragraph*{\bf b}
Throughout the period of $T$ time steps labeled {\bf b} in Fig.
\ref{fig:exampleone-3b} $\lp\approx 0.16$.  Equation
\eqref{eqn:reffolprobs_ml-3b} gives $P[A_t=-1]$ in terms of \lp\ and this is
depicted graphically in Fig. \ref{fig:erf-3b}, where we have taken values for
$n_t$ and $\sigma_t$ corresponding to their mean values over the period
described by Figs. \ref{fig:exampleone-3b} and \ref{fig:exampleone_graphs-3b}.
We can see from Fig. \ref{fig:erf-3b} that for $\lp<-0.05$ and $\lp>0.05$
$P[A_t=-1]\approx 0$ and $1$ respectively. Therefore, if the magnitude of \lp\
exceeds $0.05$ the model becomes quasi-deterministic at time step $t$. Thus,
for the period labeled {\bf b}, $A_t=-1$. The effect of this on the \zero and
\one agents is exactly the opposite of that of period {\bf a}; the virtual
points of the inactive \zero agents and the active \one agents now increases at
each time step. Once again $\ind v=T=12$ for both these populations at the end
of period {\bf b} and so the inactive \zero agents activate while the active
\one agents deactivate. Note that the inactive \zero agents activating at the
end of period {\bf b} are not just those that deactivated at the end of period
{\bf a}. Any \zero agents that were inactive at the start of period {\bf a}
would have been unaffected by the trigger sequence, however they now
activate along with those that previously deactivated. The result of this is
that $\lp$ does not return to the value of $\lp\approx 0.4$ that it had at the
end of period {\bf a}. That would have returned the model to its
non-deterministic state. Instead, however, $\lp\approx -0.06$ and so at the
start of period {\bf c} $P[A_t=-1]=0$.

\paragraph*{\bf c} 
This period of $T$ time steps in which $A_t=+1$ is identical to period {\bf a}
in terms of its effect on the \one and \zero agents. Thus, at the end of period
{\bf c} the inactive \one agents reactivate and the active \zero deactivate
resulting in the same step increase in \lp\ that we observed before. The
important difference between periods {\bf a} and {\bf c} is that the first
occurred stochastically (there was a significant non-zero probability that
$A_t=\pm1$ at each time step) whereas period {\bf c} occurs
quasi-deterministically (the probability that $A_t=+1$ for the $T$ time steps in
period {\bf c} is $P\approx1$).

\paragraph*{\bf d}
Similarly period {\bf d} is identical to period {\bf b} except that at the end 
$\lp=-0.03$. For Fig. \ref{fig:erf-3b} we can see that this gives $P[A_t=-1]>0$.

\paragraph*{\bf e}
$P[A_t=-1]$ is no longer $\approx 0$ and so the model returns to its usual
stochastic behavior, $A_t$ taking values $-1$ and $+1$ probabilistically.
This represents the end of the ELC since there is now no mechanism for the
synchronized activation and deactivation that occurs during periods {\bf a-e}.
The model now returns slowly to the \emph{equilibrium} state. In other words the
state that it is in once transients due to any ELC have died away.

\bigfloatfig[\floatplace]{fig:exampleone-3b}{
}{
The prediction performance $\eta_t$, $A_t$ and the virtual
points $\ind v$ of an agent $i$ which initially is inactive and has a gene
values of $\ind p=1.0$. The model is the \emph{memoryless} GCGM with $n=500$,
$l=0.4$, $T=12$, $r=0.2$ and $m=3$.
}{
\begin{center}
\resizebox{0.80\textwidth}{!}{\small 

}
\end{center}
}
\floatfig[\floatplace]{fig:exampleone_graphs-3b}{
}{
Various important quantities plotted over the time period corresponding to Fig.
\ref{fig:exampleone-3b}:
a) The prediction performance $\eta_t$ (measured against the left hand axis) and
the probability $P[A_t=-1]$ that the global action is $-1$ (measured against the
right hand axis). b) The deviation of \pbar\ from $l$, \lp.
}{
  \resizebox{.47\textwidth}{!}{\small 
\begingroup%
  \makeatletter%
  \newcommand{\GNUPLOTspecial}{%
    \catcode`\%=14\relax\special}%
  \setlength{\unitlength}{0.1bp}%
\begin{picture}(3600,1296)(0,0)%
\includegraphics{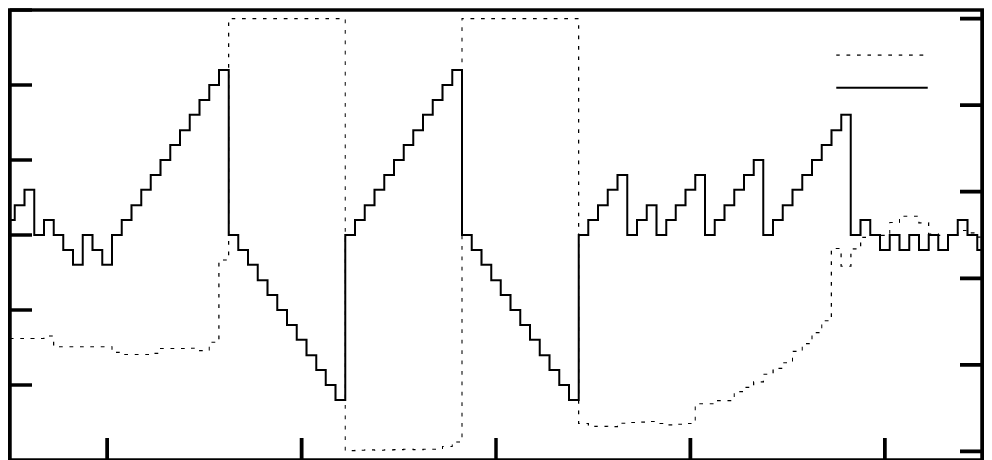}
\put(2730,1072){\makebox(0,0)[r]{$\eta_t$}}%
\put(2730,1166){\makebox(0,0)[r]{$P[A_t=-1]$}}%
\put(484,1166){\makebox(0,0)[l]{a)}}%
\put(3599,648){%
\makebox(0,0)[b]{\shortstack{$P[A_t=-1]$}}%
}%
\put(100,648){%
\makebox(0,0)[b]{\shortstack{$\eta_t$}}%
}%
\put(3250,1271){\makebox(0,0)[l]{1}}%
\put(3250,1022){\makebox(0,0)[l]{0.8}}%
\put(3250,773){\makebox(0,0)[l]{0.6}}%
\put(3250,523){\makebox(0,0)[l]{0.4}}%
\put(3250,274){\makebox(0,0)[l]{0.2}}%
\put(3250,25){\makebox(0,0)[l]{0}}%
\put(350,1296){\makebox(0,0)[r]{15}}%
\put(350,1080){\makebox(0,0)[r]{10}}%
\put(350,864){\makebox(0,0)[r]{5}}%
\put(350,648){\makebox(0,0)[r]{0}}%
\put(350,432){\makebox(0,0)[r]{-5}}%
\put(350,216){\makebox(0,0)[r]{-10}}%
\end{picture}%
\endgroup

  }\\%
  \vspace{-1pt}%
  \resizebox{.47\textwidth}{!}{\small 
\begingroup%
  \makeatletter%
  \newcommand{\GNUPLOTspecial}{%
    \catcode`\%=14\relax\special}%
  \setlength{\unitlength}{0.1bp}%
\begin{picture}(3600,1080)(0,0)%
\includegraphics{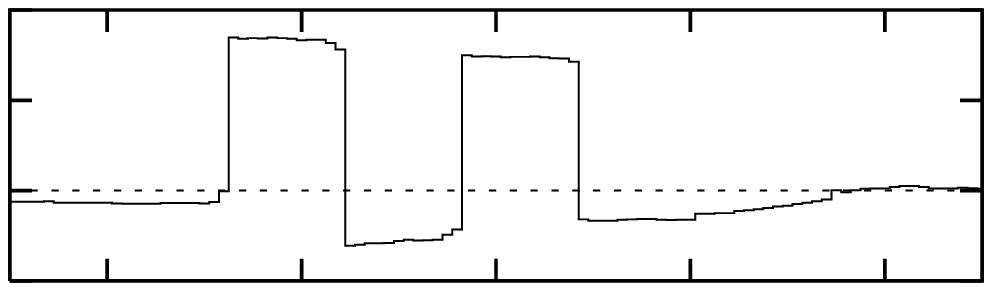}
\put(484,1002){\makebox(0,0)[l]{b)}}%
\put(1800,50){\makebox(0,0){time step: $t$}}%
\put(100,690){%
\makebox(0,0)[b]{\shortstack{\lp}}%
}%
\put(2920,200){\makebox(0,0){684140}}%
\put(2360,200){\makebox(0,0){684120}}%
\put(1800,200){\makebox(0,0){684100}}%
\put(1240,200){\makebox(0,0){684080}}%
\put(680,200){\makebox(0,0){684060}}%
\put(350,1080){\makebox(0,0)[r]{0.20}}%
\put(350,820){\makebox(0,0)[r]{0.10}}%
\put(350,560){\makebox(0,0)[r]{0.00}}%
\put(350,300){\makebox(0,0)[r]{-0.10}}%
\end{picture}%
\endgroup

  }%
}

\floatfig[\floatplace]{fig:erf-3b}{
Probability  that the global action is $-1$ in terms of \lp\ given
by Eq. \eqref{eqn:reffolprobs_ml-3b}.
}{
Probability $P[A_t=-1]$ that the global action $A_t=-1$ in terms of \lp\ given
by Eq. \eqref{eqn:reffolprobs_ml-3b} with $n_t=205$ and $\sigma=4.9$. These
values correspond to the mean values of these quantities over the time period
depicted in Figs. \ref{fig:exampleone-3b} and \ref{fig:exampleone_graphs-3b}.
}{
\begin{center}
\resizebox{0.40\textwidth}{!}{\small 
\begingroup%
  \makeatletter%
  \newcommand{\GNUPLOTspecial}{%
    \catcode`\%=14\relax\special}%
  \setlength{\unitlength}{0.1bp}%
\begin{picture}(3600,1728)(0,0)%
\includegraphics{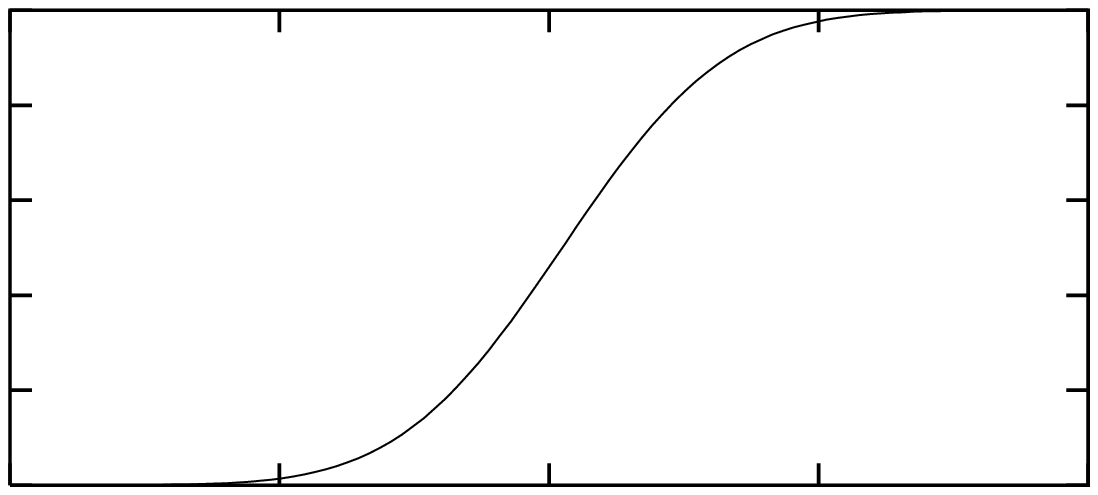}
\put(1912,45){\makebox(0,0){$\lp$}}%
\put(90,954){%
\makebox(0,0)[b]{\shortstack{$P[A_t=-1]$}}%
}%
\put(3465,180){\makebox(0,0){0.1}}%
\put(2689,180){\makebox(0,0){0.05}}%
\put(1913,180){\makebox(0,0){0}}%
\put(1136,180){\makebox(0,0){-0.05}}%
\put(360,180){\makebox(0,0){-0.1}}%
\put(315,1638){\makebox(0,0)[r]{1}}%
\put(315,1364){\makebox(0,0)[r]{0.8}}%
\put(315,1091){\makebox(0,0)[r]{0.6}}%
\put(315,817){\makebox(0,0)[r]{0.4}}%
\put(315,544){\makebox(0,0)[r]{0.2}}%
\put(315,270){\makebox(0,0)[r]{0}}%
\end{picture}%
\endgroup

}
\end{center}
}

\bigskip
From the analysis that we have presented above we might have expected that the
periodic synchronized activations and deactivations that we described above would
continue indefinitely and that the model would never return to the stochastic
state. One question that we did not address above, however, is that of how the
model manages to break out of the deterministic behavior that it exhibits
during the ELC. We have considered the effect of a period of $T$ time steps in
which $A_t=\pm 1$ on the \zero and \one agents in terms of agent activation and
deactivation. However, we have not considered agent mutation. We shall see in
what follows that it is agent mutation that allows the model to return to the
stochastic state. During periods {\bf b-d} the \one agents are only active at
time steps at which $A_t=-1$ and the \zero agents are only active when $A_t=+1$.
Because of this the scores of \zero and \one agents are decreasing functions
of time. Their scores are fixed when they are inactive and when they are active
their individual actions are the inverse of the global action: $\ind a=-A_t$.
While these agents are inactive the model is favorable to them and so after $T$
time steps they reactive. However, because the behavior of all these agents is
so highly correlated, in doing so they change the dynamic of the model so that
it is no longer favorable. This has a clear analogy with the phenomenon of
\emph{market impact} in economic systems.

Since the scores of the \zero and \one agents are decreasing functions of time
during the ELC the scores of these agents will rapidly reach the death score
$\ind s=-D$ at which they mutate. If $r\gg \delta=0.02$ (defined by Eq. 
\ref{eqn:d-3b}) then with a very high probability of $\frac{r-\delta}{r}$ a
mutating \zero or \one agent will mutate to a gene value of $\ind p>\delta$ 
or $\ind p<1-\delta$ respectively. The result of this is that the population of
\zero and \one agents that participate in the synchronized activations and
deactivations steadily decreases throughout the ELC. Eventually there are no
longer enough of these agents to maintain \lp\ at a magnitude greater than
$0.05$ and so $P[A_t=-1]\ne0$ or $1$. The model then returns to the stochastic
state.

\floatfig[\floatplace]{fig:exampleone_highlow-3b}{
Evolution of the numbers of \zero and \one agents.
}{
a) The evolution of the numbers of agents for which $\ind p<\delta$ (\zero
agents) and $\ind p>1-\delta$ (\one agents) over a time period that includes
that depicted in Figs. \ref{fig:exampleone-3b} and
\ref{fig:exampleone_graphs-3b}.
b) \lp\ over the same time period. The occurrence of ELC is indicated by spikes
in the \lp\ time series. The first ELC identifiable is that depicted in 
Figs. \ref{fig:exampleone-3b} and \ref{fig:exampleone_graphs-3b}.
}{
\resizebox{.47\textwidth}{!}{\small%
\begingroup%
  \makeatletter%
  \newcommand{\GNUPLOTspecial}{%
    \catcode`\%=14\relax\special}%
  \setlength{\unitlength}{0.1bp}%
\begin{picture}(3600,864)(0,0)%
\includegraphics{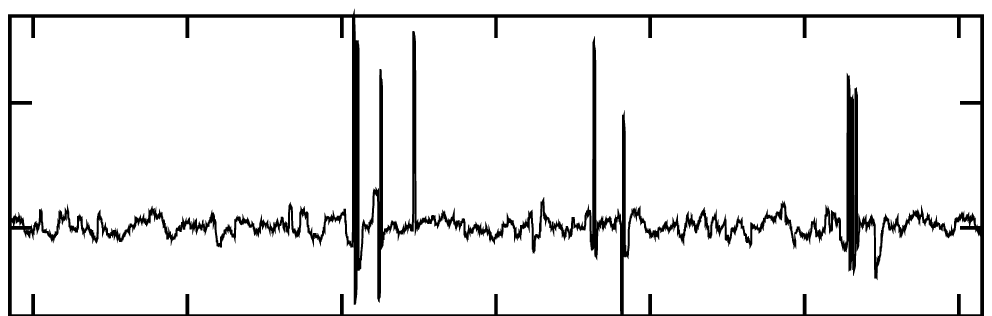}
\put(484,778){\makebox(0,0)[l]{a)}}%
\put(150,432){%
\makebox(0,0)[b]{\shortstack{\lp}}%
}%
\put(350,614){\makebox(0,0)[r]{0.10}}%
\put(350,254){\makebox(0,0)[r]{0.00}}%
\end{picture}%
\endgroup

 }\\
\vspace{-1pt}%
\resizebox{.47\textwidth}{!}{\small%
\begingroup%
  \makeatletter%
  \newcommand{\GNUPLOTspecial}{%
    \catcode`\%=14\relax\special}%
  \setlength{\unitlength}{0.1bp}%
\begin{picture}(3600,1511)(0,0)%
\includegraphics{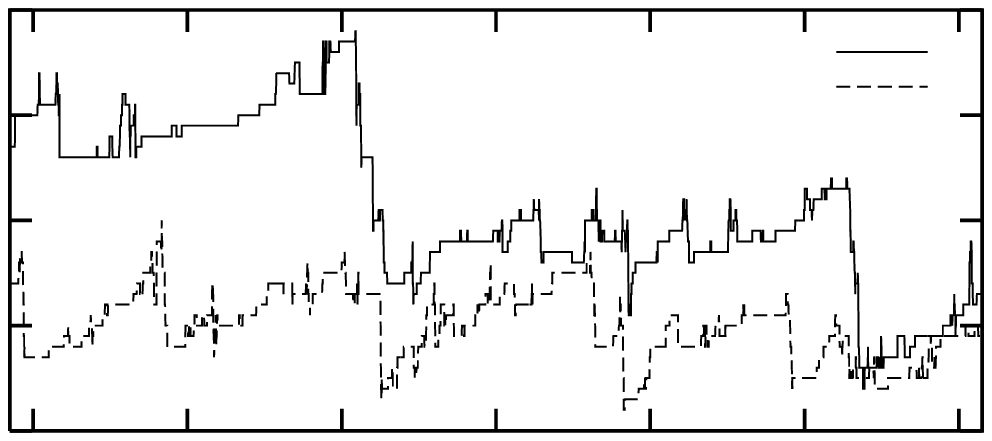}
\put(2730,1291){\makebox(0,0)[r]{\one agents}}%
\put(2730,1391){\makebox(0,0)[r]{\zero agents}}%
\put(484,1391){\makebox(0,0)[l]{b)}}%
\put(1800,50){\makebox(0,0){time step: t}}%
\put(150,906){%
\makebox(0,0)[b]{\shortstack{Number of agents}}%
}%
\put(3133,200){\makebox(0,0){688000}}%
\put(2689,200){\makebox(0,0){687000}}%
\put(2244,200){\makebox(0,0){686000}}%
\put(1800,200){\makebox(0,0){685000}}%
\put(1356,200){\makebox(0,0){684000}}%
\put(911,200){\makebox(0,0){683000}}%
\put(467,200){\makebox(0,0){682000}}%
\put(350,1512){\makebox(0,0)[r]{40}}%
\put(350,1209){\makebox(0,0)[r]{30}}%
\put(350,906){\makebox(0,0)[r]{20}}%
\put(350,603){\makebox(0,0)[r]{10}}%
\put(350,300){\makebox(0,0)[r]{0}}%
\end{picture}%
\endgroup

  }%

}

\subsubsection{Summary of ELC in the memoryless GCGM}

In this section we shall bring together the elements that we have introduced so
far in order to give a broad overview of ELC in the memoryless model. We
demonstrated in Sec. \ref{sec:suscep-3b} that in, what we now call, the
stochastic state the agents migrate towards gene values $p=0.0$ and $p=1.0$.
Therefore, while the model is in the stochastic state the number of \zero and
\one agents increases. This increases the susceptibility of the gene value
distribution $P(p)$ to any trigger sequences that might occur and also
increases the duration of the next ELC. If a trigger sequence occurs in the
evolution of the model then if there are enough \zero and \one agents an ELC
will take place as described in the previous section. One of the effects of the
ELC is to reduce the numbers of \zero and \one agents. This decreases the
probability of a subsequent ELC occurring.

In order to make this clear we show in Fig. \ref{fig:exampleone_highlow-3b}b
the numbers of \zero and \one agents during the time period leading up
to and after that depicted in Figs. \ref{fig:exampleone-3b} and
\ref{fig:exampleone_graphs-3b}. Figure \ref{fig:exampleone_highlow-3b}a shows
\lp\ over the same time period. As we saw in Sec. \ref{sec:exampleone-3b}, ELC
can be identified by the spikes that occur in the \lp\ time series. We can see
in Fig. \ref{fig:exampleone_highlow-3b} that, while the ELC that we examined
in Sec. \ref{sec:exampleone-3b} causes a large decrease in the numbers of \zero
and \one agents, significant numbers remain and so the initial ELC is followed
by several smaller ones. The most important feature of Fig.
\ref{fig:exampleone_highlow-3b} to notice, however, is that sudden decreases in
the numbers of \zero and \one agents that occur at each ELC and the steady
increase that these quantities exhibit in the periods between ELC.

So far, in order to understand the mechanism that leads to ELC, we have
concentrated on quantities which are internal to the model. In the economic
analogy these would correspond to  quantities whose values would be extremely
hard to quantify. For example the confidence of, or strategies adopted by, 
traders. However, as we remarked in Sec. \ref{sec:intro-3b}, ELC in the GCGM
also affect quantities that are directly observable and quantifiable such as the
\emph{volume} and the \emph{price}. We have plotted in Fig.
\ref{fig:exampleone_obs-3b} the volume and the price over the same time period
described by Figs. \ref{fig:exampleone-3b} and \ref{fig:exampleone_graphs-3b}.
We can see from Fig.  \ref{fig:exampleone_obs-3b} that the oscillatory 
activation and deactivation of the \zero and \one agents that we described 
in Sec. \ref{sec:exampleone-3b} leads to corresponding oscillations in the
volume. 

We saw in Sec. \ref{sec:priceml-3b} that the probability that the price falls at
time $t$ $P[\Delta\pi_t<0]$ is equal to $P[A_t=-1]$ which is in turn given in
terms of \lp\ by Eq. \eqref{eqn:reffolprobs_ml-3b}. Thus, we can see from 
the plot of $P[A_t=-1]$ in Fig. \ref{fig:exampleone_graphs-3b}a that the
oscillations in \lp\ will give rise to alternate periods in which the price
rises and falls as we see in Fig. \ref{fig:exampleone_obs-3b}a. In terms of \lp\
the ensemble average excess demand is given by:
\begin{equation}
	\ave{\Delta} = -\frac{n_t\lp}{1-l} \;.
\end{equation}
Thus we can see that the magnitude of the price changes at each time step is
proportional to $n_t|\lp|$. Therefore, the overall fall in price
depicted in Fig. \ref{fig:exampleone_obs-3b}a is due to the fact that the
magnitude of the positive excursions of \lp\ during the ELC exceeds that
of the negative excursions.

\floatfig[\floatplace]{fig:exampleone_obs-3b}{
Evolution of the volume and the price during the ELC.
}{
Evolution of the number of active agents (also know as the \emph{volume}) and
the price (defined by Eq. \eqref{eqn:price-3b}) over the same time period
described by Figs. \ref{fig:exampleone-3b} and \ref{fig:exampleone_graphs-3b}.
}{
\resizebox{.47\textwidth}{!}{\small%
\begingroup%
  \makeatletter%
  \newcommand{\GNUPLOTspecial}{%
    \catcode`\%=14\relax\special}%
  \setlength{\unitlength}{0.1bp}%
\begin{picture}(3600,1188)(0,0)%
\includegraphics{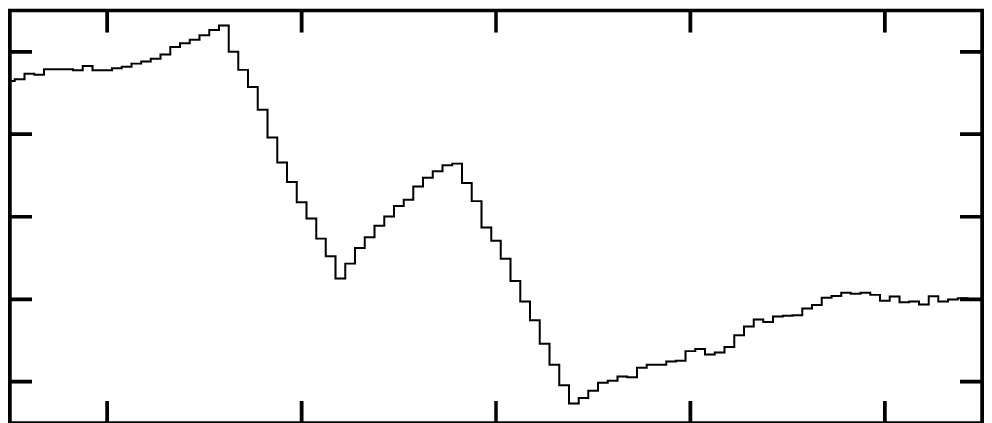}
\put(484,1069){\makebox(0,0)[l]{a)}}%
\put(50,594){%
\makebox(0,0)[b]{\shortstack{Price: $\pi_t$}}%
}%
\put(350,1069){\makebox(0,0)[r]{0}}%
\put(350,832){\makebox(0,0)[r]{-200}}%
\put(350,594){\makebox(0,0)[r]{-400}}%
\put(350,356){\makebox(0,0)[r]{-600}}%
\put(350,119){\makebox(0,0)[r]{-800}}%
\end{picture}%
\endgroup

  }\\%
\vspace{-1pt}%
\resizebox{.47\textwidth}{!}{\small%
\begingroup%
  \makeatletter%
  \newcommand{\GNUPLOTspecial}{%
    \catcode`\%=14\relax\special}%
  \setlength{\unitlength}{0.1bp}%
\begin{picture}(3600,1188)(0,0)%
\includegraphics{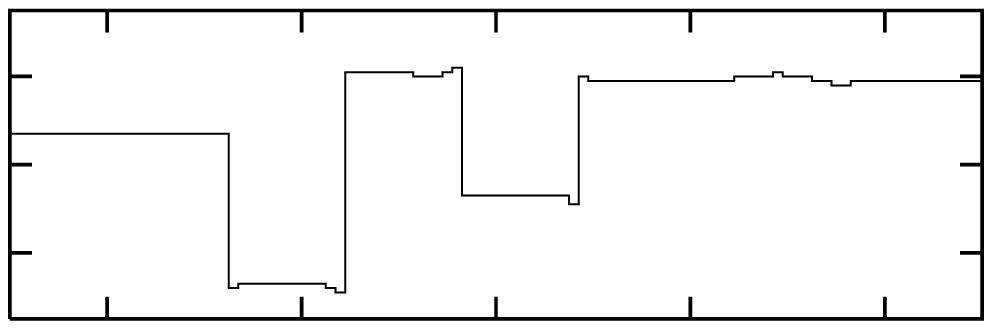}
\put(484,1099){\makebox(0,0)[l]{b)}}%
\put(1800,50){\makebox(0,0){time step: t}}%
\put(35,744){%
\makebox(0,0)[b]{\shortstack{No. active agents: $n_t$}}%
}%
\put(2920,200){\makebox(0,0){684140}}%
\put(2360,200){\makebox(0,0){684120}}%
\put(1800,200){\makebox(0,0){684100}}%
\put(1240,200){\makebox(0,0){684080}}%
\put(680,200){\makebox(0,0){684060}}%
\put(350,998){\makebox(0,0)[r]{220}}%
\put(350,744){\makebox(0,0)[r]{200}}%
\put(350,490){\makebox(0,0)[r]{180}}%
\end{picture}%
\endgroup

  }%
}

\subsubsection{Another example ELC}
\label{sec:exampletwo-3b}

In Figs. \ref{fig:exampletwo-3b} and \ref{fig:exampletwo_graphs-3b} we give
a second example of an ELC in the memoryless GCGM. We can see that the
behavior of the model depicted therein is much the same as that exhibited
in Figs. \ref{fig:exampleone-3b} and \ref{fig:exampleone_graphs-3b}.
The period labeled {\bf a} provides the trigger sequence of $T$ time steps
for which $A_t=+1$. As before \lp\ exhibits decaying step-like oscillations
during periods {\bf b,c} and {\bf e} until the model returns to dynamic
equilibrium at the end of {\bf e}. Note, however that the periods {\bf c} and
{\bf e}, for which $A_t=-1$ and $A_t=+1$ respectively, are separated by 
three time steps, {\bf d}, in which $A_t=-1,-1,+1$. Figure
\ref{fig:exampletwo-3b} demonstrates why this occurs. We can see from the figure
that during periods {\bf d} and {\bf e} $\lp=0.552$ and that this is not
sufficiently large to yield $P[A_t=-1]=1.0$. Because of this period {\bf e}, in
which $A_t=-1$, does not occur deterministically. However, since
$P[A_t=-1]=0.967$ it is more probable that $A_t=-1$ for $T$ time steps than when
the system is in dynamic equilibrium. From Fig. \ref{fig:exampletwo-3b} we see
that at the last time step of period {\bf d} $A_t=+1$ because $P[A_t=-1]\ne1.0$
however $T$ time steps then follow (period {\bf e}) in which $A_t=-1$. Thus the
model is fairly robust against stochastic fluctuations. It's not important if
there is a period like {\bf d} in which the oscillatory pattern is briefly
broken.  During this interjected period, however, agent mutation will act to
bring \lp\ closer to $0.0$. Therefore the longer the interjected period the
lower the probability that the ELC will continue. 

\bigfloatfig[\floatplace]{fig:exampletwo-3b}{
}{
The prediction performance $\eta_t$, $A_t$ and the virtual
points $\ind v$ of an agent $i$ which initially is inactive and has a gene
values of $\ind p=1.0$. The model is the \emph{memoryless} GCGM with $n=500$,
$l=0.5$, $T=12$, $r=0.2$ and $m=3$.
The symbols in the margins denote the positions of the step change of \lp\
in Fig. \ref{fig:exampletwo_graphs-3b}.
}{
\begin{center}
\resizebox{.80\textwidth}{!}{\small%

}
\end{center}
}
\floatfig[\floatplace]{fig:exampletwo_graphs-3b}{
}{
Various important quantities plotted over the time period corresponding to Fig.
\ref{fig:exampletwo-3b}:
a) The prediction performance $\eta_t$ (measured against the left hand axis) and
the probability $P[A_t=-1]$ that the global action is $-1$ (measured against the
right hand axis). b) The deviation of \pbar\ from $l$, \lp.
}{
    \resizebox{.47\textwidth}{!}{\small%
\begingroup%
  \makeatletter%
  \newcommand{\GNUPLOTspecial}{%
    \catcode`\%=14\relax\special}%
  \setlength{\unitlength}{0.1bp}%
\begin{picture}(3600,1296)(0,0)%
\includegraphics{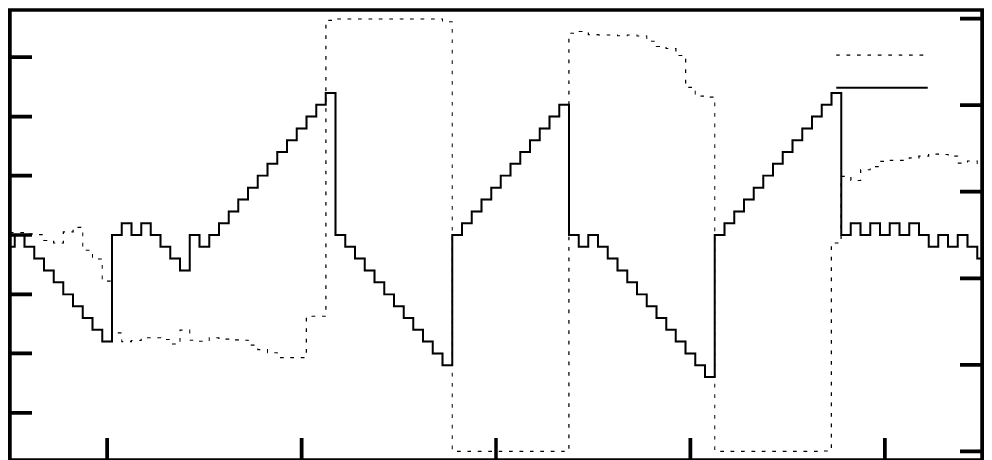}
\put(2730,1072){\makebox(0,0)[r]{$\eta_t$}}%
\put(2730,1166){\makebox(0,0)[r]{$P[A_t=-1]$}}%
\put(484,1166){\makebox(0,0)[l]{a)}}%
\put(3599,648){%
\makebox(0,0)[b]{\shortstack{$P[A_t=-1]$}}%
}%
\put(100,648){%
\makebox(0,0)[b]{\shortstack{$\eta_t$}}%
}%
\put(3250,1271){\makebox(0,0)[l]{1}}%
\put(3250,1022){\makebox(0,0)[l]{0.8}}%
\put(3250,773){\makebox(0,0)[l]{0.6}}%
\put(3250,523){\makebox(0,0)[l]{0.4}}%
\put(3250,274){\makebox(0,0)[l]{0.2}}%
\put(3250,25){\makebox(0,0)[l]{0}}%
\put(350,1160){\makebox(0,0)[r]{15}}%
\put(350,989){\makebox(0,0)[r]{10}}%
\put(350,819){\makebox(0,0)[r]{5}}%
\put(350,648){\makebox(0,0)[r]{0}}%
\put(350,477){\makebox(0,0)[r]{-5}}%
\put(350,307){\makebox(0,0)[r]{-10}}%
\put(350,136){\makebox(0,0)[r]{-15}}%
\end{picture}%
\endgroup

    }\\%
    \vspace{-1pt}%
    \resizebox{.47\textwidth}{!}{\small%
\begingroup%
  \makeatletter%
  \newcommand{\GNUPLOTspecial}{%
    \catcode`\%=14\relax\special}%
  \setlength{\unitlength}{0.1bp}%
\begin{picture}(3600,1080)(0,0)%
\includegraphics{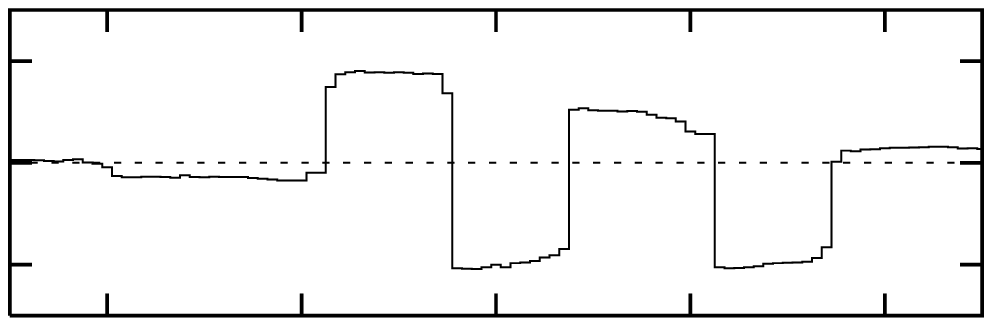}
\put(484,992){\makebox(0,0)[l]{b)}}%
\put(1800,-50){\makebox(0,0){time step: $t$}}%
\put(100,640){%
\makebox(0,0)[b]{\shortstack{\lp}}%
}%
\put(2920,100){\makebox(0,0){707200}}%
\put(2360,100){\makebox(0,0){707180}}%
\put(1800,100){\makebox(0,0){707160}}%
\put(1240,100){\makebox(0,0){707140}}%
\put(680,100){\makebox(0,0){707120}}%
\put(350,933){\makebox(0,0)[r]{0.10}}%
\put(350,640){\makebox(0,0)[r]{0.00}}%
\put(350,347){\makebox(0,0)[r]{-0.10}}%
\end{picture}%
\endgroup

    }%
}

\floatfig[\floatplace]{fig:exampletwo_highlow-3b}{
Evolution of the numbers of \zero and \one agents.
}{
a) The evolution of the numbers of agents for which $\ind p<\delta$ (\zero
agents) and $\ind p>1-\delta$ (\one agents) over a time period that includes
that depicted in Figs. \ref{fig:exampletwo-3b} and
\ref{fig:exampletwo_graphs-3b}.
b) \lp\ over the same time period. The occurrence of ELC is indicated by spikes
in the \lp\ time series. The first ELC identifiable is that depicted in 
Figs. \ref{fig:exampletwo-3b} and \ref{fig:exampletwo_graphs-3b}.
}{
    \resizebox{.47\textwidth}{!}{\small%
\begingroup%
  \makeatletter%
  \newcommand{\GNUPLOTspecial}{%
    \catcode`\%=14\relax\special}%
  \setlength{\unitlength}{0.1bp}%
\begin{picture}(3600,864)(0,0)%
\includegraphics{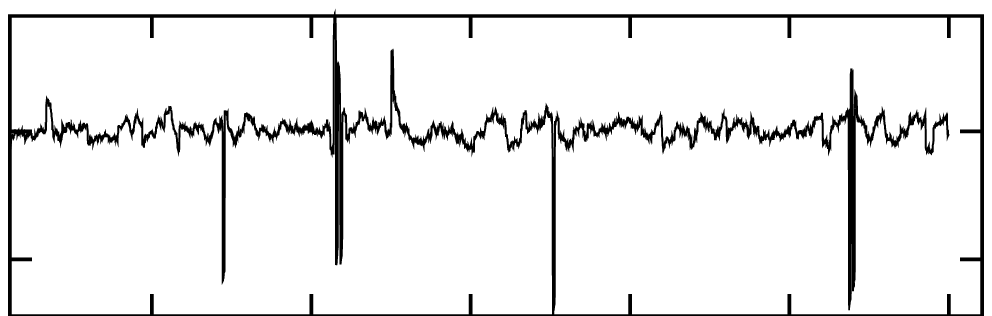}
\put(484,778){\makebox(0,0)[l]{a)}}%
\put(150,432){%
\makebox(0,0)[b]{\shortstack{\lp}}%
}%
\put(350,532){\makebox(0,0)[r]{0.10}}%
\put(350,163){\makebox(0,0)[r]{0.00}}%
\end{picture}%
\endgroup

}\\%
    \vspace{-1pt}%
    \resizebox{.47\textwidth}{!}{\small%
\begingroup%
  \makeatletter%
  \newcommand{\GNUPLOTspecial}{%
    \catcode`\%=14\relax\special}%
  \setlength{\unitlength}{0.1bp}%
\begin{picture}(3600,1511)(0,0)%
\includegraphics{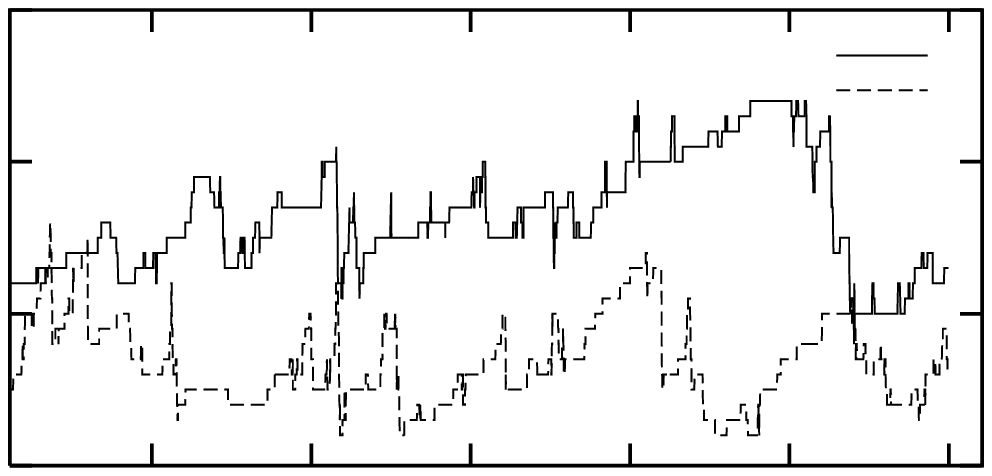}
\put(2730,1281){\makebox(0,0)[r]{\one agents}}%
\put(2730,1381){\makebox(0,0)[r]{\zero agents}}%
\put(484,1381){\makebox(0,0)[l]{b)}}%
\put(1800,-50){\makebox(0,0){time step: t}}%
\put(150,856){%
\makebox(0,0)[b]{\shortstack{Number of agents}}%
}%
\put(3104,100){\makebox(0,0){711000}}%
\put(2645,100){\makebox(0,0){710000}}%
\put(2186,100){\makebox(0,0){709000}}%
\put(1727,100){\makebox(0,0){708000}}%
\put(1268,100){\makebox(0,0){707000}}%
\put(809,100){\makebox(0,0){706000}}%
\put(350,1512){\makebox(0,0)[r]{30}}%
\put(350,1075){\makebox(0,0)[r]{20}}%
\put(350,637){\makebox(0,0)[r]{10}}%
\put(350,200){\makebox(0,0)[r]{0}}%
\end{picture}%
\endgroup

      }%
}

\floatfig[\floatplace]{fig:exampletwo_vol-3b}{
Evolution of the volume and the price during the ELC.
}{
Evolution of the number of active agents (also know as the \emph{volume}) and
the price (defined by Eq. \eqref{eqn:price-3b}) over the same time period
described by Figs. \ref{fig:exampletwo-3b} and \ref{fig:exampletwo_graphs-3b}.
}{
  \resizebox{.47\textwidth}{!}{\small%
\begingroup%
  \makeatletter%
  \newcommand{\GNUPLOTspecial}{%
    \catcode`\%=14\relax\special}%
  \setlength{\unitlength}{0.1bp}%
\begin{picture}(3600,1188)(0,0)%
\includegraphics{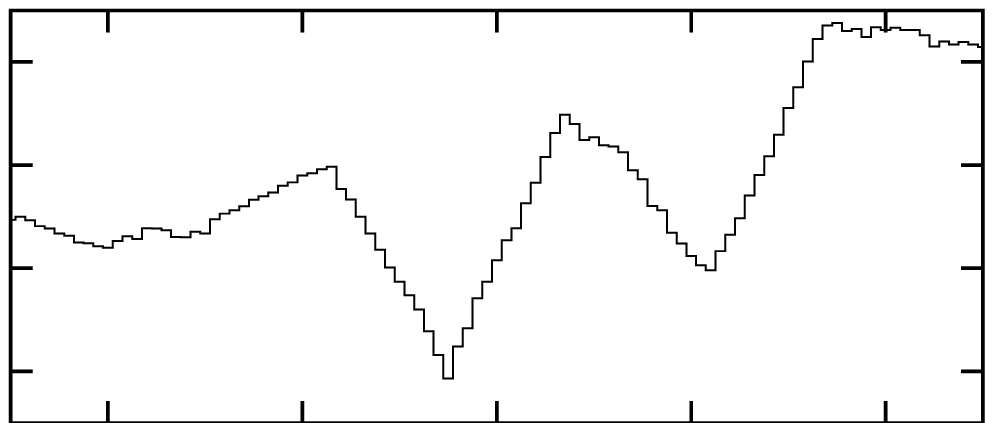}
\put(200,594){%
\makebox(0,0)[b]{\shortstack{Price: $\pi_t$}}%
}%
\put(600,1040){\makebox(0,0)[r]{-11100}}%
\put(600,743){\makebox(0,0)[r]{-11300}}%
\put(600,446){\makebox(0,0)[r]{-11500}}%
\put(600,149){\makebox(0,0)[r]{-11700}}%
\end{picture}%
\endgroup

}%
    \\%
  \vspace{-1pt}%
  \resizebox{.47\textwidth}{!}{\small%
\begingroup%
  \makeatletter%
  \newcommand{\GNUPLOTspecial}{%
    \catcode`\%=14\relax\special}%
  \setlength{\unitlength}{0.1bp}%
\begin{picture}(3600,1188)(0,0)%
\includegraphics{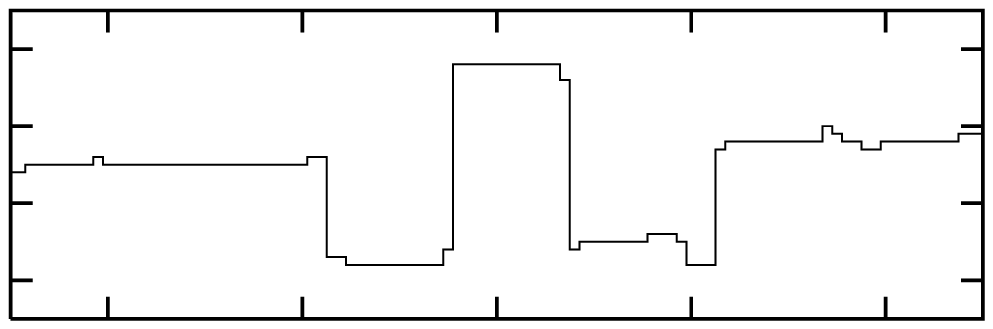}
\put(2050,50){\makebox(0,0){time step: t}}%
\put(186,744){%
\makebox(0,0)[b]{\shortstack{No. active agents: $n_t$}}%
}%
\put(3170,200){\makebox(0,0){707200}}%
\put(2610,200){\makebox(0,0){707180}}%
\put(2050,200){\makebox(0,0){707160}}%
\put(1490,200){\makebox(0,0){707140}}%
\put(930,200){\makebox(0,0){707120}}%
\put(600,1077){\makebox(0,0)[r]{220}}%
\put(600,855){\makebox(0,0)[r]{210}}%
\put(600,633){\makebox(0,0)[r]{200}}%
\put(600,411){\makebox(0,0)[r]{190}}%
\end{picture}%
\endgroup

}%
}

\subsection{ELC in the full GCGM: an idealized case}

Before we examine some examples of numerically observed ELC in the full GCGM
we shall consider a theoretically idealized case. 
In this section we will assume that in equilibrium $\ave{\pbar}_t$ is equal to
the optimal value $\opt[\pbar]$ given by Eq. \eqref{eqn:optp-3a} of $1-l$ for
$l<0.5$ and $l$ for $l>0.5$. During the ELC we will assume that \pbar\
oscillates between values that are greater and less than the equilibrium value
by a magnitude sufficiently large that $P[A_t=-h_t]$, given by Eq.
\eqref{eqn:reffolprobs-3b}, takes only the values $0$ and $1$. This yields the
following values for $P[A_t=-h_t]$,

\noindent
for $l<0.5$:
\begin{align}
\label{eqn:elcprobs-3b}
	\pbar&<\opt[\pbar]: & 
	  P[A_t=-h_t]=\begin{cases} 0 & \f h_t=-1 \\ 1 &\f h_t=+1\end{cases}
	     \notag\\
	\pbar&>\opt[\pbar]: & 
	  P[A_t=-h_t]=\begin{cases} 1 & \f h_t=-1 \\ 1 &\f h_t=+1\end{cases}
\intertext{for $l>0.5$:}
\label{eqn:elcprobs_b-3b}
	\pbar&<\opt[\pbar]: & 
	  P[A_t=-h_t]=\begin{cases} 1 & \f h_t=-1 \\ 0 &\f h_t=+1\end{cases}
	     \notag\\
	\pbar&>\opt[\pbar]: & 
	  P[A_t=-h_t]=\begin{cases} 1 & \f h_t=-1 \\ 1 &\f h_t=+1\end{cases}
\end{align}
As we saw in Sec. \ref{sec:triggers-3b} it is sequences of $T$ time steps in
which $A_t=-h_t$ or $A_t=+h_t$ which act as the triggers for ELC in the full
GCGM. 
We can see from Eqs. \eqref{eqn:elcprobs-3b} and \eqref{eqn:elcprobs_b-3b} that
$A_t=-h_t$ with probability $P=1$ when $\pbar>\opt[\pbar]$. Thus,
$\pbar>\opt[\pbar]$ leads to the sequences of time steps which have the same
effect on the \zero and \one agents as the sequences of time steps in which
$A_t=-1$ that we saw in the memoryless GCGM. The situation when
$\pbar<\opt[\pbar]$ is more complicated. We can see from the above expressions
that for $l<0.5$, $A_t=-1$ while for $l>0.5$, $A_t=+1$. Thus, for
$\pbar<\opt[\pbar]$ we expect sequences of time steps in which $A_t=-1$ and
$A_t=+1$ respectively. However, it is not immediately apparent that $A_t=+h_t$
as we might expect.

By application of the same Markovian analysis that we used in Sec.
\ref{sec:acmarkov-3a} to the $m=1$ case, we can derive the state transition
diagrams given in Fig. \ref{fig:elctrans-3b}. The state labels are as defined
by Fig. \ref{fig:allstates-3a}. We can see from Fig. \ref{fig:elctrans-3b} that
when $\pbar<\opt[\pbar]$ the transition diagrams each contain two attractor
states in which $h_t=-1$ and $+1$ for $l<0.5$ and $l>0.5$ respectively. Thus,
when the model is in these states $P[A_t=-h_t]=0$ and therefore $A_t=+h_t$. 
The reason that they can be divided into two congruent sub-diagrams is
that in each case, as we remarked above, $A_t$ only takes a single value.
Therefore, the value of the memory bit that corresponds to the opposite value of
$A_t$ has no significance. This leads to the two-fold state degeneracy that we
observe; states which only differ in the value of this attribute are equivalent.
Another feature to note in Fig.  \ref{fig:elctrans-3b} is that, depending on the
state that the model is in when it changes from $\pbar>\opt[\pbar]$ to
$\pbar<\opt[\pbar]$, it may take several time steps to reach the attractor.
Thus, unlike in the case of the memoryless GCGM, we should not expect
oscillations with period $T$. There will likely be interjected time steps while
the model finds the attractor.

We can see that this analysis will apply to the case of general $m$ by
considering Eq. \eqref{eqn:elcprobs-3b}. For $\pbar<\opt[\pbar]$, $A_t=-1$ or
$+1$ consistently. At the first time step after the activation and deactivation
of \one and \zero agents the history will contain a mixture of $-1$s and
$+1$s. However, it is clear that after $m+1$ time steps it will contain only
$-1$s and $+1$s for $l<0.5$ and $l>0.5$ respectively and the
memory bit corresponding to this history will also take the same value. These
states in which the history is $\{-1,-1,\ldots,-1\}$ and $h_t=-1$ or
$\{+1,+1,\ldots,+1\}$ and $h_t=+1$ correspond to the attractor states in Fig.
\ref{fig:elctrans-3b}.

\floatfig[\floatplace]{fig:elctrans-3b}{
Markovian transition diagrams for ELC in the full GCGM.
}{
Markovian transition diagrams for ELC in the full GCGM. The state labels are
as defined by Fig. \ref{fig:allstates-3a} and $\pm$ signs give the value
of $h_t$ in each state.
}{
\begin{center}
\resizebox{.476\textwidth}{!}{\small%
\begin{picture}(0,0)%
\includegraphics{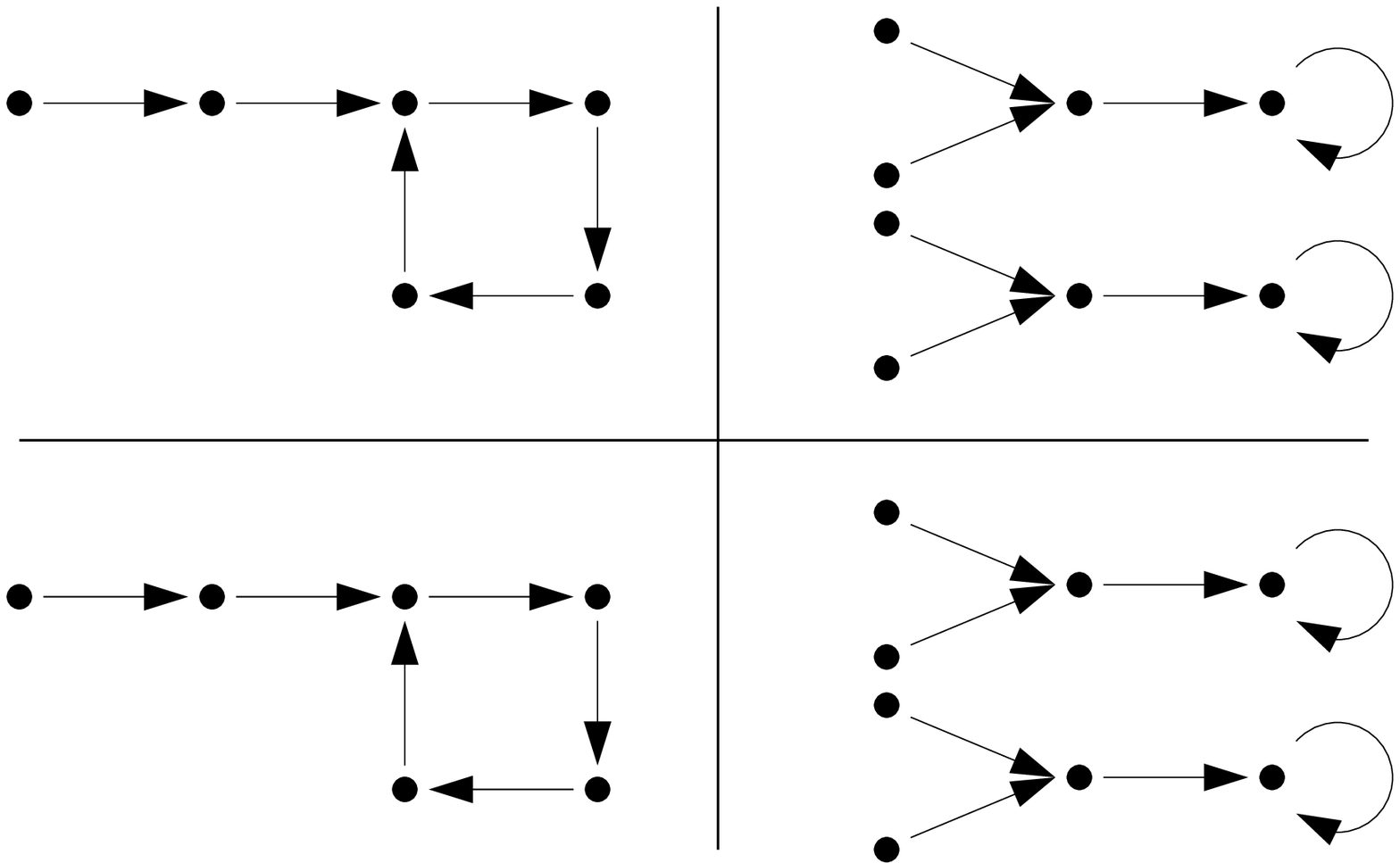}%
\end{picture}%
\setlength{\unitlength}{3947sp}%
\begingroup\makeatletter\ifx\SetFigFont\undefined%
\gdef\SetFigFont#1#2#3#4#5{%
  \fontsize{#1}{#2pt}%
  \fontfamily{#3}\fontseries{#4}\fontshape{#5}%
  \selectfont}%
\fi\endgroup%
\begin{picture}(9368,5957)(391,-5094)
\put(6376,-2986){\makebox(0,0)[b]{\smash{\SetFigFont{20}{24.0}{\familydefault}{\mddefault}{\updefault}{\color[rgb]{0,0,0}4}%
}}}
\put(6376,-3886){\makebox(0,0)[b]{\smash{\SetFigFont{20}{24.0}{\familydefault}{\mddefault}{\updefault}{\color[rgb]{0,0,0}5}%
}}}
\put(7801,-3736){\makebox(0,0)[b]{\smash{\SetFigFont{20}{24.0}{\familydefault}{\mddefault}{\updefault}{\color[rgb]{0,0,0}1}%
}}}
\put(9001,-3736){\makebox(0,0)[b]{\smash{\SetFigFont{20}{24.0}{\familydefault}{\mddefault}{\updefault}{\color[rgb]{0,0,0}0}%
}}}
\put(6376,-4186){\makebox(0,0)[b]{\smash{\SetFigFont{20}{24.0}{\familydefault}{\mddefault}{\updefault}{\color[rgb]{0,0,0}12}%
}}}
\put(6376,-5086){\makebox(0,0)[b]{\smash{\SetFigFont{20}{24.0}{\familydefault}{\mddefault}{\updefault}{\color[rgb]{0,0,0}13}%
}}}
\put(7801,-4936){\makebox(0,0)[b]{\smash{\SetFigFont{20}{24.0}{\familydefault}{\mddefault}{\updefault}{\color[rgb]{0,0,0}9}%
}}}
\put(9001,-4936){\makebox(0,0)[b]{\smash{\SetFigFont{20}{24.0}{\familydefault}{\mddefault}{\updefault}{\color[rgb]{0,0,0}8}%
}}}
\put(6376, 14){\makebox(0,0)[b]{\smash{\SetFigFont{20}{24.0}{\familydefault}{\mddefault}{\updefault}{\color[rgb]{0,0,0}2}%
}}}
\put(6376,-886){\makebox(0,0)[b]{\smash{\SetFigFont{20}{24.0}{\familydefault}{\mddefault}{\updefault}{\color[rgb]{0,0,0}3}%
}}}
\put(7801,-736){\makebox(0,0)[b]{\smash{\SetFigFont{20}{24.0}{\familydefault}{\mddefault}{\updefault}{\color[rgb]{0,0,0}10}%
}}}
\put(9001,-736){\makebox(0,0)[b]{\smash{\SetFigFont{20}{24.0}{\familydefault}{\mddefault}{\updefault}{\color[rgb]{0,0,0}11}%
}}}
\put(6376,-1186){\makebox(0,0)[b]{\smash{\SetFigFont{20}{24.0}{\familydefault}{\mddefault}{\updefault}{\color[rgb]{0,0,0}6}%
}}}
\put(6376,-2086){\makebox(0,0)[b]{\smash{\SetFigFont{20}{24.0}{\familydefault}{\mddefault}{\updefault}{\color[rgb]{0,0,0}7}%
}}}
\put(7801,-1936){\makebox(0,0)[b]{\smash{\SetFigFont{20}{24.0}{\familydefault}{\mddefault}{\updefault}{\color[rgb]{0,0,0}14}%
}}}
\put(9001,-1936){\makebox(0,0)[b]{\smash{\SetFigFont{20}{24.0}{\familydefault}{\mddefault}{\updefault}{\color[rgb]{0,0,0}15}%
}}}
\put(751,-1261){\rotatebox{90.0}{\makebox(0,0)[b]{\smash{\SetFigFont{25}{30.0}{\familydefault}{\mddefault}{\updefault}{\color[rgb]{0,0,0}$l>0.5$}%
}}}}
\put(8176,-4036){\makebox(0,0)[b]{\smash{\SetFigFont{17}{20.4}{\familydefault}{\mddefault}{\updefault}{\color[rgb]{0,0,0}$\opt[\pbar]=1-l$}%
}}}
\put(751,-3961){\rotatebox{90.0}{\makebox(0,0)[b]{\smash{\SetFigFont{25}{30.0}{\familydefault}{\mddefault}{\updefault}{\color[rgb]{0,0,0}$l<0.5$}%
}}}}
\put(4801,-5011){\makebox(0,0)[b]{\smash{\SetFigFont{20}{24.0}{\familydefault}{\mddefault}{\updefault}{\color[rgb]{0,0,0}1}%
}}}
\put(3601,-5011){\makebox(0,0)[b]{\smash{\SetFigFont{20}{24.0}{\familydefault}{\mddefault}{\updefault}{\color[rgb]{0,0,0}6}%
}}}
\put(4801,-3286){\makebox(0,0)[b]{\smash{\SetFigFont{20}{24.0}{\familydefault}{\mddefault}{\updefault}{\color[rgb]{0,0,0}5}%
}}}
\put(3601,-3286){\makebox(0,0)[b]{\smash{\SetFigFont{20}{24.0}{\familydefault}{\mddefault}{\updefault}{\color[rgb]{0,0,0}14}%
}}}
\put(2401,-3286){\makebox(0,0)[b]{\smash{\SetFigFont{20}{24.0}{\familydefault}{\mddefault}{\updefault}{\color[rgb]{0,0,0}9}%
}}}
\put(1201,-3286){\makebox(0,0)[b]{\smash{\SetFigFont{20}{24.0}{\familydefault}{\mddefault}{\updefault}{\color[rgb]{0,0,0}13}%
}}}
\put(2401,-4111){\makebox(0,0)[b]{\smash{\SetFigFont{17}{20.4}{\familydefault}{\mddefault}{\updefault}{\color[rgb]{0,0,0}$\opt[\pbar]=1-l$}%
}}}
\put(4801,-1936){\makebox(0,0)[b]{\smash{\SetFigFont{20}{24.0}{\familydefault}{\mddefault}{\updefault}{\color[rgb]{0,0,0}1}%
}}}
\put(3601,-1936){\makebox(0,0)[b]{\smash{\SetFigFont{20}{24.0}{\familydefault}{\mddefault}{\updefault}{\color[rgb]{0,0,0}6}%
}}}
\put(4801,-211){\makebox(0,0)[b]{\smash{\SetFigFont{20}{24.0}{\familydefault}{\mddefault}{\updefault}{\color[rgb]{0,0,0}5}%
}}}
\put(3601,-211){\makebox(0,0)[b]{\smash{\SetFigFont{20}{24.0}{\familydefault}{\mddefault}{\updefault}{\color[rgb]{0,0,0}14}%
}}}
\put(2401,-211){\makebox(0,0)[b]{\smash{\SetFigFont{20}{24.0}{\familydefault}{\mddefault}{\updefault}{\color[rgb]{0,0,0}9}%
}}}
\put(1201,-211){\makebox(0,0)[b]{\smash{\SetFigFont{20}{24.0}{\familydefault}{\mddefault}{\updefault}{\color[rgb]{0,0,0}13}%
}}}
\put(8176,-1036){\makebox(0,0)[b]{\smash{\SetFigFont{17}{20.4}{\familydefault}{\mddefault}{\updefault}{\color[rgb]{0,0,0}$\opt[\pbar]=l$}%
}}}
\put(1201,-3811){\makebox(0,0)[b]{\smash{\SetFigFont{20}{24.0}{\familydefault}{\mddefault}{\updefault}{\color[rgb]{0,0,0}$+$}%
}}}
\put(2401,-3811){\makebox(0,0)[b]{\smash{\SetFigFont{20}{24.0}{\familydefault}{\mddefault}{\updefault}{\color[rgb]{0,0,0}$+$}%
}}}
\put(3901,-3811){\makebox(0,0)[b]{\smash{\SetFigFont{20}{24.0}{\familydefault}{\mddefault}{\updefault}{\color[rgb]{0,0,0}$-$}%
}}}
\put(4501,-3811){\makebox(0,0)[b]{\smash{\SetFigFont{20}{24.0}{\familydefault}{\mddefault}{\updefault}{\color[rgb]{0,0,0}$+$}%
}}}
\put(4501,-4486){\makebox(0,0)[b]{\smash{\SetFigFont{20}{24.0}{\familydefault}{\mddefault}{\updefault}{\color[rgb]{0,0,0}$+$}%
}}}
\put(3901,-4486){\makebox(0,0)[b]{\smash{\SetFigFont{20}{24.0}{\familydefault}{\mddefault}{\updefault}{\color[rgb]{0,0,0}$-$}%
}}}
\put(1201,-736){\makebox(0,0)[b]{\smash{\SetFigFont{20}{24.0}{\familydefault}{\mddefault}{\updefault}{\color[rgb]{0,0,0}$+$}%
}}}
\put(2401,-736){\makebox(0,0)[b]{\smash{\SetFigFont{20}{24.0}{\familydefault}{\mddefault}{\updefault}{\color[rgb]{0,0,0}$+$}%
}}}
\put(3901,-736){\makebox(0,0)[b]{\smash{\SetFigFont{20}{24.0}{\familydefault}{\mddefault}{\updefault}{\color[rgb]{0,0,0}$-$}%
}}}
\put(4501,-736){\makebox(0,0)[b]{\smash{\SetFigFont{20}{24.0}{\familydefault}{\mddefault}{\updefault}{\color[rgb]{0,0,0}$+$}%
}}}
\put(4501,-1411){\makebox(0,0)[b]{\smash{\SetFigFont{20}{24.0}{\familydefault}{\mddefault}{\updefault}{\color[rgb]{0,0,0}$+$}%
}}}
\put(3901,-1411){\makebox(0,0)[b]{\smash{\SetFigFont{20}{24.0}{\familydefault}{\mddefault}{\updefault}{\color[rgb]{0,0,0}$-$}%
}}}
\put(9301,-3436){\makebox(0,0)[b]{\smash{\SetFigFont{20}{24.0}{\familydefault}{\mddefault}{\updefault}{\color[rgb]{0,0,0}$-$}%
}}}
\put(7801,-3211){\makebox(0,0)[b]{\smash{\SetFigFont{20}{24.0}{\familydefault}{\mddefault}{\updefault}{\color[rgb]{0,0,0}$+$}%
}}}
\put(6601,-3211){\makebox(0,0)[b]{\smash{\SetFigFont{20}{24.0}{\familydefault}{\mddefault}{\updefault}{\color[rgb]{0,0,0}$-$}%
}}}
\put(6601,-3661){\makebox(0,0)[b]{\smash{\SetFigFont{20}{24.0}{\familydefault}{\mddefault}{\updefault}{\color[rgb]{0,0,0}$+$}%
}}}
\put(6601,-4861){\makebox(0,0)[b]{\smash{\SetFigFont{20}{24.0}{\familydefault}{\mddefault}{\updefault}{\color[rgb]{0,0,0}$+$}%
}}}
\put(6601,-4411){\makebox(0,0)[b]{\smash{\SetFigFont{20}{24.0}{\familydefault}{\mddefault}{\updefault}{\color[rgb]{0,0,0}$-$}%
}}}
\put(7801,-4411){\makebox(0,0)[b]{\smash{\SetFigFont{20}{24.0}{\familydefault}{\mddefault}{\updefault}{\color[rgb]{0,0,0}$+$}%
}}}
\put(9301,-4636){\makebox(0,0)[b]{\smash{\SetFigFont{20}{24.0}{\familydefault}{\mddefault}{\updefault}{\color[rgb]{0,0,0}$-$}%
}}}
\put(9301,-436){\makebox(0,0)[b]{\smash{\SetFigFont{20}{24.0}{\familydefault}{\mddefault}{\updefault}{\color[rgb]{0,0,0}$+$}%
}}}
\put(7801,-211){\makebox(0,0)[b]{\smash{\SetFigFont{20}{24.0}{\familydefault}{\mddefault}{\updefault}{\color[rgb]{0,0,0}$-$}%
}}}
\put(6601,-211){\makebox(0,0)[b]{\smash{\SetFigFont{20}{24.0}{\familydefault}{\mddefault}{\updefault}{\color[rgb]{0,0,0}$-$}%
}}}
\put(6601,-661){\makebox(0,0)[b]{\smash{\SetFigFont{20}{24.0}{\familydefault}{\mddefault}{\updefault}{\color[rgb]{0,0,0}$+$}%
}}}
\put(6601,-1861){\makebox(0,0)[b]{\smash{\SetFigFont{20}{24.0}{\familydefault}{\mddefault}{\updefault}{\color[rgb]{0,0,0}$+$}%
}}}
\put(6601,-1411){\makebox(0,0)[b]{\smash{\SetFigFont{20}{24.0}{\familydefault}{\mddefault}{\updefault}{\color[rgb]{0,0,0}$-$}%
}}}
\put(7801,-1411){\makebox(0,0)[b]{\smash{\SetFigFont{20}{24.0}{\familydefault}{\mddefault}{\updefault}{\color[rgb]{0,0,0}$-$}%
}}}
\put(9301,-1636){\makebox(0,0)[b]{\smash{\SetFigFont{20}{24.0}{\familydefault}{\mddefault}{\updefault}{\color[rgb]{0,0,0}$+$}%
}}}
\put(2401,-1036){\makebox(0,0)[b]{\smash{\SetFigFont{17}{20.4}{\familydefault}{\mddefault}{\updefault}{\color[rgb]{0,0,0}$\opt[\pbar]=l$}%
}}}
\put(3001,539){\makebox(0,0)[b]{\smash{\SetFigFont{25}{30.0}{\familydefault}{\mddefault}{\updefault}{\color[rgb]{0,0,0}$\pbar>\opt[\pbar]$}%
}}}
\put(7951,539){\makebox(0,0)[b]{\smash{\SetFigFont{25}{30.0}{\familydefault}{\mddefault}{\updefault}{\color[rgb]{0,0,0}$\pbar<\opt[\pbar]$}%
}}}
\end{picture}
}
\end{center}
}

\subsubsection{Example ELC}

Now we shall look at a numerical example. Figure \ref{fig:examplethree-3b}
gives the values of the global action $A_t$, the prediction $h_t$ and the
prediction performance $\eta_t$. Once again we have also included the virtual
points $\ind v$ of an agent $i$ which is inactive and has $\ind p=1.0$ at the
beginning of the time period shown. Figure \ref{fig:examplethree-3b} is
equivalent to Figs. \ref{fig:exampleone-3b} and \ref{fig:exampletwo-3b} except
that we have additionally included the value of the prediction $h_t$. 
Figure \ref{fig:examplethree_graphs-3b} shows \lp, $\eta_t$ and $P[A_t=-h_t]$
given by Eq. \eqref{eqn:reffolprobs-3b}. The paragraph labels below
correspond to the labels in Fig. \ref{fig:examplethree-3b}.

\paragraph*{\bf a} 
Period {\bf a}, in which $A_t=+h_t$, provides the trigger which causes the
activation and deactivation of \zero and \one agents.

\paragraph*{\bf b,c} 
The (de)activation that is the result of {\bf
a} causes a step increase in \lp\ as we expect. However, the magnitude of \lp\
during {\bf b} and {\bf c} is not sufficiently large that $P[A_t=-h_t]=1$ when
$h_t=+1$.  In fact, as we see from Fig. \ref{fig:examplethree_graphs-3b}a,
$P[A_t=-h_t]\approx0.9$ when $h_t=+1$. Thus we can see that in this regard the
realization of an ELC described here is not ideal in the sense discussed in the
previous section. It is because $P[A_t=-h_t]\ne1$ that for the second time
step of {\bf b} $A_t=+h_t$ resulting in these two interjected time steps.
Throughout period {\bf c} $A_t=-h_t$ as expected. 

\paragraph*{\bf d,e,f}  
The interjected period {\bf d} corresponds to the model finding one of the
attractor states in which $A_t=+h_t$.
During period {\bf e}, $A_t=+h_t$ once again. However, after the reactivation and
deactivation at the end of {\bf e} $\lp\approx0$. Therefore, by chance
$A_t=+h_t$ for the next three time steps as well, resulting in the interjected
period {\bf f}. This extra long period in which $A_t=+h_t$ allows some \zero and
\one agents who had failed to (de)activate during the $T$ time steps {\bf e} to
do so. Thus, $\lp$ increases and so therefore does $P[A_t=-h_t]$ (for $h=+1$).

\paragraph*{\bf g,h,i} 
During {\bf g},
$A_t=-h_t$ despite the fact that $P[A_t=-h_t]\approx0.7$ (for $h=+1$). {\bf h}
once again corresponds the model finding the attractor state. {\bf i} represents
the final period in which $A_t=+h_t$ before $\lp$ returns to approximately the
equilibrium value and the ELC comes to an end. 

\medskip

Note that the periods like {\bf c} and {\bf g}, in which the magnitude of $\lp$
is not sufficiently large that $P[A_t=-h_t]=1$ when $h_t=+1$, occur with a much
greater probability that they do in the memoryless model. The reason for this
is that $P[A_t=-h_t]=1$ when $h_t=-1$, unless the magnitude of the oscillations
in \pbar\ is so great that $\lm\approx0$. Therefore, for any time steps during
periods like {\bf c} and {\bf g} for which $h_t=-1$, $A_t=-h_t$ with probability
$P=1$. We can see this clearly in the plot of $P[A_t=-h_t]$ in Fig.
\ref{fig:examplethree_graphs-3b}.

\bigfloatfig[\floatplace]{fig:examplethree-3b}{
}{
The prediction performance $\eta_t$, $A_t$ and the virtual
points $\ind v$ of an agent $i$ which initially is inactive and has a gene
values of $\ind p=1.0$. The model is the \emph{full} GCGM with $n=500$,
$l=0.5$, $T=12$, $r=0.2$ and $m=3$.
The symbols in the margins denote the positions of the step change of \lp\
in Fig. \ref{fig:exampletwo_graphs-3b}.
}{
\begin{center}
\resizebox{.70\textwidth}{!}{\small%

}
\end{center}
}
\floatfig[\floatplace]{fig:examplethree_graphs-3b}{
}{
Various important quantities plotted over the time period corresponding to Fig.
\ref{fig:exampletwo-3b}:
a) The prediction performance $\eta_t$ (measured against the left hand axis) and
the probability $P[A_t=-1]$ that the global action is $-1$ (measured against the
right hand axis). b) The deviation of \pbar\ from $l$, \lp.
}{
  \resizebox{.47\textwidth}{!}{\small%
\begingroup%
  \makeatletter%
  \newcommand{\GNUPLOTspecial}{%
    \catcode`\%=14\relax\special}%
  \setlength{\unitlength}{0.1bp}%
\begin{picture}(3600,1296)(0,0)%
\includegraphics{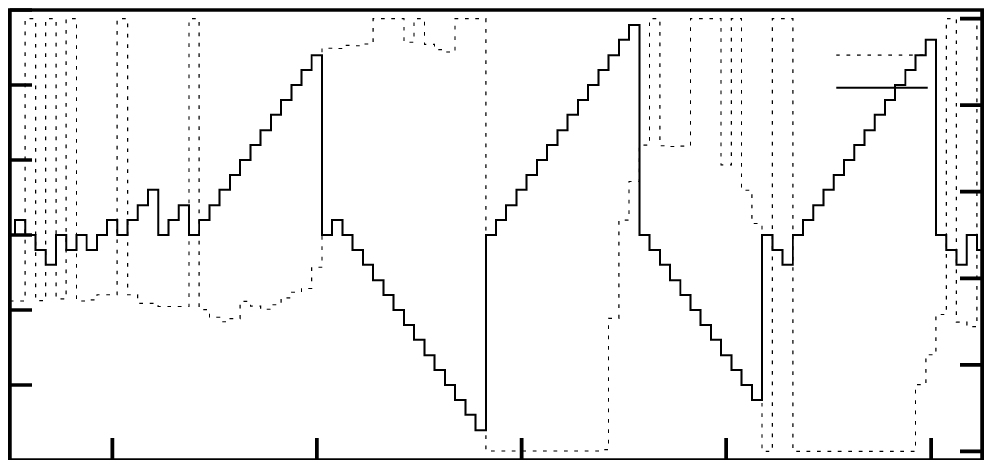}
\put(2730,1072){\makebox(0,0)[r]{$\eta_t$}}%
\put(2730,1166){\makebox(0,0)[r]{$P[A_t=-h_t]$}}%
\put(484,1166){\makebox(0,0)[l]{a)}}%
\put(3599,648){%
\makebox(0,0)[b]{\shortstack{$P[A_t=-h_t]$}}%
}%
\put(100,648){%
\makebox(0,0)[b]{\shortstack{$\eta_t$}}%
}%
\put(3250,1271){\makebox(0,0)[l]{1}}%
\put(3250,1022){\makebox(0,0)[l]{0.8}}%
\put(3250,773){\makebox(0,0)[l]{0.6}}%
\put(3250,523){\makebox(0,0)[l]{0.4}}%
\put(3250,274){\makebox(0,0)[l]{0.2}}%
\put(3250,25){\makebox(0,0)[l]{0}}%
\put(350,1296){\makebox(0,0)[r]{15}}%
\put(350,1080){\makebox(0,0)[r]{10}}%
\put(350,864){\makebox(0,0)[r]{5}}%
\put(350,648){\makebox(0,0)[r]{0}}%
\put(350,432){\makebox(0,0)[r]{-5}}%
\put(350,216){\makebox(0,0)[r]{-10}}%
\end{picture}%
\endgroup

}\\%
  \vspace{-1pt}%
  \resizebox{.47\textwidth}{!}{\small%
\begingroup%
  \makeatletter%
  \newcommand{\GNUPLOTspecial}{%
    \catcode`\%=14\relax\special}%
  \setlength{\unitlength}{0.1bp}%
\begin{picture}(3600,1080)(0,0)%
\includegraphics{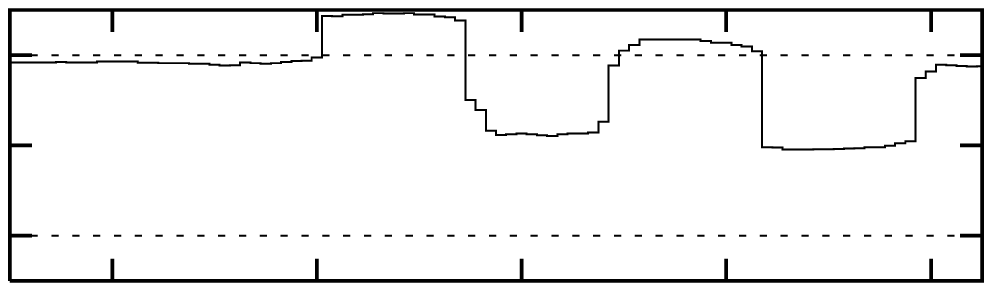}
\put(484,1002){\makebox(0,0)[l]{b)}}%
\put(1800,50){\makebox(0,0){time step: $t$}}%
\put(100,690){%
\makebox(0,0)[b]{\shortstack{\lp}}%
}%
\put(3053,200){\makebox(0,0){691040}}%
\put(2463,200){\makebox(0,0){691020}}%
\put(1874,200){\makebox(0,0){691000}}%
\put(1284,200){\makebox(0,0){690980}}%
\put(695,200){\makebox(0,0){690960}}%
\put(350,950){\makebox(0,0)[r]{0.00}}%
\put(350,690){\makebox(0,0)[r]{-0.10}}%
\put(350,430){\makebox(0,0)[r]{-0.20}}%
\end{picture}%
\endgroup

}%
}

\floatfig[\floatplace]{fig:examplethree_vol-3b}{
Evolution of the volume and the price during the ELC.
}{
Evolution of the number of active agents (also known as the \emph{volume}) and
the price (defined by Eq. \eqref{eqn:price-3b}) over the same time period
described by Figs. \ref{fig:examplethree-3b} and
\ref{fig:examplethree_graphs-3b}.  
}{
  \resizebox{.47\textwidth}{!}{\small%
\begingroup%
  \makeatletter%
  \newcommand{\GNUPLOTspecial}{%
    \catcode`\%=14\relax\special}%
  \setlength{\unitlength}{0.1bp}%
\begin{picture}(3600,1188)(0,0)%
\includegraphics{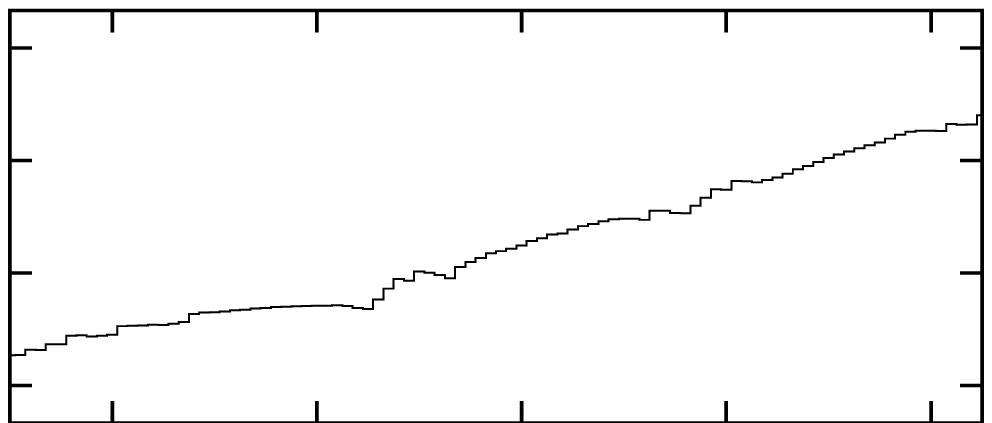}
\put(484,1069){\makebox(0,0)[l]{a)}}%
\put(100,594){%
\makebox(0,0)[b]{\shortstack{Price: $\pi_t$/100}}%
}%
\put(350,1080){\makebox(0,0)[r]{3375}}%
\put(350,756){\makebox(0,0)[r]{3360}}%
\put(350,432){\makebox(0,0)[r]{3345}}%
\put(350,108){\makebox(0,0)[r]{3330}}%
\end{picture}%
\endgroup

}%
    \\%
  \vspace{-1pt}%
  \resizebox{.47\textwidth}{!}{\small%
\begingroup%
  \makeatletter%
  \newcommand{\GNUPLOTspecial}{%
    \catcode`\%=14\relax\special}%
  \setlength{\unitlength}{0.1bp}%
\begin{picture}(3600,1188)(0,0)%
\includegraphics{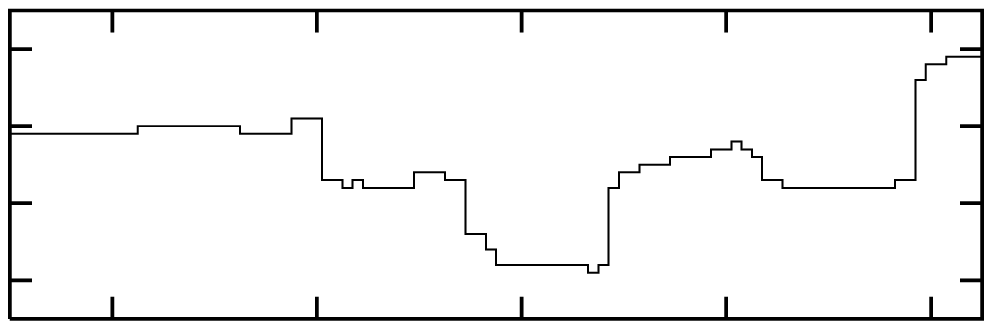}
\put(484,1099){\makebox(0,0)[l]{b)}}%
\put(1800,50){\makebox(0,0){time step: t}}%
\put(100,744){%
\makebox(0,0)[b]{\shortstack{No. active agents: $n_t$}}%
}%
\put(3053,200){\makebox(0,0){691040}}%
\put(2463,200){\makebox(0,0){691020}}%
\put(1874,200){\makebox(0,0){691000}}%
\put(1284,200){\makebox(0,0){690980}}%
\put(695,200){\makebox(0,0){690960}}%
\put(350,1077){\makebox(0,0)[r]{220}}%
\put(350,855){\makebox(0,0)[r]{210}}%
\put(350,633){\makebox(0,0)[r]{200}}%
\put(350,411){\makebox(0,0)[r]{190}}%
\end{picture}%
\endgroup

}%
}

In Fig. \ref{fig:examplethree_vol-3b} we show once again the volume $n_t$
and the price $\pi_t$ over the same period described by Figs. 
\ref{fig:examplethree-3b} and \ref{fig:examplethree_graphs-3b}. In Fig.
\ref{fig:examplethree_vol-3b}a we can clearly see the effect of the
discussion in Sec. \ref{sec:pricets-3b}. Each time that $h_t=-1$ the price 
rises with probability $P=1$. Furthermore the result of the deviation of the
equilibrium value of $\pbar$ from $\opt[\pbar]$ (see Sec. \ref{sec:hbar-3a})
is that it is more probable that the price will rise rather than fall when
$h_t=+1$.  These two effects ensure that in equilibrium (between ELC) the price
$\pi_t$ is an increasing function of time. We can see in Fig.
\ref{fig:examplethree_vol-3b} that one of the effects of the ELC described in
this section is to halt and even briefly reverse this continuous price rise.

The fundamental reason for this behavior is that, as we remarked in
Sec. \ref{ch:gm-memory}, the agents in the GCGM are unable to control directly
whether they will buy or sell at each time step. They can only control the
probability with which they will follow the prediction $h_t$. As we saw
in Sec. \ref{ch:gm-memory} the optimum behavior for the agents is to evolve
such that the excess demand will be zero when $h_t$ takes it's most common 
value ($h_t=+1$ in the case of $l>0.5$). Therefore, in the asymmetric case
of $l\ne0.5$, the magnitude of the excess demand $\Delta$ will be large
when $h_t$ takes the opposite value. This represents the agents mistakenly
believing the prediction which in turn leads to an excess of buyers or sellers
for $l<0.5$ and $l>0.5$ respectively.

Johansen and Sornette have provided evidence\cite{drawdowns,nasdaq} that large
price changes in financial markets are \emph{outliers}. By this it is meant that
the frequency with which large price changes occur cannot be predicted using
the distribution of smaller price changes. From the results presented in this
section, we can see that the large changes that occur in the volume during ELC
are also outliers in this sense. The distribution of volume changes between ELC
is such that changes of the magnitude observed during an ELC occur with a very
small probability. As we saw in Sec. \ref{sec:overview-3b} a different mechanism
(I.e. that of the susceptibility of $P(p)$ and the occurrence of triggers in
$A_t$) is responsible of the occurrence of ELC, which therefore occur with a
much greater probability.

\subsection{Summary of ELC Characteristics}

We have seen how ELC can occur in the GCGM as a result of the
susceptibility of the gene value distribution $P(p)$ to triggers in the global
action time series $A_t$. We also saw that between ELC the self-segregation of
the agents increases the susceptibility of $P(p)$ while this process is reversed
during an ELC. Furthermore we saw that an ELC in the the memoryless GCGM leads
to an approximately periodic oscillation in a derived price time series. In
contrast the price time series in the full GCGM is a divergent quantity
resulting from the inability of the agents to control \fol\ directly. In both
models ELC lead to approximately periodic volume $n_t$ oscillations.

\section{Conclusion and Future Directions}

In conclusion, we have presented a detailed discussion of the role played by
memory, and the nature of self-induced shocks, in an evolutionary population
competing for limited resources. We have left open several interesting questions
that we hope will be addressed by future work. First of all in Sec.
\ref{ch:large-changes} we have only considered the case in which the death score
$D$ is less than the confidence interval $T$. The result of this is that agents
mutate over a shorter time scale than the period of oscillation of an ELC. We
would expect that values of $D>T$ would lead to ELC that persisted for many more
periods.

We have taken the ratio of the number of points gained by an agent
when $\ind a=+A_t$ to that lost when $\ind a=-A_t$ to be unity.
Hod, Nakar and Burgos \ea\ Refs. \cite{emg-cluster, ceva-mem, ceva-quench}
demonstrated that for values of this \emph{prize-to-fine} ratio $R<1$ the
self-segregation of the gene value distribution observed by \jea\cite{self-seg}
is replaced by clustering behavior. In this case the agents tend to evolve
towards $\ind p=0.5$ in equilibrium. This has implications for the
occurrence of ELC since, as we saw in Sec. \ref{sec:suscep-3b} this clustered
gene value distribution is not susceptible to the trigger sequences that cause
ELC. Therefore we expect that reducing the prize-to-fine ratio would suppress
the occurrence of ELC in the GCGM.

Perhaps most importantly, we showed in Secs. \ref{sec:hbar-3a} and
\ref{sec:dev-3a} that for the original Genetic Model in equilibrium \pbar\
deviates from the optimal value given in Eq. \eqref{eqn:optp-3a}. This
apparently small deviation has some important consequences as it determines the
values taken by $\ave{h_t}_t$ and the magnitude of the autocorrelations observed
in the $h_t$ time series. We have shown that since this effect is not present in
the memoryless Genetic Model it does not result from the finite standard
deviation of the gene value distribution as we might expect but must instead
result from the action of the global memory. We hope that future work will
provide clarification of this point.

\vfill

\begin{acknowledgements}
We are extremely grateful to P.M. Hui and T.S. Lo (Chinese University of Hong
Kong) for detailed discussions, and for sharing their results with us.  RK is
financed by an EPSRC studentship.
\end{acknowledgements}

\end{document}